%% file: 0_main.tex
\documentclass[twoside,leqno,twocolumn]{article}
\usepackage{ltexpprt}
\usepackage{algorithm}
\usepackage{algorithmicx}
\usepackage[compatible]{algpseudocode}
\usepackage{amsmath}
\usepackage{times}
\usepackage[numbers,sort&compress]{natbib}

\usepackage{booktabs} 

\usepackage{graphicx}
\DeclareGraphicsExtensions{.pdf,.png,.eps}
\usepackage{xfrac}
\usepackage{dcolumn}
\usepackage[colorinlistoftodos]{todonotes}
\usepackage[colorlinks=true,linkcolor=blue,citecolor=blue,urlcolor=blue]{hyperref}
\usepackage{bm}
\definecolor{OliveGreen}{cmyk}{0.64,0,0.95,0.40}
\usepackage{subcaption}

\newcommand{\refalg}[1]{Algorithm~\ref{#1}}
\newcommand{\reffig}[1]{Figure~\ref{#1}}

\newcommand{\refeqb}[1]{Equation~\ref{#1}}

\usepackage{newfloat}
\usepackage[size=small]{caption}
\usepackage{etoolbox}
\DeclareCaptionFormat{algorithms}{\vskip-15pt\hrulefill\par#1#2#3\vskip-6pt\hrulefill}
\newcommand{\COMMENTc}[1]{{\scriptsize\color{OliveGreen}\COMMENT{#1}}}

\newcommand{\ie}{i.e. }
\newcommand{\eg}{e.g. }
\newcommand{\aka}{a.k.a. }

\usepackage{lipsum}

\let\OLDthebibliography\thebibliography
\renewcommand\thebibliography[1]{
  \OLDthebibliography{#1}
  \setlength{\parskip}{0pt}
  \setlength{\itemsep}{0pt plus 0.3ex}
}

\begin{document}



\title{\Large Modular Networks for Validating Community Detection Algorithms\vspace{-5pt}
 \thanks{Available at: \href{https://github.com/rabbanyk/FARZ}{https://github.com/rabbanyk/FARZ}}
 }
\author{Justin \textbf{F}agnan, Afra \textbf{A}bnar, Reihaneh \textbf{R}abbany, and Osmar R. \textbf{Z}a\"{\i}ane\\
\normalsize{Department of Computing Science, University of Alberta} \\ 
\small{
\textit{\{fagnan, aabnar, rabbanyk, zaiane\}@ualberta.ca}
}
}





\date{}

\maketitle

\begin{abstract}
How can we accurately compare different community detection algorithms? 
These algorithms cluster nodes in a given network, and their performance is often validated on benchmark networks with explicit ground-truth communities. 
Given the lack of cluster labels in real-world networks, a model that generates realistic networks is required for accurate evaluation of these algorithm. 
In this paper, we present a simple, intuitive, and flexible benchmark generator to generate intrinsically modular networks for community validation. 
We show how the generated networks closely comply with the characteristics
observed for real networks; whereas their characteristics could be directly controlled to match wide range of real world networks. 
We further show how common community detection algorithms rank differently when being evaluated on these benchmarks compared to current available alternatives.

\end{abstract}



\input{1_intro}
\input{2_related}
\section{Generalized 3-Pass Model}
 We generalize and modify the original LFR benchmarks: \emph{1)} to start with any network model, so that it could be plugged in with more realistic network models;  
  and more importantly, \emph{2)} to assign nodes to communities in a more efficient way, so that the resulted assignments require far fewer rewirings, hence keeping the properties of the original network intact.  

The generalized benchmark has three phases: first, realize a network according to network model $\mathcal{M}$, and given parameter set $\theta^{G}$
; second, create communities based on the given parameter set, $\theta^{C}$, and assign the nodes to these communities
; and third, overlay the community structure on the network, to satisfy the constraints given in $\theta^{C}$. 
In the original LFR, the network model is CF hence we have $\mathcal{M}=CF$; whereas $\theta^G= \{N, k_{avg}, k_{max}, \gamma\}$, which are respectively:  number of nodes in the graph, average degree, maximum degree and  exponent of power law degree distribution. 
These parameters are basically used to determine the degree sequence of $G$, from which the graph is then synthesized using the CF model. 
%

\begin{figure}
 \centering
\includegraphics[width=.99\linewidth,height=.9\linewidth]{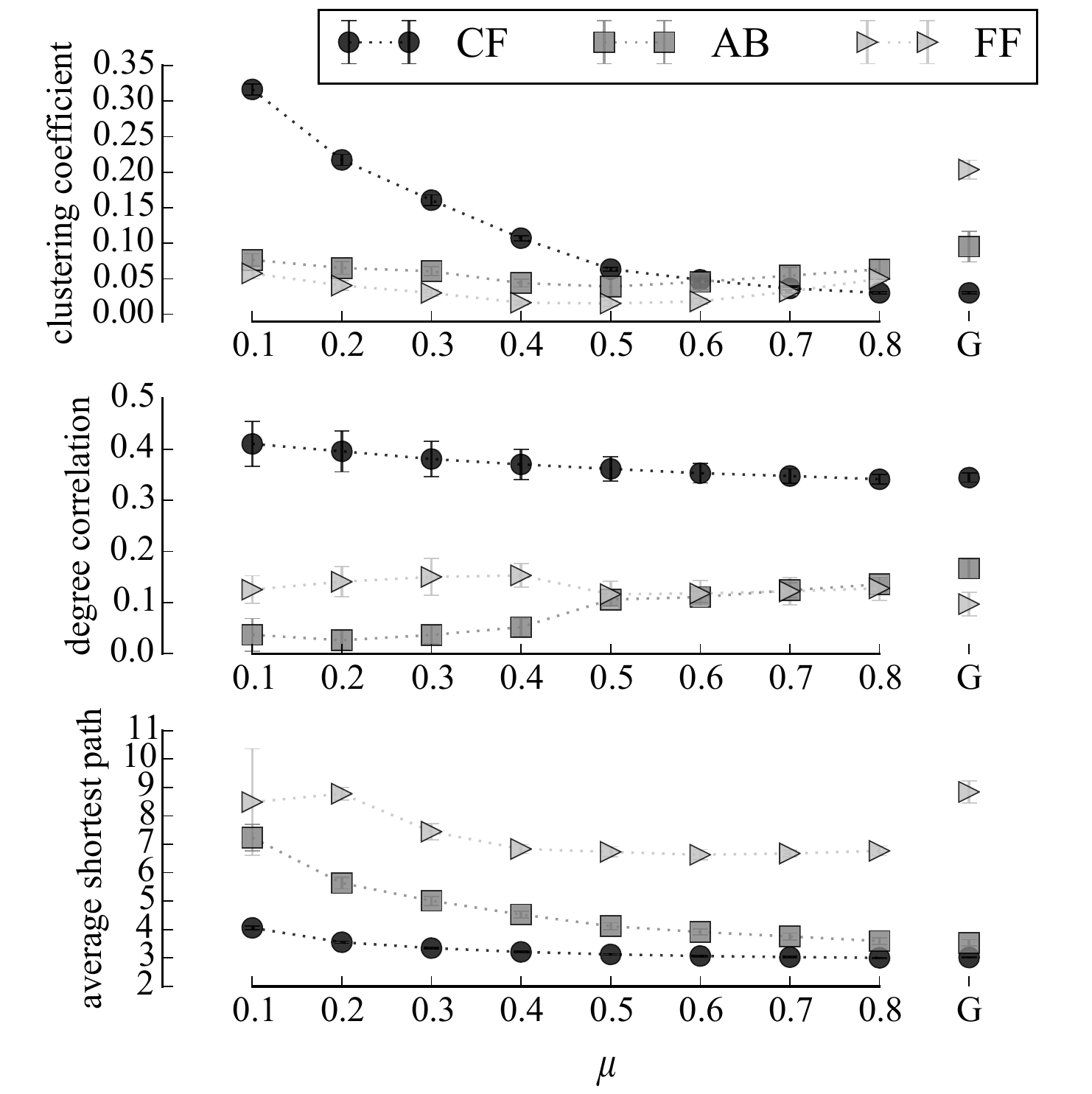}
\caption{Benchmarks created by the generalized 3-pass model using different start network models. 
 {\small Properties of the synthesized networks are plotted as a function of mixing parameter $\mu$.
The properties are also reported for the start network (marked by $G$), i.e. before overlay and rewiring phases.
Results are averaged over 10 simulations. }
} \vspace{-10pt}
\label{fig:propnets}
\end{figure}
\textbf{First}, we substitute the CF model with two well-known network models: the \textsl{\citeauthor{Barabasi99} (BA)} model \cite{Barabasi99} and \textsl{Forest Fire (FF) model} by \citet{Leskovec05ForestFire,leskovec2007graph}.  
 \reffig{fig:propnets} compares three basic properties of the synthesized benchmarks using these alternative network models. 
The properties compared are average clustering coefficient of nodes, degree correlation coefficient (Pearson correlation for degrees of connecting nodes), and the average shortest path distances between nodes. The parameters for CF, BA, and FF models, 
$\theta^{G}$, are respectively $\{N:1000,\, k_{avg}:15,\, k_{max}:50,\, \gamma:3\}$, $\{N:1000,\, m:2\}$, and $\{N:1000,\, p:0.1,\, rp:0.0\}$; which yields initial networks with similar degree distributions. 
%
Here, we can see that the clustering coefficient of the CF model is almost zero for the initial network, and the rewiring actually brings some modularity to the network and increases the average clustering coefficient, but only for small mixing parameters  --when communities are well-separated and many links are rewired to be inside communities. However for larger values of $\mu$ --more interesting problems, there is no clustering in the generated network. This is also true if we start with a network with high clustering coefficient, such as $FF$, as the network structure is extensively changed after the rewiring phase to overlay the communities.
        
 \begin{figure}
\centering
 \includegraphics[width=0.99\linewidth]{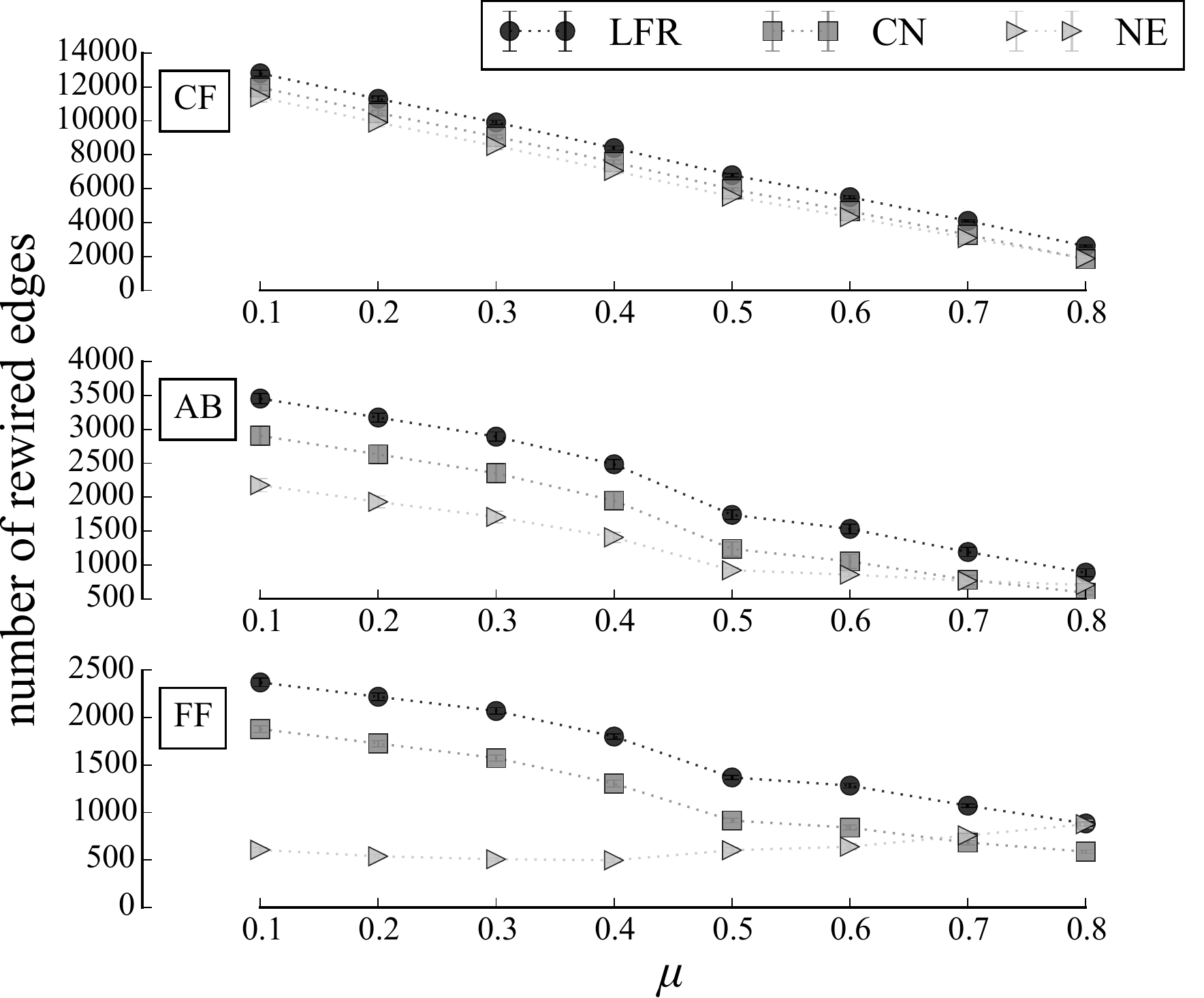}
\caption{Comparing the number of edges rewired using each of the three node assignment variations, i.e. LFR (original), CN (common neighbour), and NE (neighbour expansion). The subplots correspond to different network models.}
\label{fig:rewperC}\vspace{-10pt}
\end{figure}
%
\textbf{Second}, in the original rewiring/overlay phase, the nodes are assigned to communities uniformly at random; here we propose two modified variations that result in far less rewiring in the subsequent overlay procedure. More specifically, we examine \emph{1)} Common neighbour ($CN$) assignment, i.e. probability of joining a community is proportional to the neighbours a node has in that community; \emph{2)} Neighbour expansion ($NE$) assignment, i.e. after assigning a node to a community chosen uniformly at random, also assign all of its neighbours to the same community, continue until the community is full, according to its size predetermined based on $\theta^{C}$. 
\reffig{fig:rewperC} shows the amount of edges rewired using these three node assignment approaches, as a function of mixing parameter $\mu$, i.e. the constraint used in the rewiring/overlay phase. The overall parameters used are $\theta^{C}=\{\mu, \beta:2,\, c_{min}:20,\, c_{max}:50\}$.  The latter three determine the capacity of communities; which are respectively: exponent of power law distribution, minimum, and maximum for community sizes. 

\begin{figure}[]
\centering
 \includegraphics[width=.99 \linewidth]{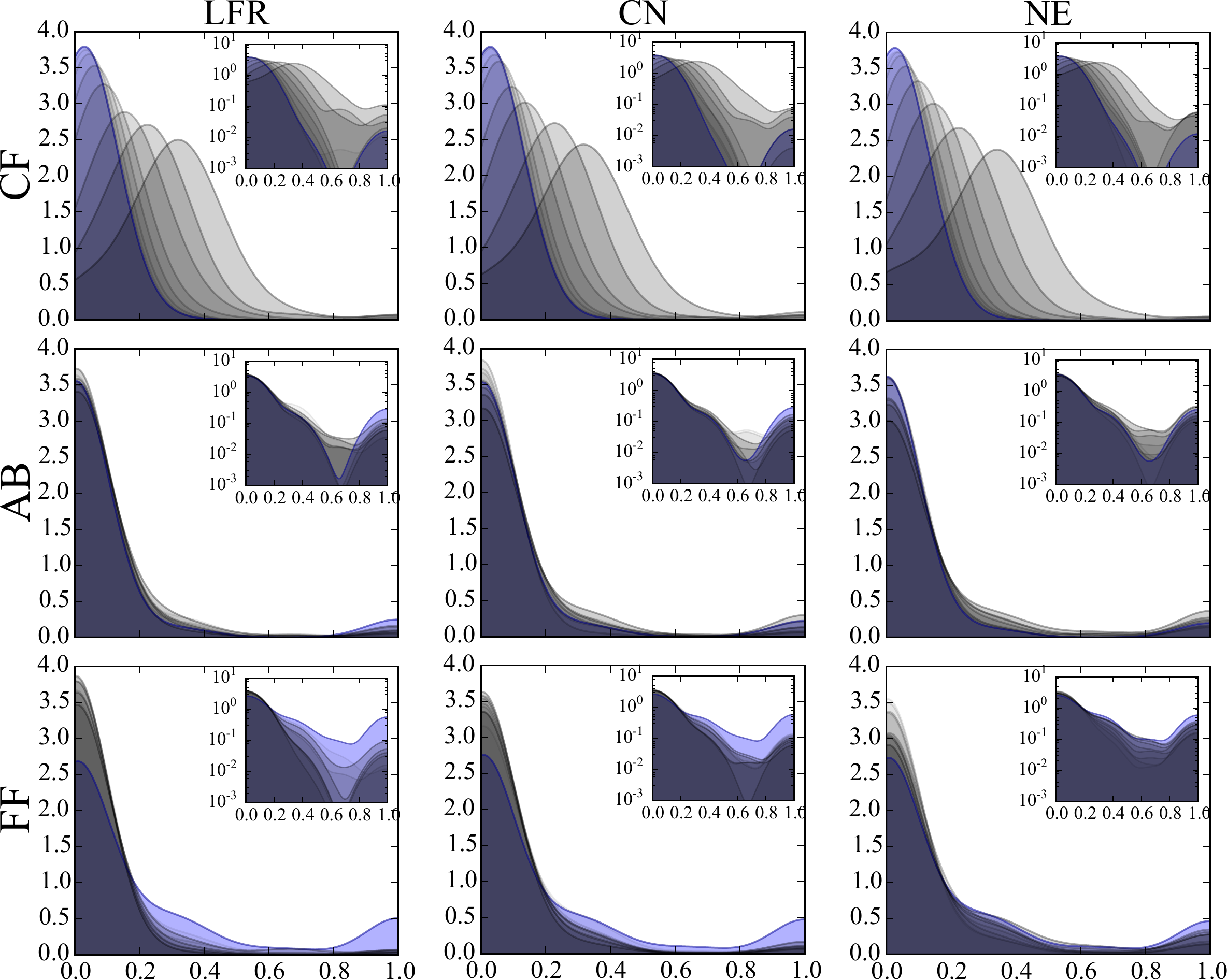}
\caption{The effect of rewiring on the probability density function (pdf) of the \textsl{clustering coefficient}. The blue pdf shows the distribution of clustering coefficients of all nodes in the initial network. The grey pdfs correspond to different values of $\mu$. The insets represent these
 log scale y-axis.}
\label{fig:ccall}
\end{figure}

In the last subplot of \reffig{fig:rewperC}, which shows the amount of rewiring using three assignment variations when the network model is \textsl{FF}, we can see that the NE assignment significantly reduces the amount of rewiring required to reach the constraint $\mu$. This is however only evident if the initial network model is realistic. 
In other words, the improvement over original LFR requires both \textsl{a more realistic model and a better assignment approach}. 
In particular, \reffig{fig:ccall} illustrates the effect of changing $\mu$ on the clustering coefficient of nodes. 
We can see that the distribution of clustering coefficient of nodes is better preserved when the starting network has more clustering ($FF$)
, and the assignment of nodes preserves those clustered nodes ($NE$). 

\begin{figure}[t!]
\includegraphics[width=.99\linewidth,trim={0.0cm 0.0cm 0.0cm 0.0cm},clip]{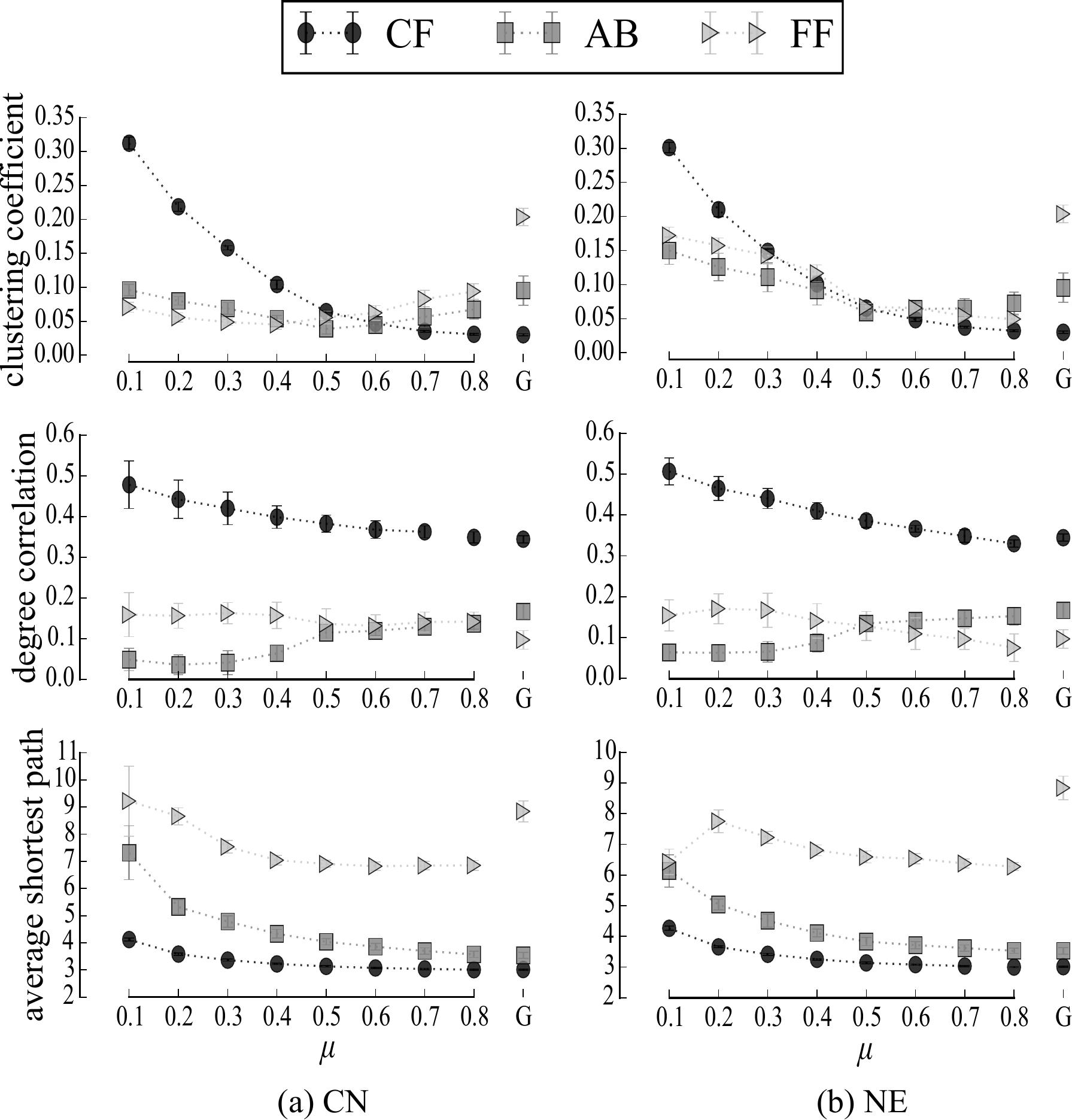}
\caption{Comparing the properties of networks using the two assignment variations, CN on the left and NE on the right; which compares against the \reffig{fig:propnets} that uses the LFR's original assignment approach.}
\label{fig:propnetsCNE}
\vspace{-10pt}
\end{figure}

On par with  \reffig{fig:propnets}, \reffig{fig:propnetsCNE}  compares the 
properties for the synthesized networks when using these two assignment variations.
We can see that the CN better preserves the properties of the network compared to the original LFR. 
For the  clustering coefficient in particular, we see improvement over larger values of $\mu$. However, this does not hold as $\mu$ decreases and rewiring become more intrusive. 
The NE assignment variation performs better, where the clustering coefficient of the original network is preserved and increases as communities become denser, i.e. $\mu$ decreases.   

Although exhibiting more realistic properties, the benchmarks generated by these variations, similar to the original LFR benchmarks,  enforce communities later on the network; which is in contrary to their definition as the natural structure underlying the networks. 
Another issue with the LFR benchmarks is their expressiveness and flexibility. More details on this generalization can be found in \cite{rabbanyThesis}.
Although they have many parameters, these parameters are not relevant in most cases, and almost all works that use LFR in their evaluation, stick to the setting first used when introducing these benchmarks in \cite{Lancichinetti08LFR}.

In the following we present a simple \textsl{alternative generator, called FARZ}. 
Similar to classical network models, FARZ follows a growth patterns, \ie it gradually expands the network following different evolution patterns, while also evolving its modular structure. 

 \input{3_method}

\input{4_exps}

 \section{Conclusion}
In this paper we first discussed the shortcomings of the popular LFR network generator which is widely used for validating and comparing community detection methods, and then introduced extensions to improve upon these shortcomings.  
We showed how these LFR extensions refine the generated networks towards more lifelike networks, while still are suffering from  the restrictive generative process used. 
Therefore, we next introduced a simple and flexible benchmark generator, called FARZ which, similar to LFR, generates networks with built-in community structure, that can be compared, as a ground truth, against the results of different community mining algorithms. FARZ produces truthful networks, in a sense that the characteristics of the networks and communities synthesized by FARZ are similar to what is observed in real world networks. 
FARZ also incorporates intuitive parameters, which have meaningful interpretation and are easy to tune to directly control the properties of the synthesized network.
More precisely, FARZ has three input parameters, $FARZ(n,m,k)$, which respectively determine the number of nodes, the (half of) average degree, and the number of communities. It also has four intuitive control parameters, $\beta$, $\alpha$, $\gamma$, and $\phi$; which respectively control the strength of the community structure, the clustering coefficient, the degree correlation,
and the distribution of the community sizes. In our experiment we showed how tuning these parameters provides means to generate a variety of realistic networks and presents different settings for comparing community detection algorithms. 

\subsection{Brief Review of Related Network Models}
The \textsl{BA model} evolves one node at a time; 
whereas each newly entered node forms $m$ connections with the existing nodes. Connections are formed according to \textit{preferential attachment} (a.k.a. accumulative advantage, Yule process, Matthew effect, or rich get richer), which states that the probability of forming a connection to an existing node is proportional to its degree, i.e. \(p_i = \sfrac{k_i}{\sum_j k_j }\).
The networks generated with this model are analytically shown to have a power law degree distribution
, small average path length,  assortative mixing (of degrees), and transitivity higher than random graphs \cite{albert2002statistical}. 
%
However, the evolution of networks in this model is not realistic \cite{leskovec2008microscopic}. 
 The second model, \textsl{FF}, has similar properties, while it is also designed to  follow the empirically observed evolution trends in real social networks --become denser over time, with the average degree increasing, and the diameter decreasing \cite{Leskovec05ForestFire,leskovec2007graph}. This model grows one node at a time, where every new node, first connects to an existing node called ambassador, chosen uniformly at random. 
Then, the new node recursively forms a random number of connections with the neighbours of every node it connects to --outlinks to specific number of inlink and outlink neighbours, drawn from geometric distributions with means of $p/(1-p)$ and $rp/(1-rp)$ respectively, where $p$/$rp$ is called forward/backward burning probability. 



There also exists a family of generative models which are used to learn the latent parameters of the model given real-world data, and can be then used to simulate similar networks \cite{seshadhri2012community,Gong12evolution,Yang2013,kolda2014scalable,leskovec2010kronecker,akoglu2009rtg}. Between these models, only the Block Two-level Erd\H{o}s-R\'{e}nyi (BTER) \cite{kolda2014scalable} is used for benchmarking. 
Where they present ways to randomly generate the degree and clustering coefficient distributions, which are required by their model, to be matched against and generate sample networks. 
 The main idea of FARZ, presented in this paper, is similar to the BTER model, i.e.  community structure is present from the start and affects how edges are formed. However, FARZ directly extends the network evolution models by incorporating the extra factor of communities. 
It is defined based on relevant and intuitive parameters that directly control different growth factors in networks. 
This  provides flexibility and expressiveness, and makes FARZ an perceptive and simple alternative benchmark generator for community evaluation.  
One notable class of synthetic generators are the mathematical tractable models, such as the Stochastic {Kronecker Graph} model~\cite{leskovec2010kronecker}, and {Multifractal Network} model~\cite{benson2014learning,palla2010multifractal}. These models generate networks with realistic properties, \ie heavy-tailed degree distributions and high clustering coefficient, which can analytically confirmed. 
It is however not straightforward how to set the initial conditions to obtain modular graphs, and how to extract these modules for benchmarking purpose of this paper. 
For a general surveys on generative models for real world networks refer to \cite{newman2010networks,Goldenberg10survey,Chakrabarti2006,Newman2002}. 

%
{\scriptsize

}


\input{farz_res_extended}

\end{document}

%% file: 1_intro.tex
\section{Introduction} 

Networks model the relationships in complex systems, \eg biological interactions between proteins and genes, hyperlinks between web pages, co-authorships between research scholars.  Although drawn from a wide range of domains, these networks exhibit similar properties 
and evolution patterns. 
One fundamental property of real-world networks is that they tend to organize according to an \textit{underlying modular structure}, commonly referred to as community structure \cite{Newman06modularity,bianconi2014triadic,Lancichinetti08LFR,zuev2015emergence,danon2005comparing,Gustafsson06Comparison,Lancichinetti09Comparison}. Many algorithms have been proposed to detect communities in a given network; whereas a community is defined as a group of nodes that have relatively more links between themselves than to the rest of the network. 

Community detection methods are commonly validated  %
and compared based on their performance on benchmark datasets for which the true communities are known \cite{danon2005comparing,Gustafsson06Comparison,Lancichinetti09Comparison}. 
These benchmarks are obtained from either synthetic generators which generate networks with built-in communities \cite{Girvan02Betweenness,Lancichinetti08LFR}, or large real-world networks with explicit or predefined communities \cite{yang2012defining,kloumann2014community,kloster2014heat,shao2015community}. 
In the latter, properties such as user memberships in a social network, venues in a scholarly collaboration network, or product categories in an online co-purchasing network are considered as true communities. 
In general, there exists an interplay between the characteristics of nodes and the structure of the networks \cite{crandall2008feedback,LaFond2010Randomization}, and in some contexts the characteristics of nodes act as the primary organizing principle of the underlying communities \cite{traud2011comparing}. However, this notion of ground-truth communities is weak \cite{jeub2015think}, and these nodal characteristics should be considered rather as attributes correlated with the underlying communities \cite{rabbany15evalu}. 

On the other hand, the GN benchmarks by \citet{Girvan02Betweenness}, and LFR benchmarks  by \citet*{Lancichinetti08LFR},  which generate networks with built-in communities, and are the current gold standards in the evaluation and comparison of community detection methods \cite{duan2014community,chakraborty2014permanence,shao2015community}, fail to exhibit some basic characteristics of real networks \cite{ruths2014control,zuev2015emergence}. 
This is critical since the evaluation is built upon the \textit{assumption} that performance of an algorithm on these benchmarks is a good predictor for its performance when applied to real world networks, and for this assumption to hold, these benchmarks should be similar to real world networks and comply with their observed characteristics. 
%
%
%
%

In an attempt to provide better benchmarks for the community detection task, here we first examine the current generators, discuss their shortcomings and limitations, and propose alterations to improve them. 
First, we experimentally show that the desired network properties can not be fully achieved, even with the proposed improvements, since the current generator frameworks are  inherently restrictive. 
Then, we present a simple alternative benchmark generator, called FARZ
\footnote{FARZ, based on transliteration, means sorting, division, or assess in Arabic; and assumption, or agile in Persian.  }%
, which follows the evolution patterns and characteristics of real networks, and hence is more suitable for validation of community detection algorithms. 
In FARZ, communities are defined as the natural structure underlying the networks, which is not the case in the previous benchmark generators, 
where a community structure is overlaid  on an existing network graph imposing a rewiring of multiple connections.  
Moreover, FARZ incorporates relevant intuitive parameters which could be used to generate a wide range of experimental settings, and hence enables a more thorough comparison of community detection algorithms. 

%% file: 2_related.tex
\section{Benchmark Generator Models}
The \textsl{\citeauthor{Girvan02Betweenness} (GN)} model \cite{Girvan02Betweenness} is the first community detection benchmark used to generate synthetic networks with planted community structure. It is built upon the classic \textsl{\citeauthor{Erdos60} (ER)} model \cite{Erdos60}; which generates random graphs of a given size, $n$, whereas edges are generated independently and with equal probability, $p$. In GN model, nodes in the same community link with probability of $p_{in}$, and nodes from different communities link with probability of $1 - p_{in}$. Unlike real world networks which exhibit heavy tail degree distributions (\aka scale-free), the graphs generated with the ER model have binomial degree distribution, which converges to a Poisson degree distribution for large values of $n$. Moreover, GN creates networks of only $128$ nodes, which are divided into four groups of equal sizes. The sizes of communities in real networks, however, do not have any reason to be equal in size \cite{Newman04}, and in many cases are observed to follow a power law distribution \cite{clauset2004finding}. 
Moreover, unlike real networks, the synthesized GN networks exhibit low transitivity, measured by clustering coefficient, \ie the proportion of closed triplets.

The \textsl{\citeauthor*{Lancichinetti08LFR} (LFR)} model \cite{Lancichinetti08LFR} amends the GN model by considering power law distributions for the degrees of nodes and community sizes. In more details, it first samples the degree sequence and community sizes from power law distributions. Then, it randomly assigns each node (sampled degree) to a community, and links the nodes to create a network. Finally, it rewires the links such that for each node, a fraction, $\mu$, of its links  go outside its community, while the rest, $1-\mu$, are inside its community.
%
%
%
The LFR benchmark is built upon the \textsl{configuration} (CF) model \cite{newman2003structure}; which generates random graphs from a given degree sequence, by fixing the degree of each node, and connecting the available edge stubs uniformly at random. The networks generated with CF are known to exhibit low transitivity. In LFR this is dealt with by a post-processing rewiring step. This issue could be improved upon by using a better and  more realistic start model. 
However, the LFR requires an extensive rewiring process, which changes the network structure chaotically --as confirmed in our experiments. 
Hence, even starting with a realistic network model, the properties can not be preserved.
%

The LFR benchmarks are the current gold standard in community evaluation, e.g. see the evaluation in \cite{duan2014community,chakraborty2014permanence,shao2015community}.
LFR is lately extended for hierarchical and overlapping communities \cite{Lancichinetti09LFR}, where the generation process is modified so that it generates the within and between links separately, instead of realizing the whole network at once. 
In more detail, after sampling the degree sequence $\mathcal{D}$, 
the within and between degree sequences are derived as $(1-\mu)\mathcal{D}$ and $\mu \mathcal{D}$ respectively. 
Then, the CF is used to generate a subgraph per community from the derived within degree sequence. 
There is still however  the need for an extensive rewiring step for forming the external edges, as the derived between degree sequence is first used by the CF model to generate a set of edges. Then those edges that fall within communities are rewired until none of them is a within link. 
%
%
Similar to the original model, this generation process also uses CF model which is an unrealistic network model. 
 On the other hand, since it directly depends on the degree sequence, it is less trivial how to substitute the CF model in this modified extension. Furthermore, unlike the original model, it results in all nodes having the exact same fraction of within/between edges, 
  which is artificial.

%% file: 3_method.tex
\section{FARZ Benchmark Model}
FARZ expands  the network one node at a time. 
Each node $i$ added to the network is immediately assigned to $r$ communities, where $r=1$ in case of non-overlapping communities.
The probabilities of these assignments are proportional to the (current) sizes of the communities. 
This would apply a preferential attachment mechanism and ensure the heavy tail distribution for the community sizes.
More formally, the probability of node $i$ joining community $u$ is: 
\begin{equation}
\label{eq:assig}
p(u) = \frac{|u|+\phi}{\sum\limits_{v} (|v|+\phi)}
\end{equation}
%
where  the denominator is a normalizing factor that sums over sizes of all communities; and $\phi=1$ ensures that empty communities also have a chance to recruit.
Moreover, it controls the effect of preferential attachment: 
as $\phi$ increases, the distribution for sizes of communities becomes closer to uniform (equal sized communities). 
%

After node $i$ joins the selected community or communities, it gets connected to the network by forming an edge (\refalg{alg:farzConn}), to ensure that there are no singletons, i.e. unconnected nodes with the degree of zero. 
Then, $m-1$ nodes, from the existing nodes within the network, are randomly selected and get a chance to also form connections.
These new connections may or may not involve the newly added node $i$.
Adding edges at each round results in an accumulative advantage for the nodes added earlier, since they get more chances to get selected and form connections, which naturally enforces the heavy tail degree distribution observed in real networks. 
FARZ is summarized in \refalg{alg:farzMain}:

\begin{algorithm}
\vspace{3pt}
 \caption{FARZ Generator (n, m, k)}
 \label{alg:farzMain}
\begin{algorithmic}[1]
\STATE $G \leftarrow Graph()$
\STATE $C \leftarrow \{c_1=\emptyset, c_2 =\emptyset \ldots c_k =\emptyset \} $\COMMENTc{initialize}
\FOR{$i \in [ 1\dots n]$}
\STATE $G.add\_node(i)$\COMMENTc{add node $i$ }
\STATE $assign(i,C)$\COMMENTc{assign $i$ to communities}
\STATE $connect(i, G, C)$\COMMENTc{add an edge from node $i$}
\FOR{$[2\dots m]$}\COMMENTc{add $m-1$ edges}
\STATE $j\leftarrow select(G.nodes)$\COMMENTc{select node $j$ from $G$}
\STATE $connect(j,  G, C)$\COMMENTc{add an edge from node $j$}
\ENDFOR
\ENDFOR
\STATE \textbf{return} G, C
\end{algorithmic}
 \end{algorithm}

In \refalg{alg:farzMain}, the input parameters of $n$, and $k$ respectively determine the total number of nodes, and the number of communities. 
Whereas $m$ determines the number of edges added at each step, which controls the total number of edges ($nm$), overall density of the networks ($2m/n$), and the average degree ($2m$).   
The function $assign()$, in line 5, and $select()$, in line 8, are straightforward. 
The latter selects a node uniformly at random; whereas the former randomly chooses community assignments based on the probabilities in \refeqb{eq:assig}.
%
%
%
Function ${connect()}$ enforces the community structure and controls the  edge formations.   
\refalg{alg:farzConn} describes the function ${connect()}$, called in line 6 and 9 of \refalg{alg:farzMain}. 
This function enforces the community structure and controls the  edge formations.   
The function $choose()$, in line 5 of \refalg{alg:farzConn}, determines the probability of forming an edge from node $i$ to node $j$.
\begin{algorithm}[]
 \caption{FARZ Connect (i, G, C)}
 \label{alg:farzConn}
\begin{algorithmic}[1]
\IF{random  $<\beta$}\COMMENTc{ choose a community from}
\STATE $c \leftarrow select(\{c, \;\forall c\in C \land i \in c\}) $\COMMENTc{memberships of i}
\ELSE
\STATE $c \leftarrow select(\{c, \;\forall c\in C \land i \notin c\}) $\COMMENTc{other communities}
\ENDIF
\STATEx \hspace{5pt} \COMMENTc{choose a node within the selected community}
\STATE $j \leftarrow choose(\{j, \;\forall j\in c \land j \neq i \land (i,j)\notin G.edges\})$
\STATE $G.add\_edge(i,j)$
%
\end{algorithmic}
 \end{algorithm}

%
When forming an edge, a node first selects a community, and then connects to a node within that community. 
More specifically, node $i$ forms its connection within the communities that it is a member of, with probability $\beta$, and connects to nodes from other communities with probability $1-\beta$. The control parameter $\beta$ hence determines the strength of the overall community structure, and is analogous with the mixing parameter $\mu$ in the $LFR$ model. 
On the other hand, the probability of forming an edge from node $i$ to node $j$, depends on two driving factors: the number of their common neighbours (\refeqb{eq:cn}), and the similarity of their degrees (\refeqb{eq:cd}), i.e.
  \begin{align}
  p_{ij} \propto & \;  \sum^n_{k=1} w_{ik}w_{jk}\quad \label{eq:cn} \\
  p_{ij} \propto & \;  (d_i -  d_j)^2 \label{eq:cd}
  \end{align} 
 where $w_{ij}$ represents the edge weight between node $i$ to node $j$, and $d_i = \sum^n_{k=1} w_{ik}$.  
%
%
\refeqb{eq:cn}  enforces ``triadic closure'', which is known as a natural mechanism for edge formation in real networks \cite{bianconi2014triadic}, and results in the high clustering coefficient observed in real networks. 
\refeqb{eq:cd} implements the assortative mixing, i.e. tendency of similar nodes to connect. Here we consider degree assortativity, measured by degree correlation. 
\begin{figure}[th!]
\vspace{-10pt}
\centering
\includegraphics[trim={0.4cm 0.4cm 0.4cm 0.4cm},clip,width=.26\linewidth]{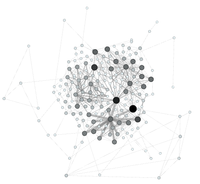}
\includegraphics[trim={0.4cm 0.4cm 0.4cm 0.4cm},clip,width=.26\linewidth]{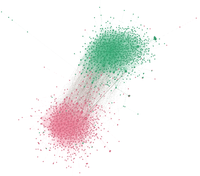}
\includegraphics[trim={0.4cm 0.4cm 0.4cm 0.4cm},clip,width=.20\linewidth]{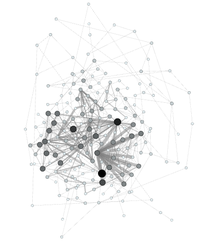}
\includegraphics[trim={0.4cm 0.4cm 0.4cm 0.4cm},clip,width=.26\linewidth]{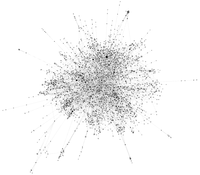}
\\
\includegraphics[trim={3cm 0.3cm 3cm 0cm},clip,width=.99\linewidth]{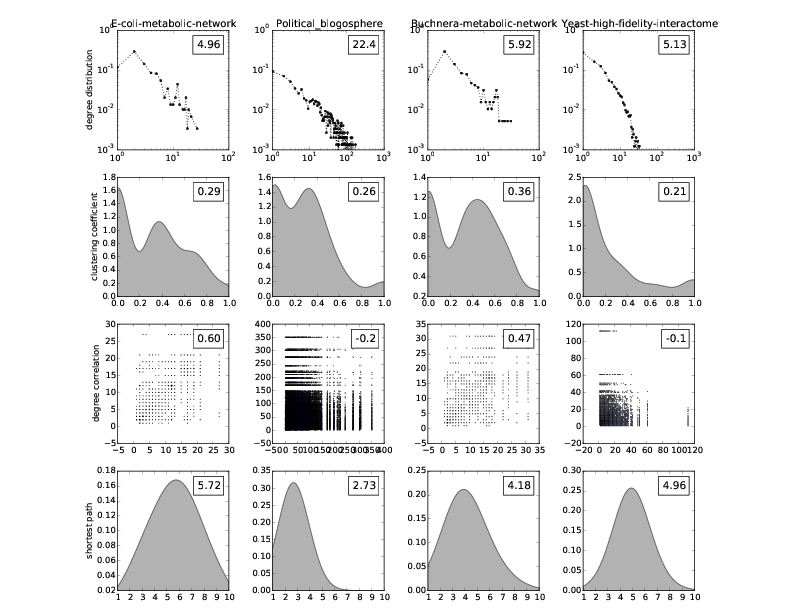}
\caption{Basic properties of four example real world networks with positive and negative degree correlations.
The insets respectively show the average degree, average clustering coefficient, degree correlation (Pearson correlation between the degrees of connected nodes), and the average shortest paths. The corresponding graphs are also visualized at the top; 
colours represent the available community labels.  }
\label{fig:realProps}
\vspace{-10pt}
\end{figure}
Real networks exhibit both negative and positive degree correlation \cite{newman2010networks}. 
In social networks, for instance, a positive degree correlation is often observed, which indicates that nodes with similar degrees tend to connect to each other; 
whereas some biological networks are known to be disassortative, i.e. hubs with high degrees often connect to nodes with small degrees; see \reffig{fig:realProps} where properties of four widely studied real networks are reported; where all exhibit strong degree correlation. 
To cover both assortative and disassortative cases, we define a control parameter $\gamma$ which indicates whether the degree correlation should be positive or negative in the generated network, i.e. whether larger $\Delta d_{ij}$ decreases  or increases $p_{ij}$, respectively. 
To combine the effect of \refeqb{eq:cn} and \refeqb{eq:cd}, a function $\varphi(x,y)$ should be used. Here we simply use \(\varphi (x,y) = x^{\alpha} y^{-\gamma}\) to have both factors in effect; whereas similar to $\gamma$, $\alpha$ controls the effect of \refeqb{eq:cn}. The overall probabilities are hence computed as:
\begin{equation}
p_{ij} =  ( \sum^n_{k=1} w_{ik}w_{jk})^{\alpha} ((d_i -  d_j)^2 +1)^{-\gamma} + \epsilon 
\end{equation}
 where $\epsilon$ is a small number that accounts for unlikely edges, and is particularly required at the initial stages.
Different choices of $\varphi$ result in structurally different networks, however all these generated networks would have the heavy tail distributions for the degree of nodes and community sizes, and a built-in modular structure. 
 
 \subsection{Comparing Properties of Networks}
\begin{figure}[t!]
 \centering
\hspace{10pt}
\includegraphics[width=.25\linewidth]{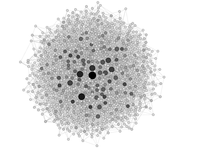}\hspace{10pt}
\includegraphics[width=.25\linewidth]{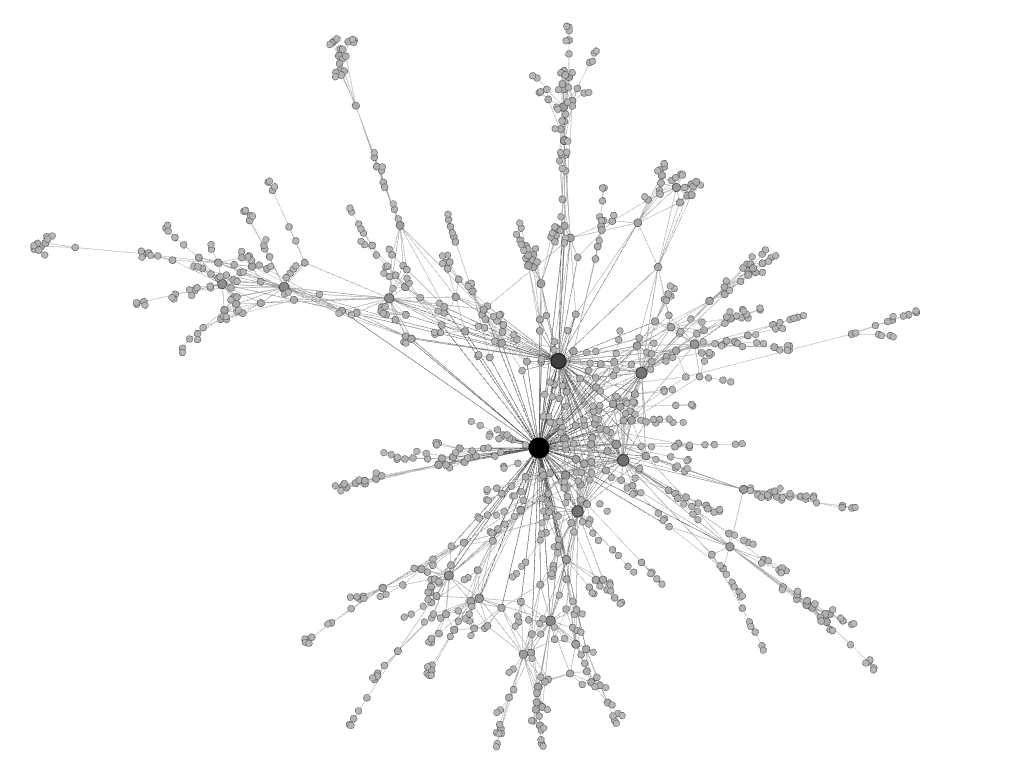}\hspace{10pt}
\includegraphics[width=.25\linewidth]{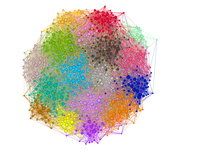} 
\includegraphics[trim={0cm 3cm 0.1cm 3cm},clip,width=.95\linewidth]{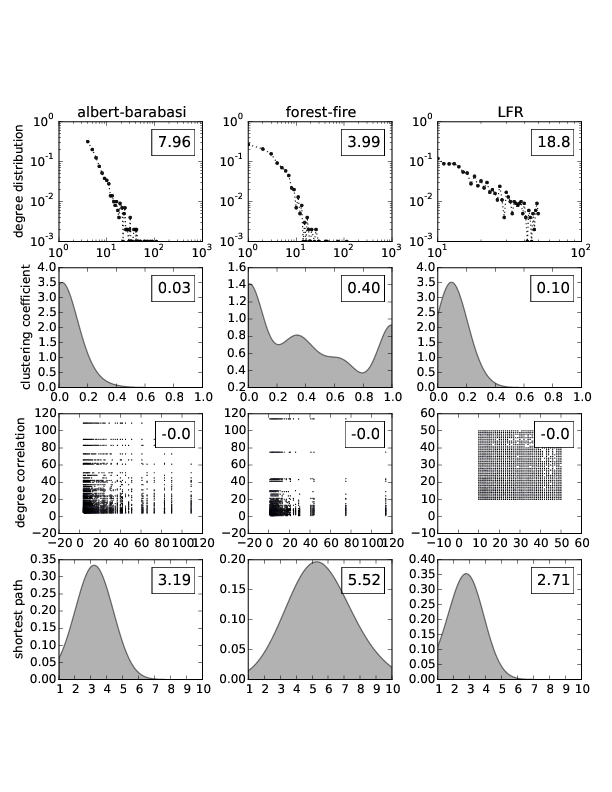}
\caption{Properties of four synthetic networks, with 1000 nodes. 
The network is generated with $p_{ij} = 0.01$ for $ER$, and with $m=4$ for AB. The FF parameters are $\{p:0.4, rp:0.2\}$, and for LFR we used the original implementation with commonly used setting of $\{k:20,\; k_{max}:50,\; t_1:2,\; t_2: 1,\; \mu:0.4,\; c_{min}:20,\; c_{max}:100\}$.}  
\label{fig:synthProps}
\end{figure}

\begin{figure}[t!]
\centering
\includegraphics[trim={0.3cm 0.3cm 0.3cm 0.3cm},clip,width=.24\linewidth]{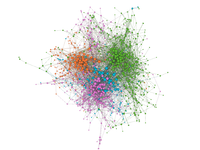}
\includegraphics[trim={0.3cm 0.3cm 0.3cm 0.3cm},clip,width=.24\linewidth]{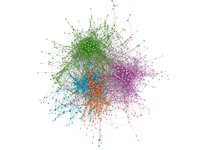}
\includegraphics[trim={0.3cm 0.3cm 0.3cm 0.3cm},clip,width=.24\linewidth]{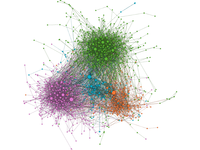}
\includegraphics[trim={0.3cm 0.3cm 0.3cm 0.3cm},clip,width=.24\linewidth]{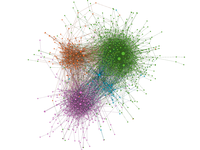}
\includegraphics[trim={.1cm 4cm 0.1cm 5cm},clip,width=.99\linewidth]{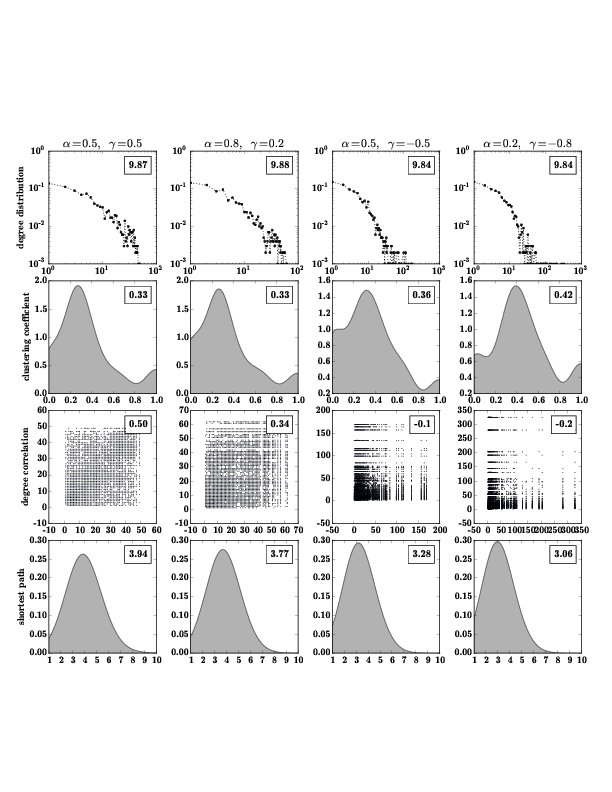}
\caption{Properties of four sample networks generated by FARZ. The parameters are the same for the networks, except $\alpha$ and $\gamma$ which are reported at the top. The exact setting is $\{n$:1000, $k$:4, $m$:5, $\beta$:0.8, $\phi$:1, $r$:1, $\epsilon:1e-07\}$. 
}
\label{fig:passdisass}
\end{figure}

\begin{figure}[h!]
 \centering
\includegraphics[trim={.1cm 1cm .1cm .1cm},clip,width=.99\linewidth]{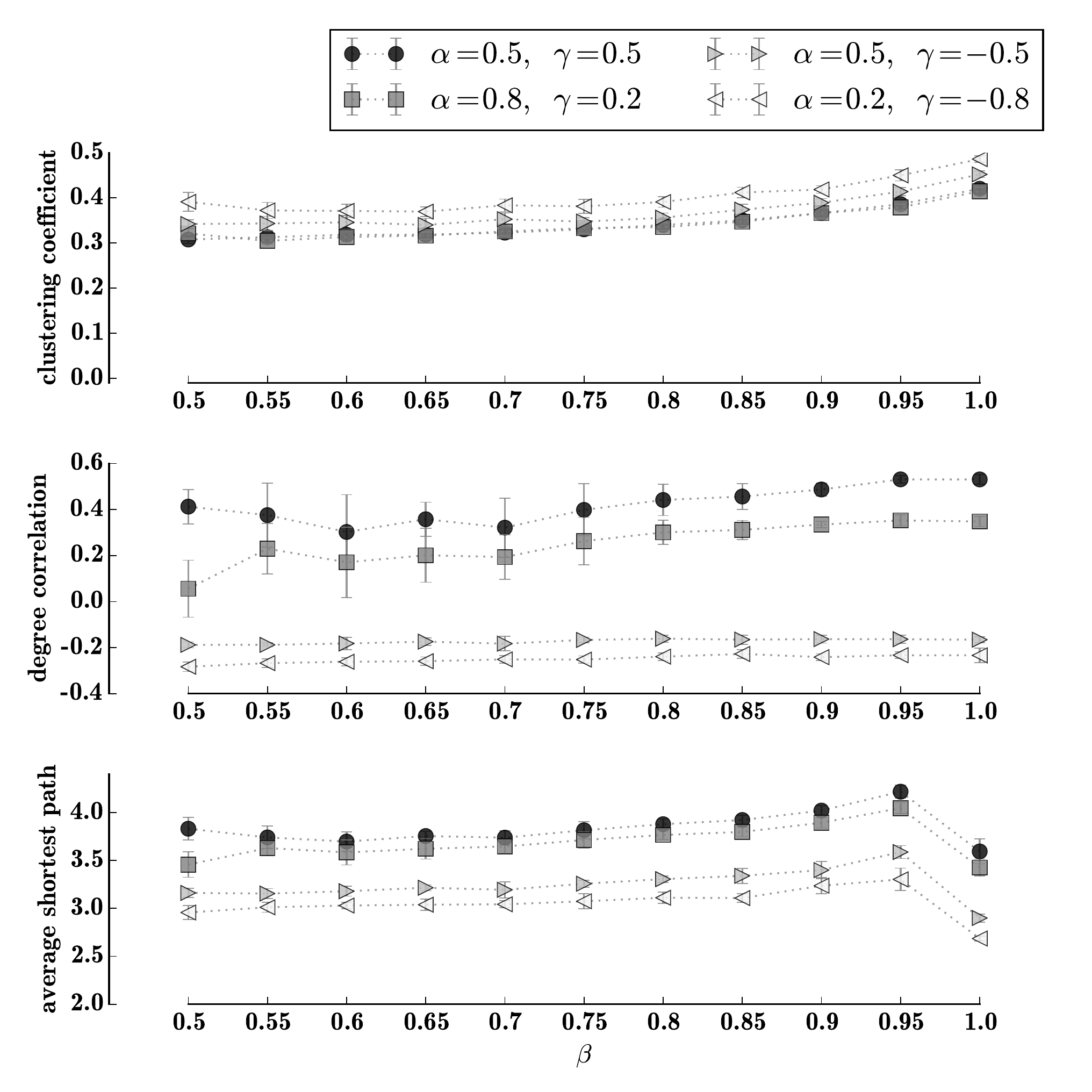}
\caption{Properties of the synthesized FARZ networks plotted as a function of $\beta$, i.e. the probability of edges to form within the communities. }
\label{fig:farzavg}
\end{figure}

 \reffig{fig:synthProps} and \ref{fig:passdisass} illustrate basic properties for sample synthetic networks, which correspond to the properties reported for real networks in \reffig{fig:realProps}. 
In  \reffig{fig:synthProps}, we observe zero degree correlation for networks generated with ER, AB, FF, and LFR models; and zero or small clustering coefficient for ER, AB, LFR models; which are inconsistent with the patterns observed for real world networks in \reffig{fig:realProps}.  
In \reffig{fig:passdisass}, we show that the sample networks generated by FARZ comply well with the properties of real networks.
They have small diameter, heavy tail degree distribution, and high clustering coefficient; whereas they can exhibit positive or negative degree correlations based on the parameter setting, which is controlled  by the parameter $\gamma$. 
\reffig{fig:farzavg} reports the average of properties for the synthesized FARZ networks, which is plotted as a function of $\beta$ -- strength of the community structure. 
 Here, we compare the four parameter settings of \reffig{fig:passdisass}, when $\beta$ varies, and the results are averaged over $10$ realizations of the networks for each $\beta$. 
We plot the results for $\beta \in [0.5, 1]$, that is where a community structure exists within the network --the chances of edge formation is higher within the communities that outside of them.   
We can see in this plot that the FARZ benchmarks are consistent, as oppose to the LFR (as seen in \reffig{fig:propnets}), i.e. all the networks synthesized by FARZ exhibit degree correlation and clustering coefficient, regardless of the strength of the underlying community structure. 

 \subsection{Comparing Properties of Communities}
\begin{figure}[h!]
\centering
\begin{subfigure}[b]{0.34\textwidth}
\includegraphics[trim={0cm 0.5cm 0cm 0cm},clip,width=\linewidth,height=.35\linewidth]{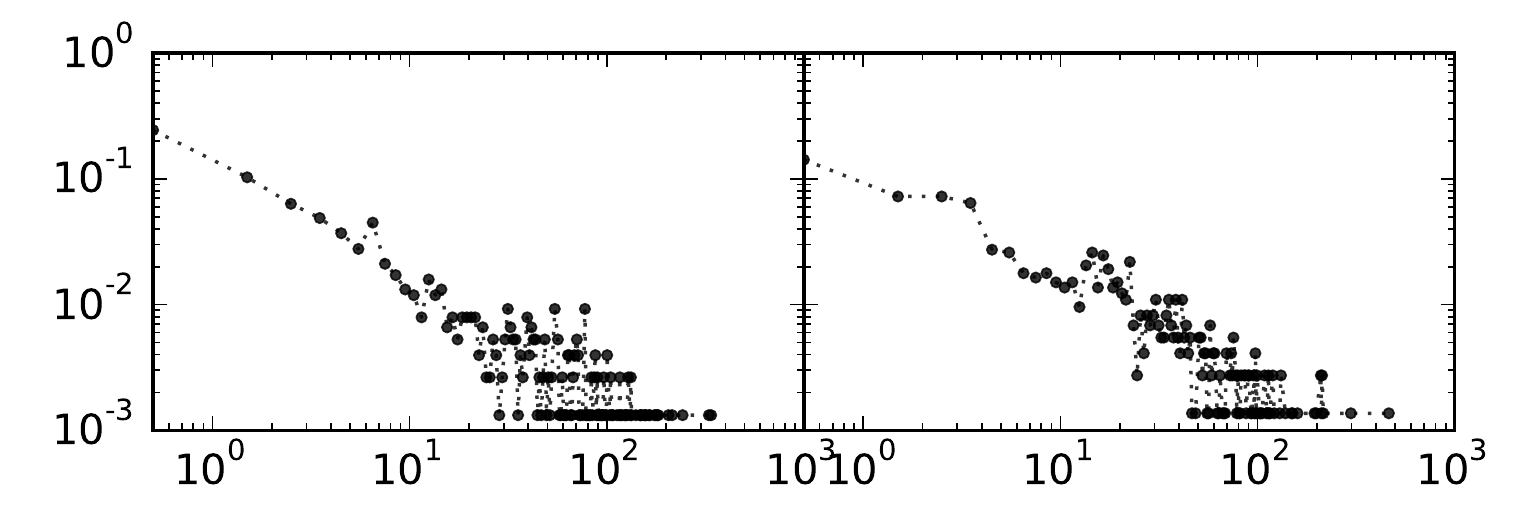}
\caption{\scriptsize Political blogosphere}
\label{fig:ddperCa}
\end{subfigure}
\begin{subfigure}[b]{0.5\textwidth}
\includegraphics[trim={2cm 0.5cm 0cm 0cm},clip,width=\linewidth,height=.26\linewidth]{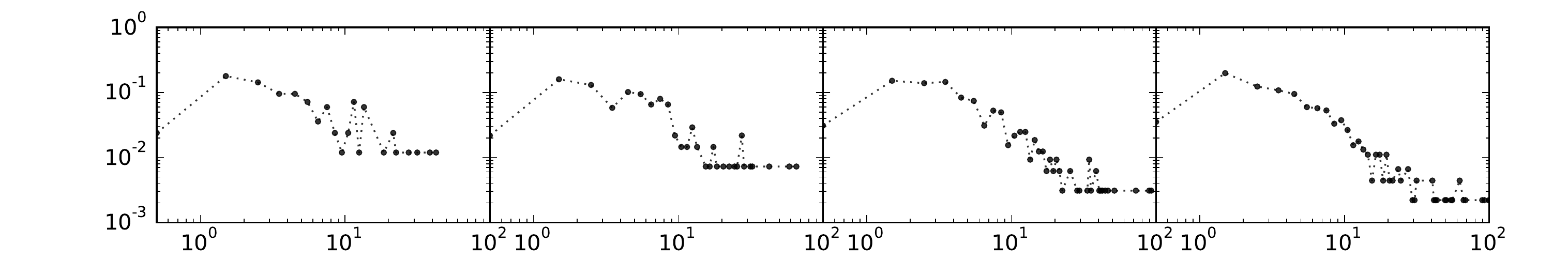}
\caption{\scriptsize FARZ}
\label{fig:ddperCb}
\end{subfigure}
\begin{subfigure}[b]{0.5\textwidth}
\includegraphics[trim={2cm 1.3cm 0cm 0cm},clip,width=\linewidth,height=.44\linewidth]{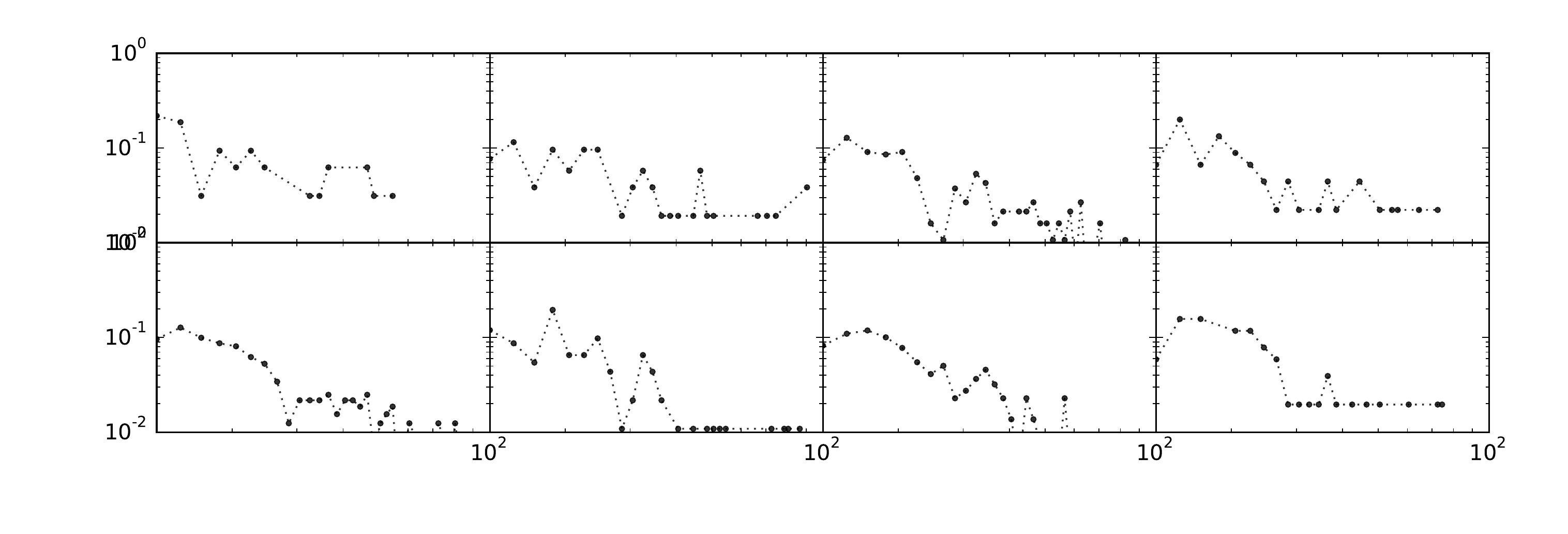}
\caption{\scriptsize LFR}
\label{fig:ddperCc}
\end{subfigure}
\caption{Degree distributions per community for an example real network (top), and two synthetic networks generated by FARZ (middle) and LFR (bottom).
Each subplot reports the degree distribution inside a community. 
}
\label{fig:ddperC}\vspace{-10pt}
\end{figure}

 All the figures plotted above compare the general properties of the synthesized networks. 
We can further look at the properties in each community, and compare the patterns with what is observed in the real networks. 
More specifically, in \reffig{fig:synthProps} and \ref{fig:passdisass}, we see that networks generated with both the LFR and FARZ model have heavy tail degree distributions.
In \reffig{fig:ddperC}, we compare the  degree distributions in each community of these benchmarks. 
We can see that in the example real world network, for which the community labels are available, the degree distributions per community follows the same heavy tail distribution as the overall network (\reffig{fig:ddperCa}).
The communities generated by FARZ benchmark, comply with this pattern and follow a  heavy tail degree distribution, similar to what is observed in real networks (\reffig{fig:ddperCb}).
However, we do not observe a clear heavy tail trend for the communities generated by the LFR benchmark (\reffig{fig:ddperCc}). 

In \reffig{fig:ddperC}, the FARZ network  (\reffig{fig:ddperCb}) corresponds to the third column in \reffig{fig:passdisass}. 
Other settings in \reffig{fig:passdisass} result in similar plots. 
The LFR network  (\reffig{fig:ddperCc}) has similar parameters as the network reported in the fourth column of \reffig{fig:synthProps}, except the maximum community size increased to $500$ to get a smaller number of communities, and to plot it. 
The plot for the exact network of \reffig{fig:synthProps}, which has 17 communities, shows similar patterns but would requires more space for plotting. 
\begin{figure}[h!]
\centering
\begin{subfigure}[b]{0.34\textwidth}
\includegraphics[trim={0cm 0.5cm 0cm 0cm},clip,width=\linewidth,height=.25\linewidth]
{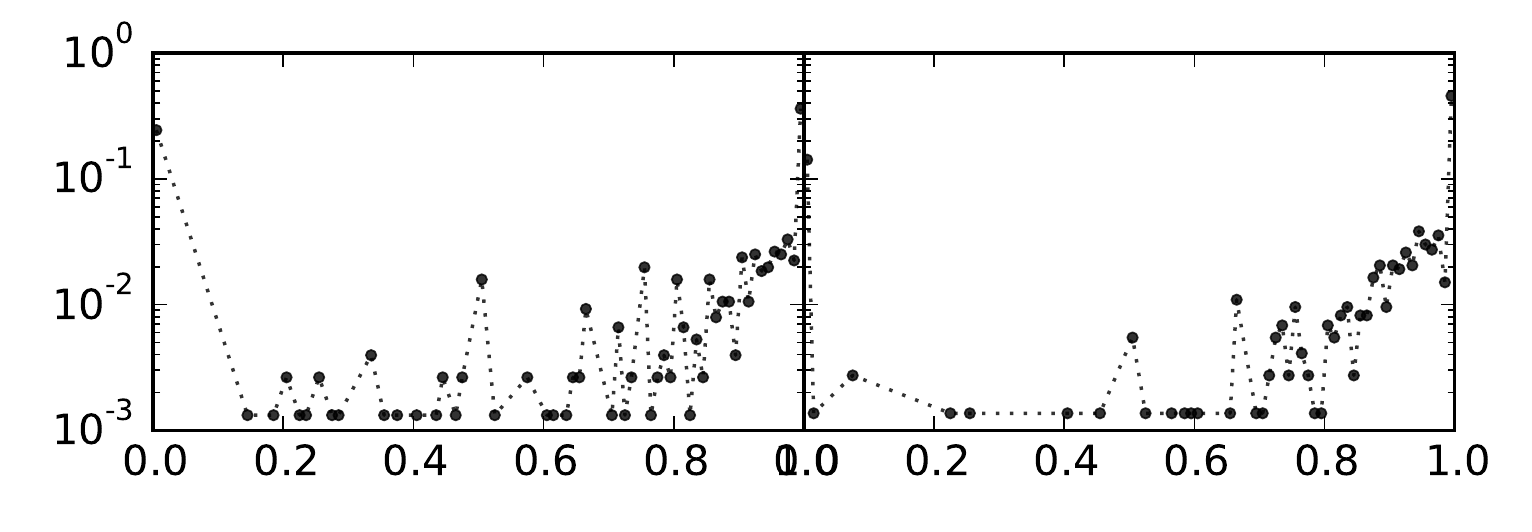}
\caption{\scriptsize Political blogosphere}
\label{fig:muperCa}
\end{subfigure}
\begin{subfigure}[b]{.5\textwidth}
\includegraphics[trim={2cm 0.5cm 0cm 0cm},clip,width=\linewidth]
{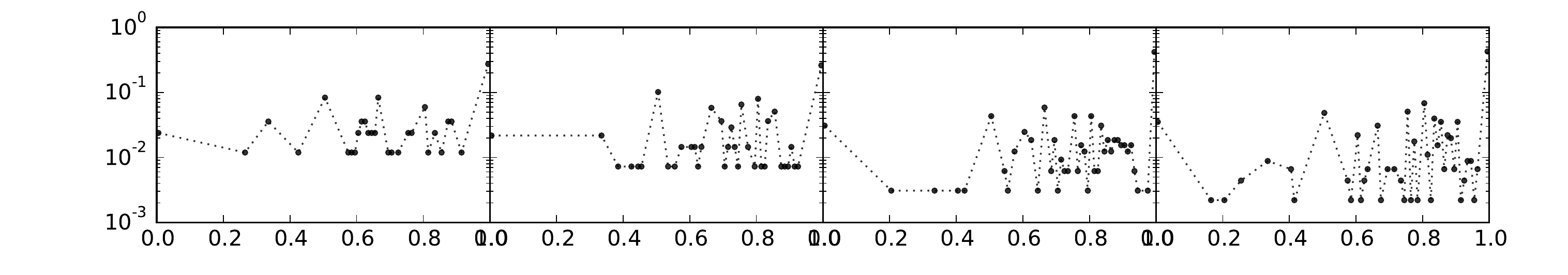}
\caption{\scriptsize FARZ}
\label{fig:muperCb}
\end{subfigure}
\begin{subfigure}[b]{0.5\textwidth}
\includegraphics[trim={2cm 1.3cm 0cm 0cm},clip,width=\linewidth]
{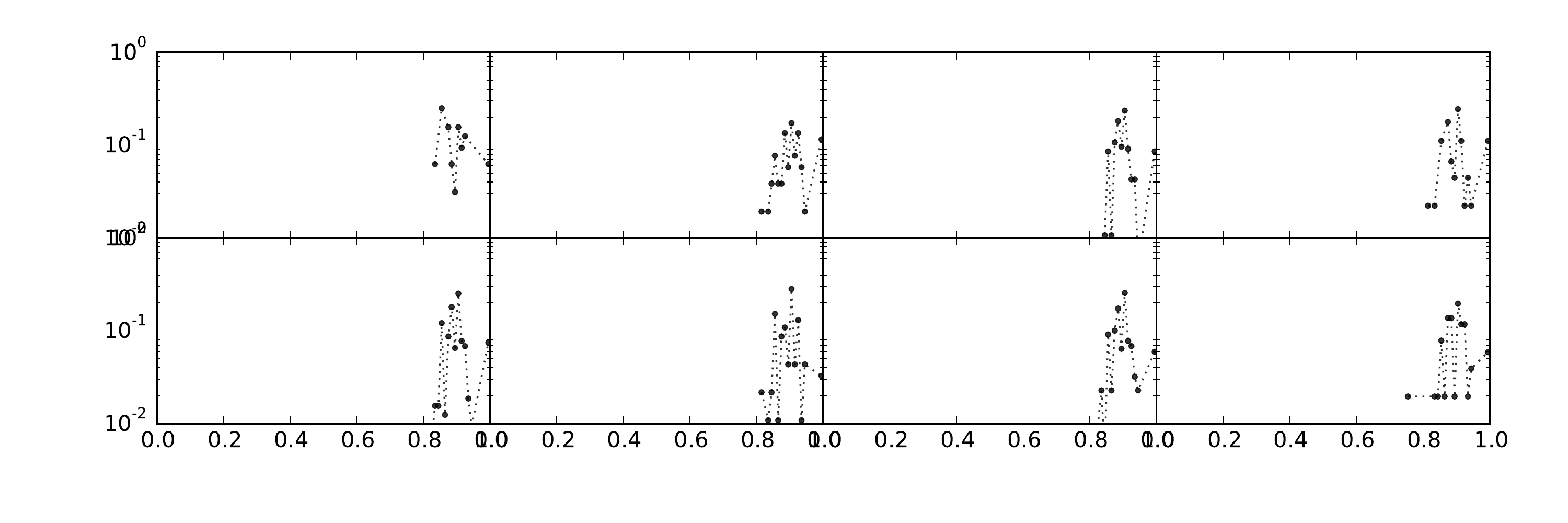}
\caption{\scriptsize LFR}
\label{fig:muperCc}
\end{subfigure}
\caption{Distributions of within to total edges for the nodes in each community; for the networks of \reffig{fig:ddperC}. 
}
\label{fig:muperC}\vspace{-10pt}
\end{figure}
In \reffig{fig:muperC}, we compare the ratio of within to total connections for the nodes in each community,  i.e. the degrees of the nodes within their community divided by their degree in the whole network; this corrsponds to $1-\mu$ in the LFR.  
Here we can see that for the real world network example (\reffig{fig:muperCa}), as well as the networks synthesized with FARZ, this ratio of within to total edges varies for the nodes inside the community between $0.0$ and $1.0$. 
Which is not the case in the LFR example. LFR gets this ratio, $\mu$, as an input parameter. 
In LFR, all the nodes within the community have the same degree of membership, which is artificial and unlike the observed pattern in real networks.    


%% file: 4_exps.tex
 \section{Application and Flexibility}
In this section we show the application of FARZ in validating and comparing community mining algorithms. 
More specifically, we compare and rank selected community mining algorithms on the benchmarks generated by FARZ,
where we tune its flexible parameters to rank the algorithms in different and meaningful experimental settings. 
We rank the algorithms based  on the the agreement of their results with the built-in community structure of FARZ benchmarks. The selected algorithms are InfoMap \cite{Rosvall08Infomap}, WalkTrap ~\cite{pons2005computing}, Louvain \cite{Vincent08Louvain}, and FastModularity  \cite{newman2004fast}. 
The agreements (higher is better) are measured and reported with both $ARI$ (Adjusted Rand Index) and $NMI$ (Normalized Mutual Information) \cite{rabbany15DAMI}.

\subsection{Effect of the Degree Assortativity}
\begin{figure}[h!]
\centering
\begin{subfigure}[b]{0.5\textwidth}
\includegraphics[trim={1cm 0.1cm 0cm 0cm},clip,width=.98\linewidth]{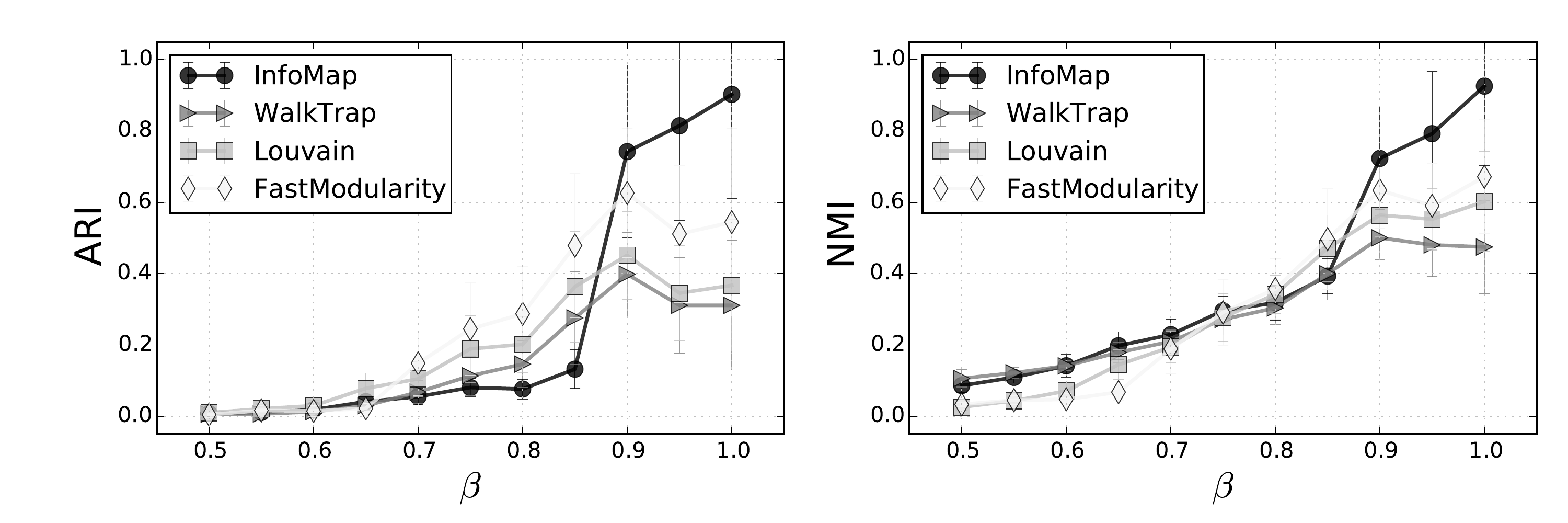}
\caption{assortative benchmarks, $\gamma = 0.5 $}
\label{fig:bass}
\end{subfigure}
\begin{subfigure}[b]{0.5\textwidth}
\includegraphics[trim={1cm 0.1cm 0cm 0cm},clip,width=.98\linewidth]{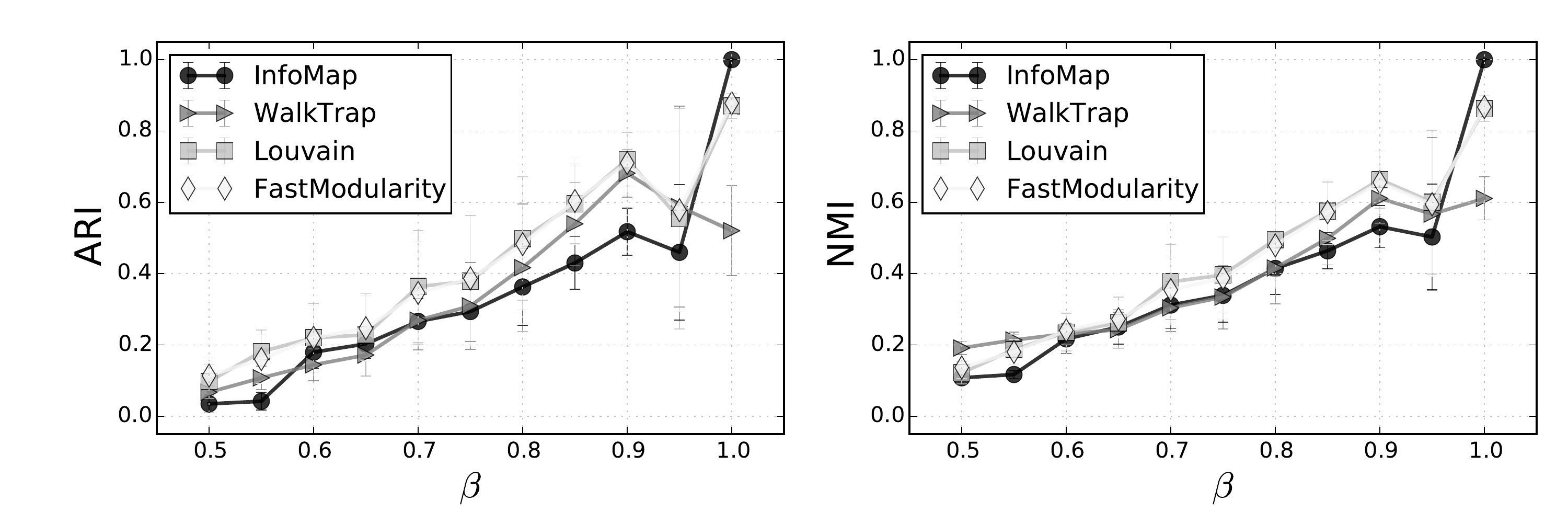}
\caption{disassortative benchmarks, $\gamma =-0.8$}
\label{fig:bdiss}
\end{subfigure}
\caption{Performance of community mining algorithms on benchmarks with \textbf{degree assortativity v.s. degree disassortativity}; plotted as a function of the strength of the built-in community structure, i.e. determined by $\beta$. Results are averaged over 10 runs. 
The parameter settings correspond to the first (\ref{fig:bass}) and last (\ref{fig:bdiss}) columns of \reffig{fig:passdisass}. }
\label{fig:bassdiss}\vspace{-10pt}
\end{figure}

Here, we compare performance of the four community detection algorithms on the benchmarks with degree assortativity, i.e. positive degree correlation (common in social networks); and degree disassortativity, i.e. negative degree correlation (common in biological networks).  
\reffig{fig:bassdiss} shows the comparison results. 
The selected algorithms overall perform better on disassortative benchmarks. 
In the case of assortative networks,  FastModularity outperforms the other three methods when communities are not predominant, i.e. for $\beta<0.9$. From $\beta=0.9$, Infomap becomes the best performing method, which is after a sharp transition from its poor performance for the less predominant communities.
In case of disassortative networks, the performance of FastModularity  is on a par with Louvain, which are superior to InfoMap until communities are well separated, i.e. $\beta=1$.
These results are interesting since the InfoMap algorithm is known to be the best performing method from the selected set when evaluated on the LFR benchmarks \cite{rabbany15DAMI,Lancichinetti09Comparison}.

\subsection{Effect of the Number of Communities}
\begin{figure}[h!]
\centering
\begin{subfigure}[b]{0.5\textwidth}
\includegraphics[trim={1cm 0.1cm 0cm 0cm},clip,width=.98\linewidth]{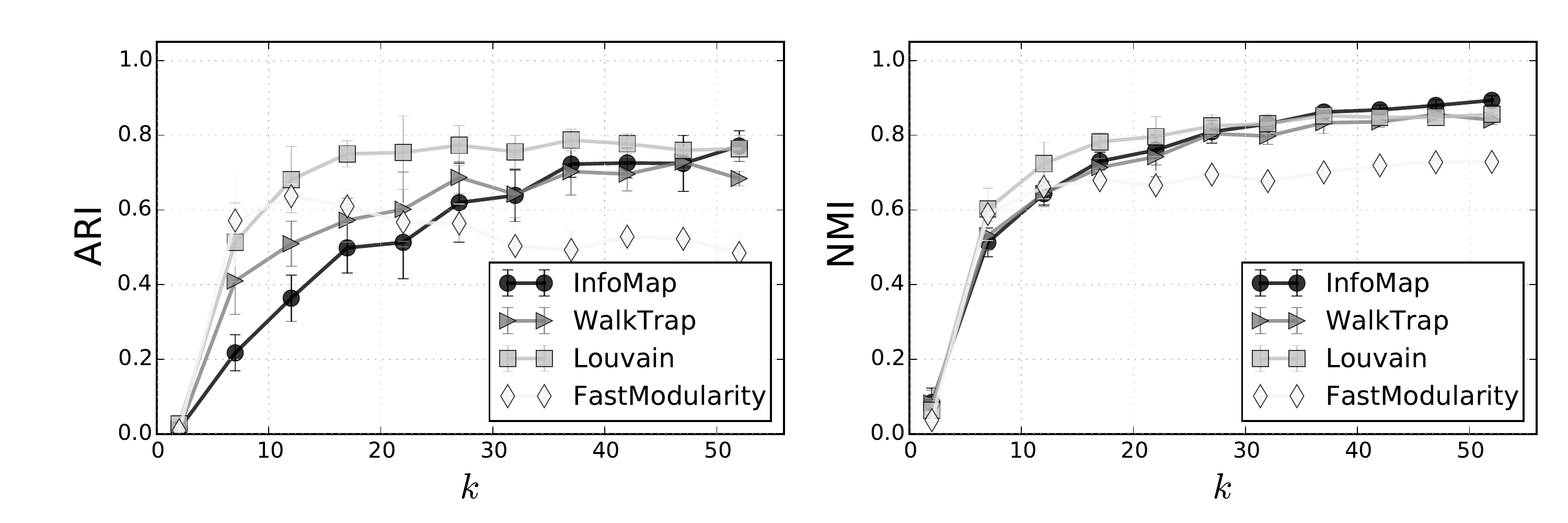}
\caption{assortative benchmarks, $\gamma = 0.5 $}
\label{fig:kass}
\end{subfigure}
\begin{subfigure}[b]{0.5\textwidth}
\includegraphics[trim={1cm 0.1cm 0cm 0cm},clip,width=.98\linewidth]{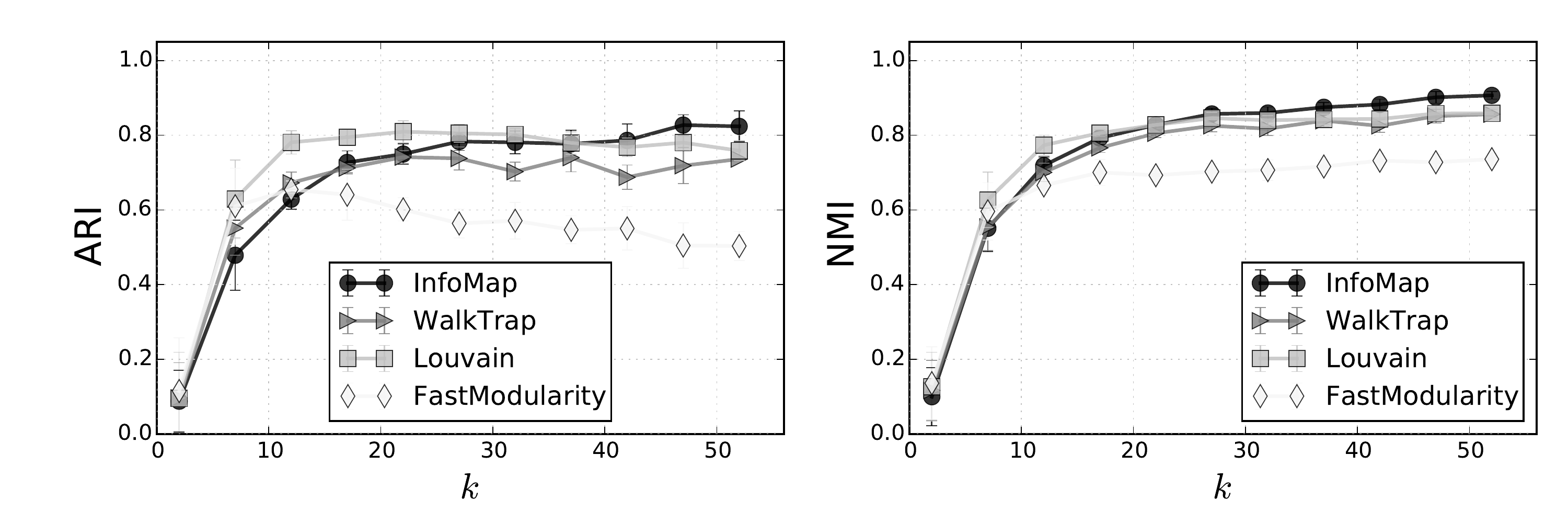}
\caption{disassortative benchmarks, $\gamma =-0.8$}
\label{fig:kdiss}
\end{subfigure}
\begin{subfigure}[b]{0.5\textwidth}
\includegraphics[trim={0cm 0.1cm 0cm 0cm},clip,width=.48\linewidth]{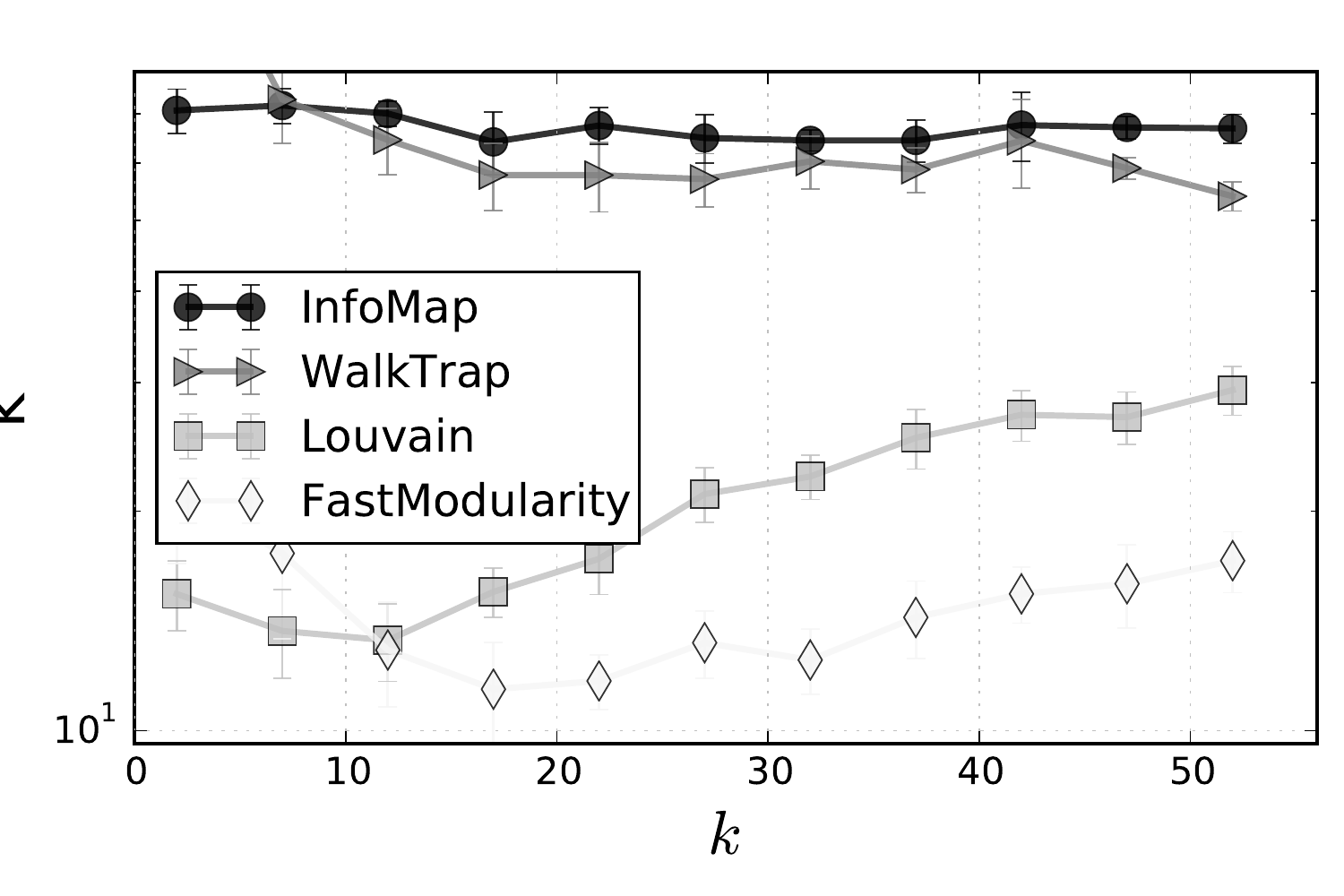}
\includegraphics[trim={0cm 0.1cm 0cm 0cm},clip,width=.48\linewidth]{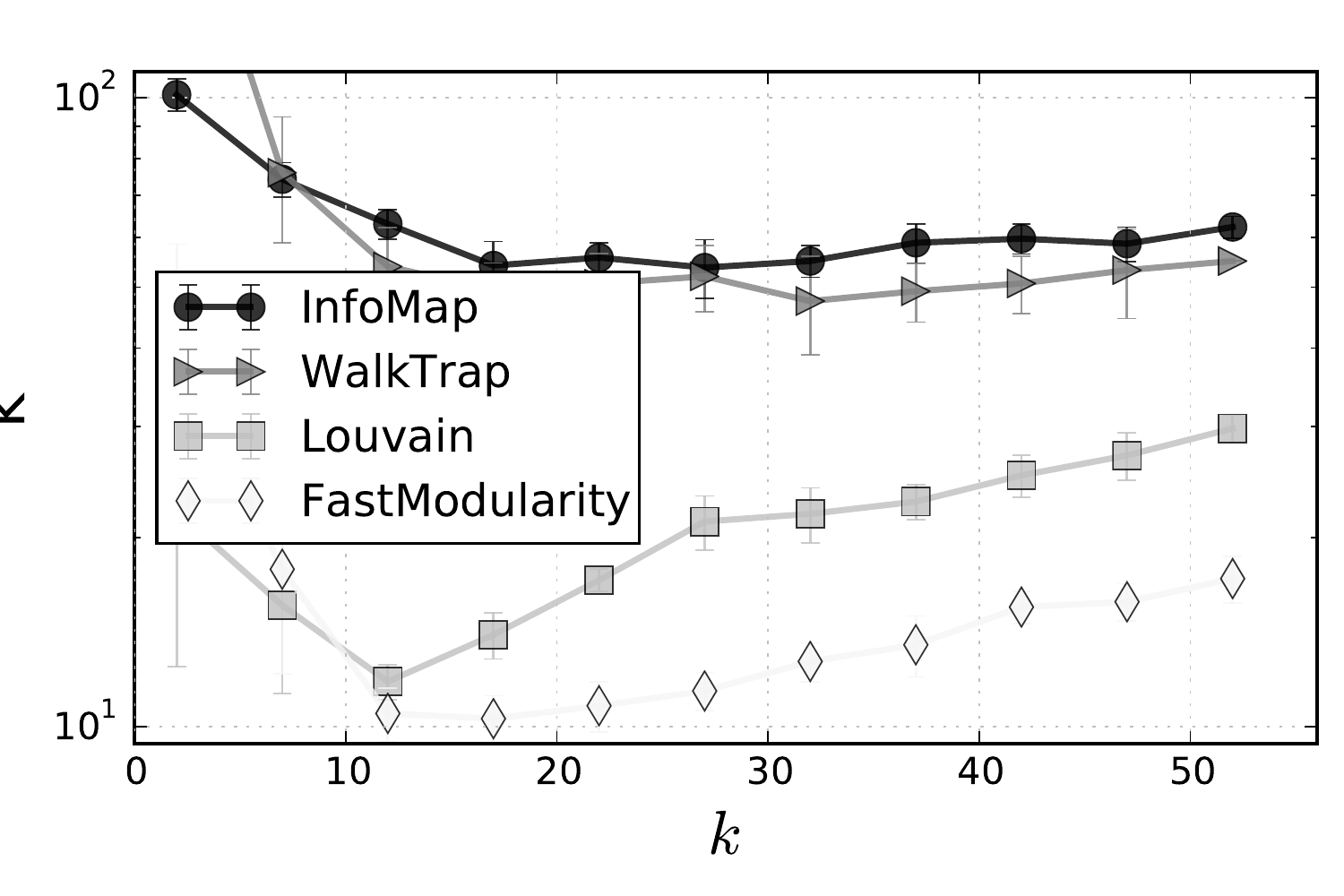}
\caption{number of communities in the results, $k'$, v.s. the true number of communities, $k$. }
\label{fig:kk}
\end{subfigure}
\caption{Performance of community mining algorithms on benchmarks with different \textbf{number of built-in communities}. Settings correspond to the \reffig{fig:bassdiss}, and $\beta$ is fixed to 0.8. }
\label{fig:kassdiss}\vspace{-10pt}
\end{figure}

Here we compare the algorithms on benchmarks with different number of built-in communities, by changing the parameter $k$. \reffig{fig:kassdiss} shows the results.
Here we see that all the algorithm have difficulty when the number of communities is small, i.e. $k<10$.  
Unlike other algorithms, the performance of FastModularity also drops as the number of communities increases. 
In the assortative networks in particular, the Louvain method is more consistent to the change of the number of communities. 
In \reffig{fig:kk}, we can also observe that all these method fail to detect the true number of communities in the ground-truth, and the number of detected communities($k'$) seems to be independent of the true number of communities in the ground-truth($k$), particularly for InfoMap and WalkTrap and when $k$ is large.  

\subsubsection{Effect of the Density of Networks}
\begin{figure}[h!]
\centering
\begin{subfigure}[b]{0.5\textwidth}
\includegraphics[trim={1cm 0.1cm 0cm 0cm},clip,width=.98\linewidth]{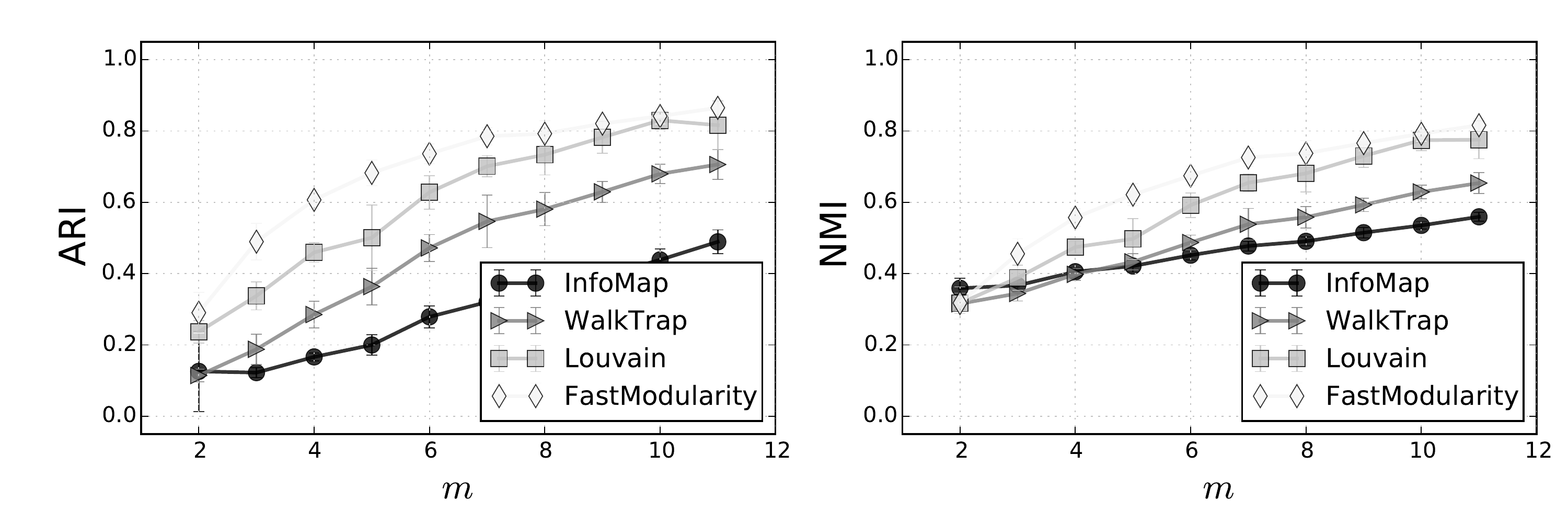}
\caption{assortative benchmarks, $\gamma = 0.5 $}
\label{fig:mass}
\end{subfigure}
\begin{subfigure}[b]{0.5\textwidth}
\includegraphics[trim={1cm 0.1cm 0cm 0cm},clip,width=.98\linewidth]{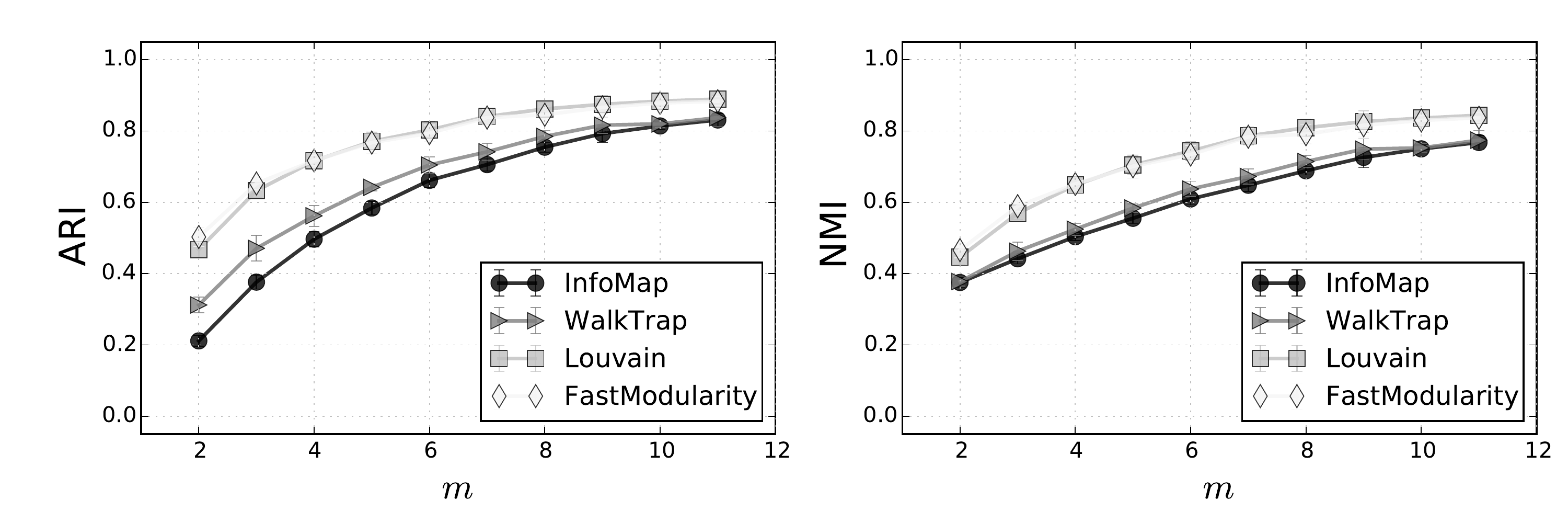}
\caption{disassortative benchmarks, $\gamma =-0.8$}
\label{fig:mdiss}
\end{subfigure}
\caption{Performance of community mining algorithms on benchmarks with different \textbf{density}. Settings correspond to the \reffig{fig:bassdiss}, $\beta$ is 0.8. }
\label{fig:massdiss}\vspace{-10pt}
\end{figure}

Here, we tune the parameter $m$ to change how many connections nodes have on average, i.e. move from sparse to less sparse networks and examine how the performance of different algorithms are affected by changing the density of benchmarks. In the results reported in \reffig{fig:massdiss} we see that  
overall the algorithms  perform better as networks become denser, as the average degree of nodes increases, i.e. as $m$ increases. 
The performance boost is more significant for FastModularity, Louvain, and WalkTrap algorithms, and particularly in the assortative setting. 

\subsubsection{Effect of Variation in Community Sizes}
\begin{figure}[h!]
\centering
\begin{subfigure}[b]{0.5\textwidth}
\includegraphics[trim={1cm 0.1cm 0cm 0cm},clip,width=.98\linewidth]{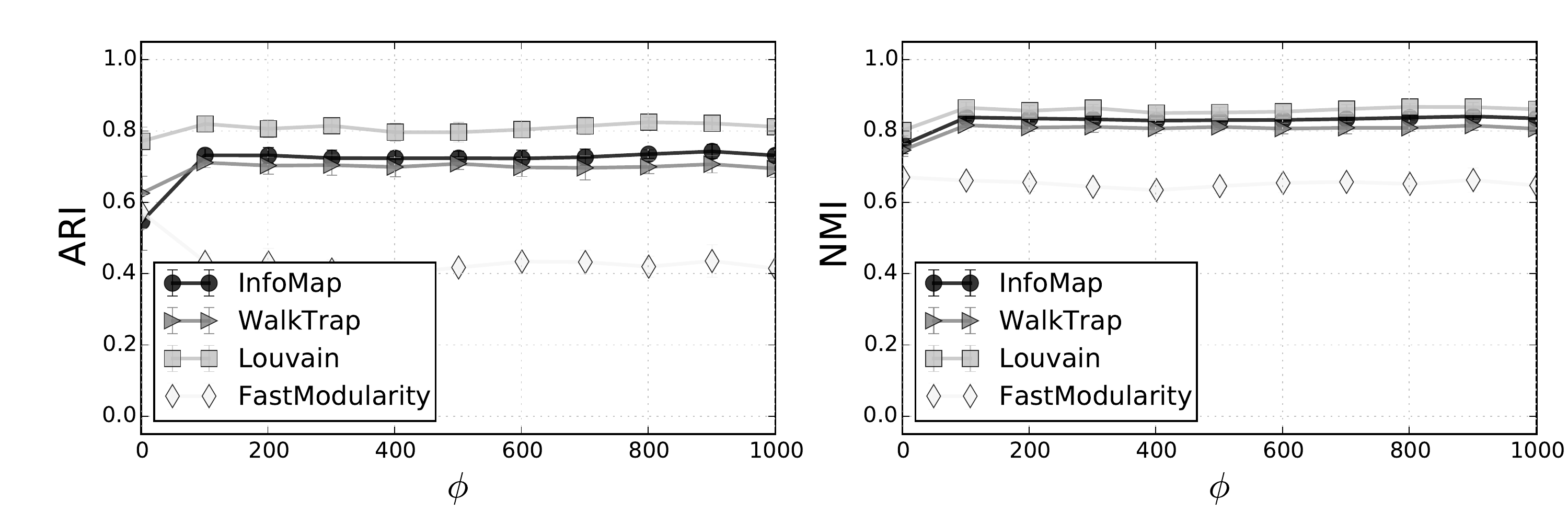}
\caption{assortative benchmarks, $\gamma = 0.5$}
\label{fig:pass}
\end{subfigure}
\begin{subfigure}[b]{0.5\textwidth}
\includegraphics[trim={1cm 0.1cm 0cm 0cm},clip,width=.98\linewidth]{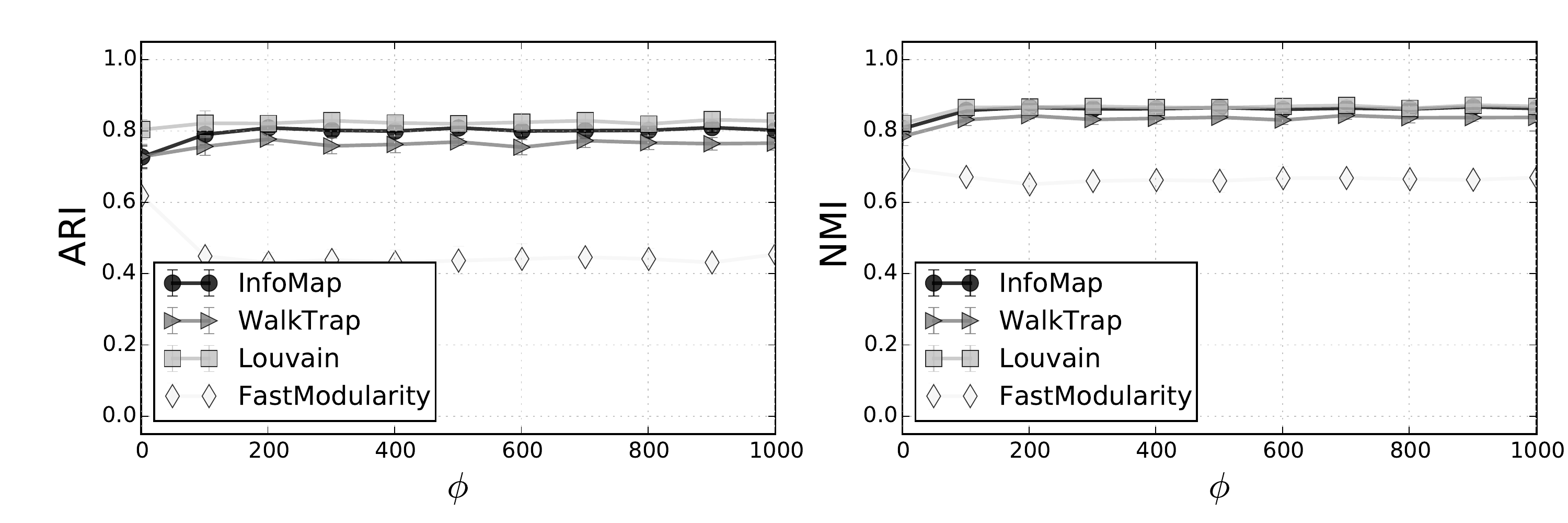}
\caption{disassortative benchmarks, $\gamma =-0.8$}
\label{fig:pdiss}
\end{subfigure}
\caption{Performance of community mining algorithms as a function of how equal are the \textbf{sizes of communities}. Settings correspond to the \reffig{fig:bassdiss}, except  $k$ that is increased to $20$ to have more community sizes. }
\label{fig:passdiss}\vspace{-10pt}
\end{figure}

Here, we tune the parameter $\phi$ to change how well-balanced communities are in sizes, i.e. move the distribution of community sizes form heavy tail to uniform. \reffig{fig:passdiss} shows the comparison results. Similar to the effect of number of communities, FastModularity seems to be the least consistent method when the distribution of community sizes changes. While Lovain seems to be the superior method particularly in the assortative setting.

\subsection{Effect of the Overlap}
\begin{figure*}[h!]
\includegraphics[trim={1cm 0.1cm 0cm 0cm},clip,width=.6\linewidth]{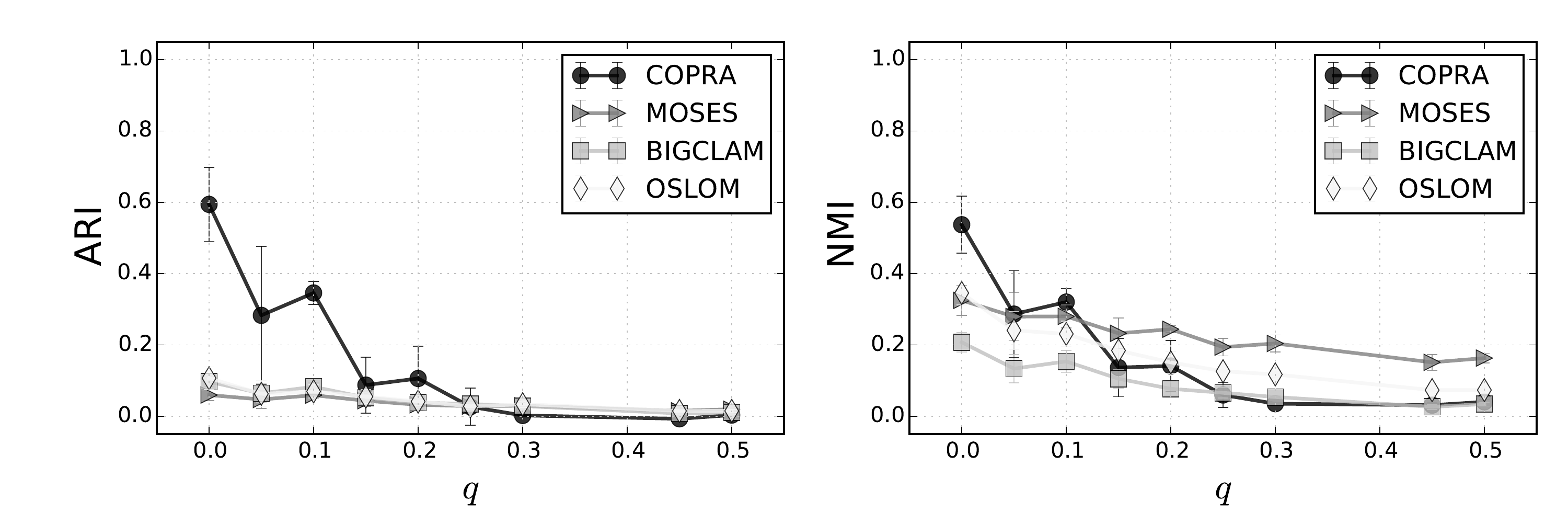}
\includegraphics[trim={0cm 0.1cm 0cm 0cm},clip,width=.3\linewidth]{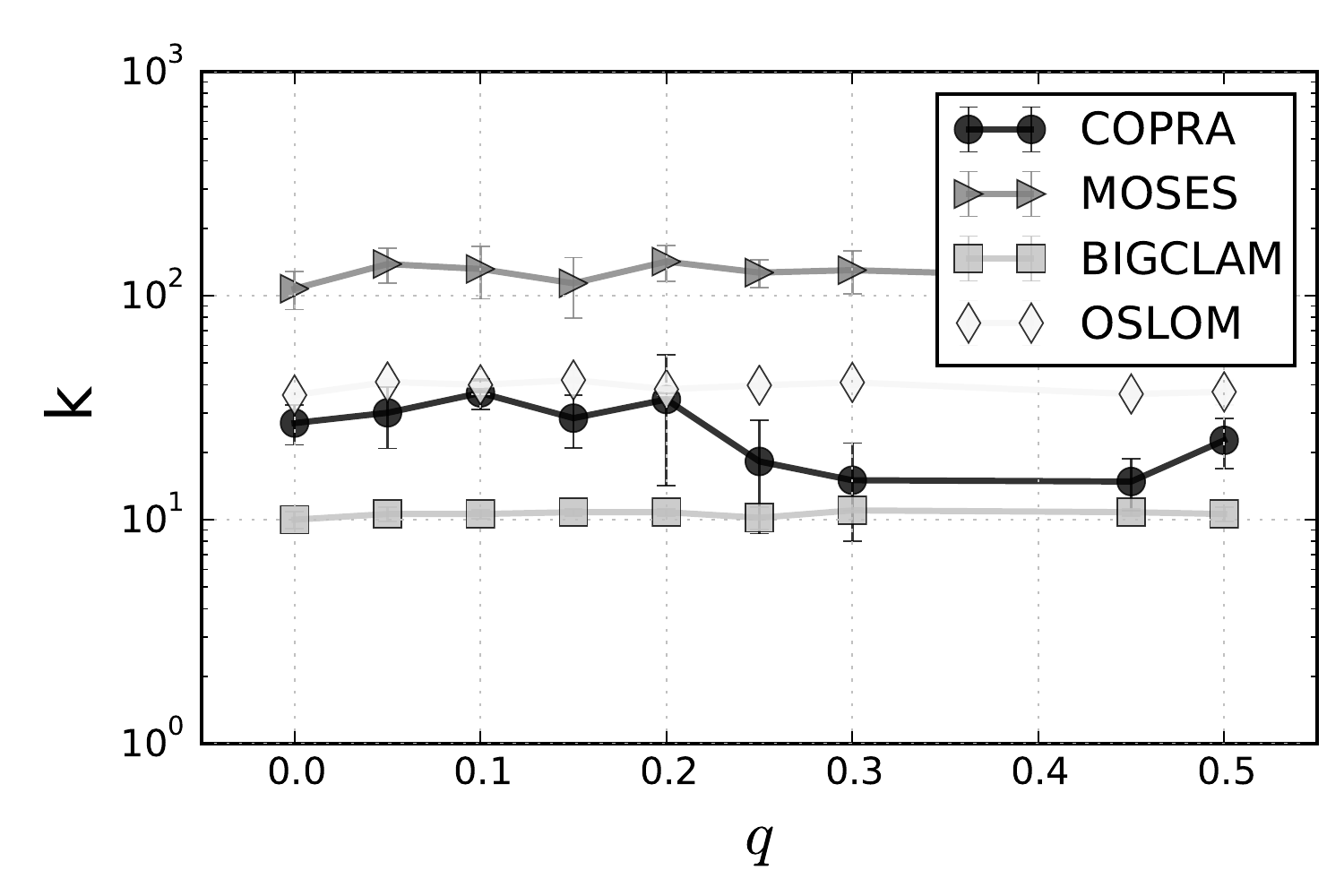}
\caption[Performances as a function of the fraction of overlapping nodes]{Performances as a function of the fraction of overlapping nodes, for the setting of $\alpha=0.5$, $\beta=0.5$, $\gamma = 0.8$, where the number of communities that each node can belong to is fixed to 3, and the portion of overlapping nodes ($q$) is varied from 0.0 (no overlap), to 0.5 (half of the nodes are overlapping). }
\label{fig:overlapfracfar}
\end{figure*}

\reffig{fig:overlapfracfar} compares the performance of four overlapping community detection methods on the FARZ benchmarks with overlapping communities.  
We can see that all methods perform poorly, except COPRA, which is able to detect communities when the portion of overlapping nodes is small enough, i.e. $q<0.2$. 
This is also interesting since these methods are shown to perform reasonably good on the overlapping extensions of LFR.

%% file: farz_res_extended.tex
\newpage
\section{Appendix: Extended Results}
\label{farzres}
\begin{figure}[h!]
\caption{Degree distributions per community for synthetic networks generated by FARZ for 4 different settings of Figure 7. The first plot reports the degree
distribution for the overall network, and the subsequent subplots
show the degree distribution per community.
}
\centering
\begin{subfigure}[b]{0.45\textwidth}
\caption{$\alpha = 0.2, \; \gamma = -0.8, \; m = 5,\; k=4$}
\includegraphics[trim={0cm 0cm 0cm 0cm},clip,width=1\linewidth]{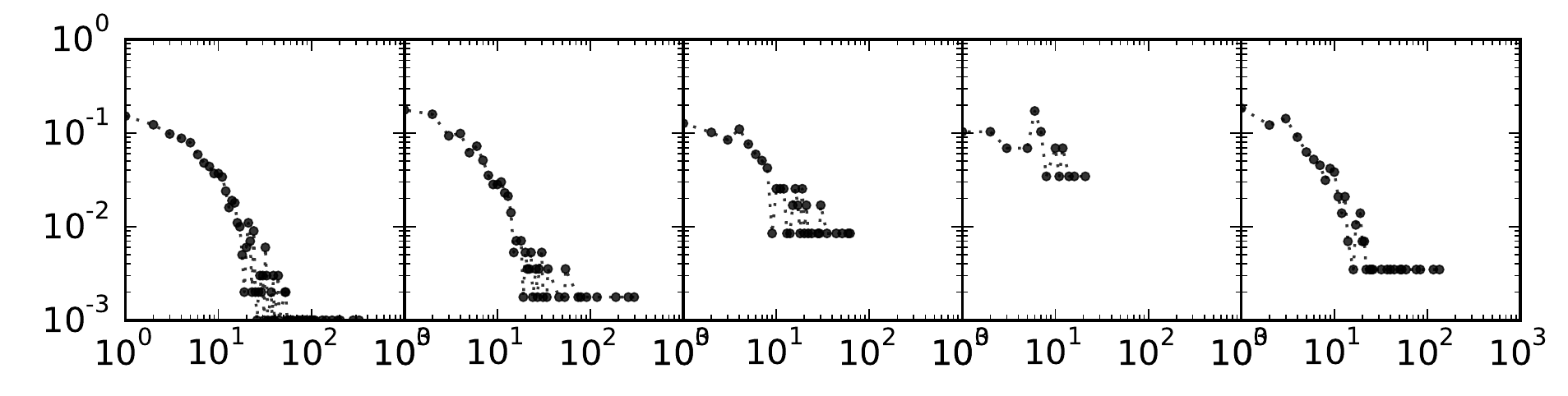}
\end{subfigure}
\begin{subfigure}[b]{0.45\textwidth}
\caption{$\alpha = 0.5, \; \gamma = -0.5, \; m = 5,\; k=4$}
\includegraphics[trim={0cm 0cm 0cm 0cm},clip,width=1\linewidth]{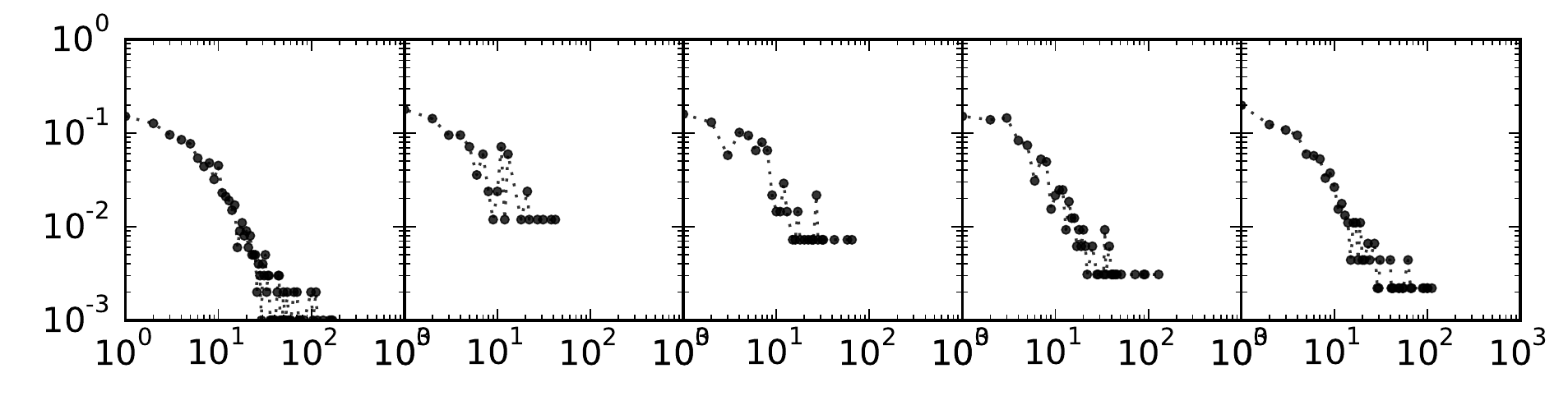}
\end{subfigure}
\begin{subfigure}[b]{0.45\textwidth}
\caption{$\alpha = 0.5, \; \gamma = 0.5, \; m = 5,\; k=4$}
\includegraphics[trim={0cm 0cm 0cm 0cm},clip,width=1\linewidth]{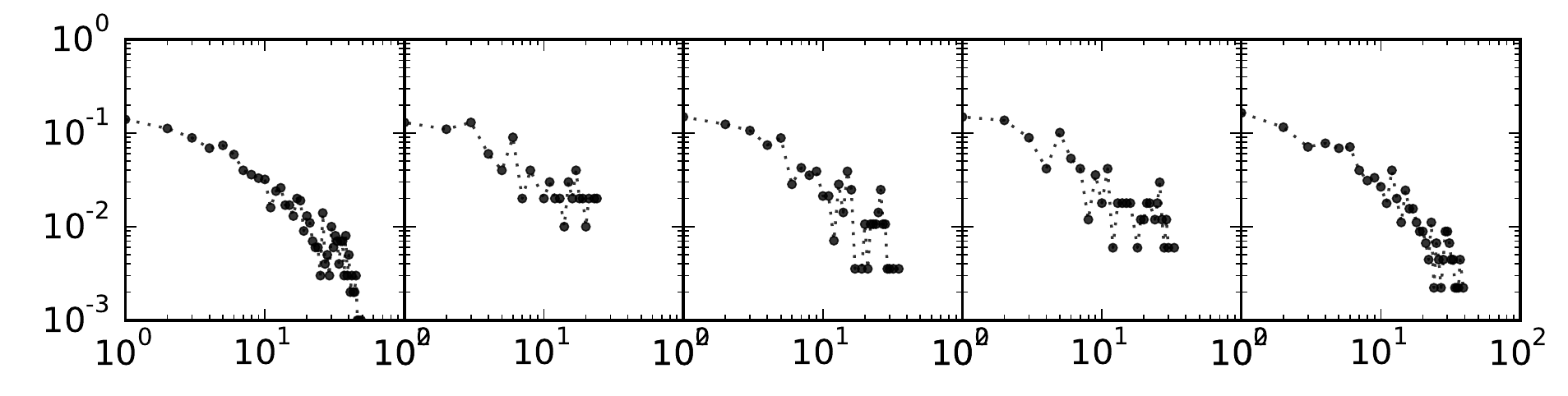}
\end{subfigure}
\begin{subfigure}[b]{0.45\textwidth}
\caption{$\alpha = 0.8, \; \gamma = -0.2, \; m = 5,\; k=4$}
\includegraphics[trim={0cm 0cm 0cm 0cm},clip,width=1\linewidth]{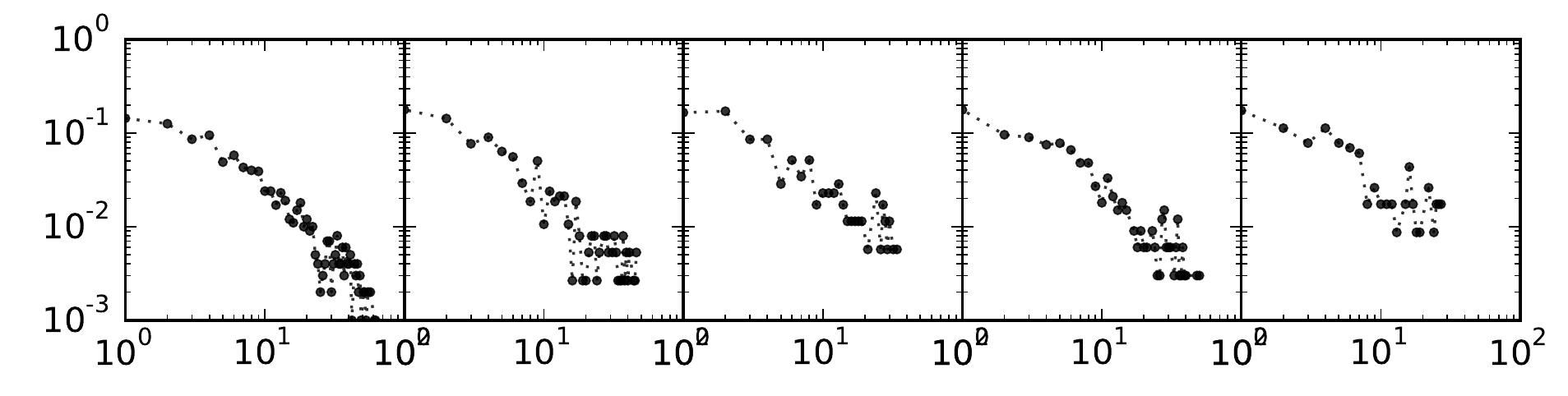}
\end{subfigure}
\label{fig:beta}
\end{figure}


\begin{figure}[t]
\centering
\caption{Degree distributions per community for synthetic networks generated by LFR of Figure 6 in the paper.}
\begin{subfigure}{0.5\textwidth}
\includegraphics[trim={0cm 0cm 0cm 0cm},clip,width=1\linewidth]{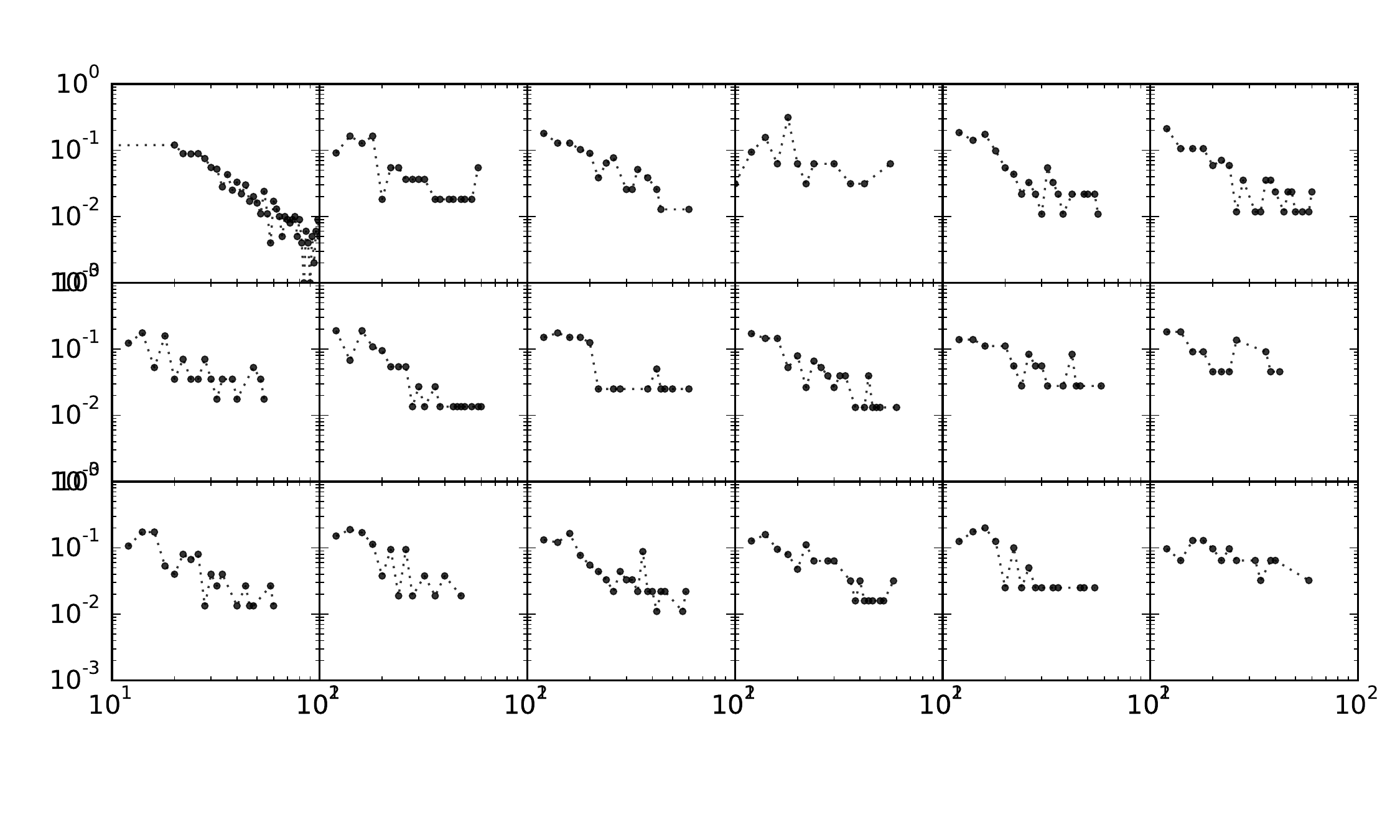}
\end{subfigure}
 \end{figure}
 
\begin{figure*}
\centering
\caption{Comparing performance of community mining algorithms on  benchmarks with  \textbf{positive and negative degree correlation}, all the four settings. Also reporting the number of clusters found by each method. }
\begin{subfigure}[b]{1\textwidth}
\caption{$\alpha = 0.5, \; \gamma = 0.5, \; m = 5,\; k=4$}
\includegraphics[width=.3\linewidth]{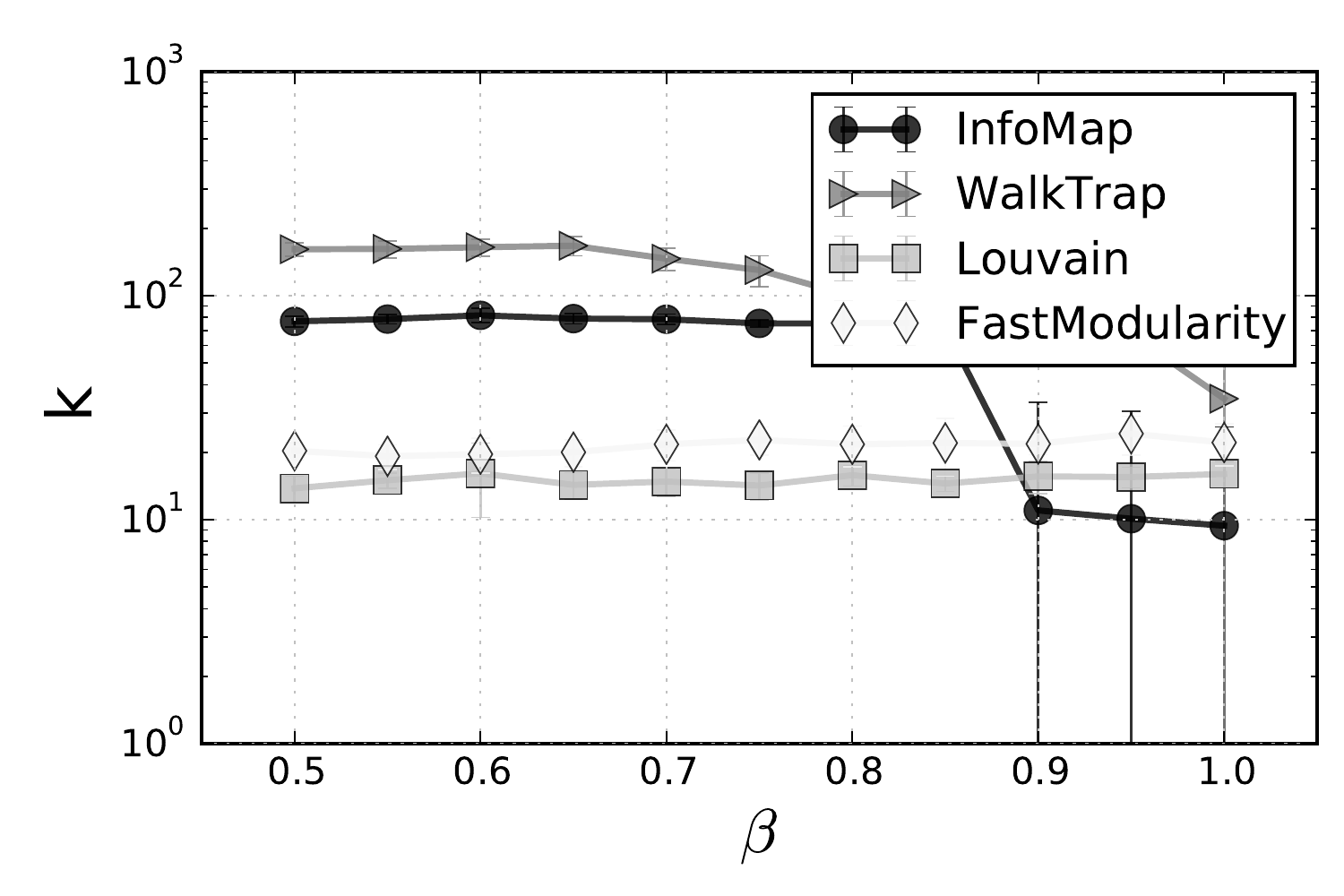}
\includegraphics[clip,width=.6\linewidth]{figs/resvbeta55}
\end{subfigure}
\begin{subfigure}[b]{1\textwidth}
\caption{$\alpha = 0.8, \; \gamma = 0.2, \; m = 5,\; k=4$}
\includegraphics[width=.3\linewidth]{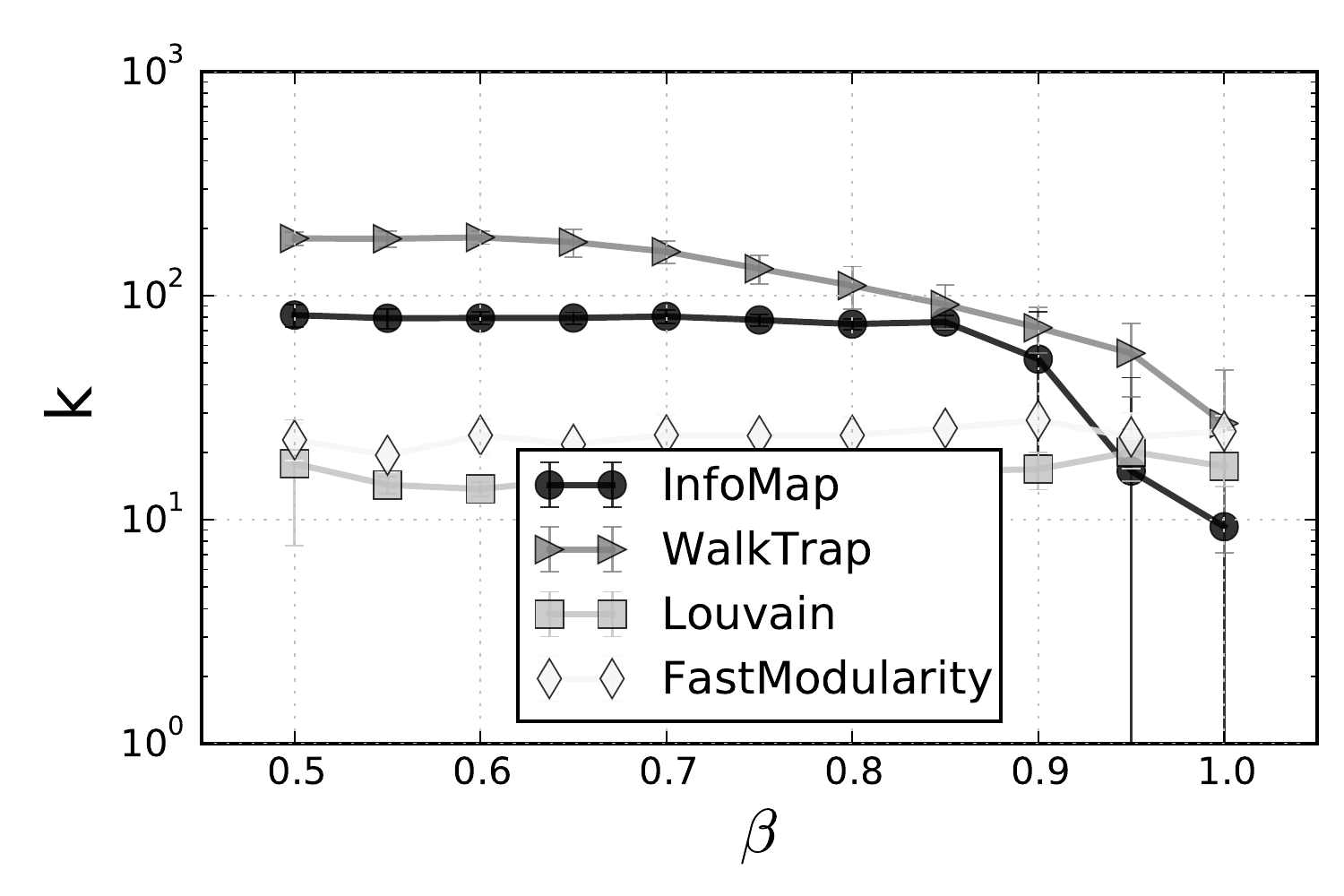}
\includegraphics[width=.6\linewidth]{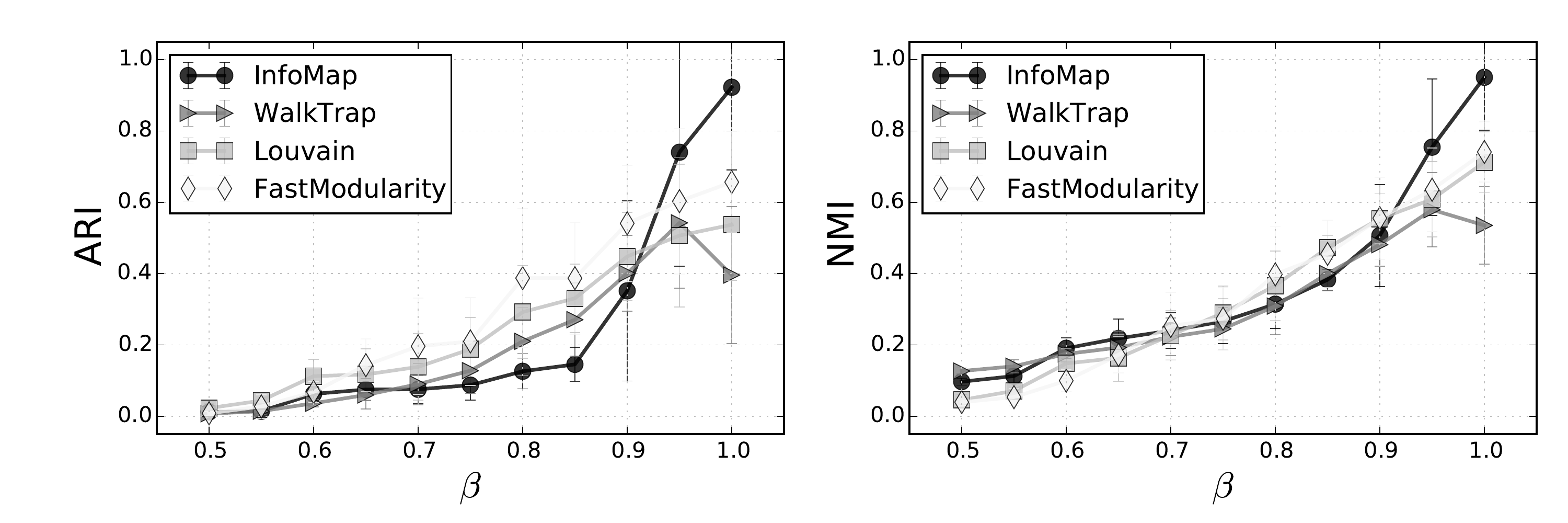}
\end{subfigure}
\begin{subfigure}[b]{1\textwidth}
\caption{$\alpha = 0.2, \; \gamma = -0.8, \; m = 5,\; k=4$}
\includegraphics[width=.3\linewidth]{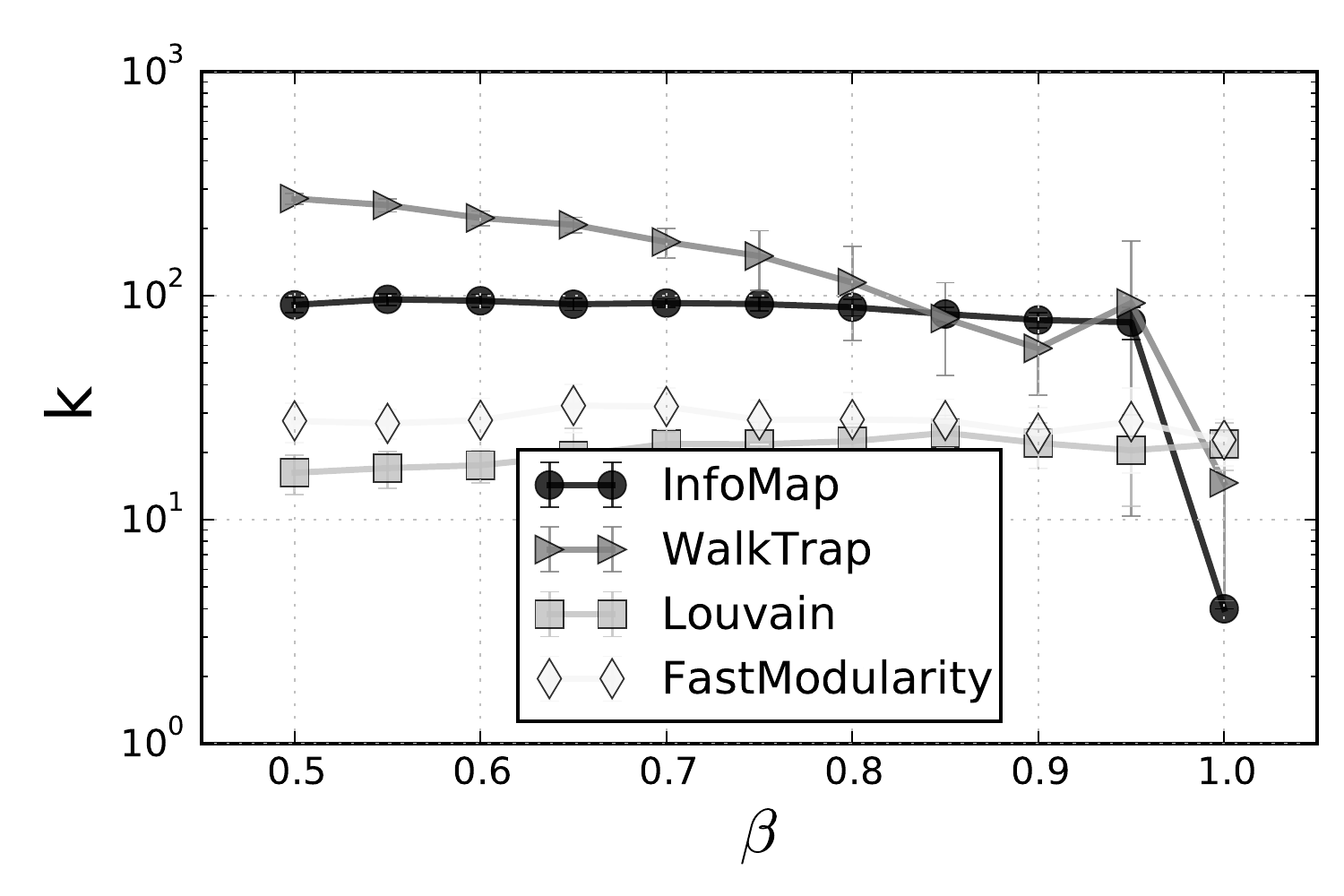}
\includegraphics[width=.6\linewidth]{figs/resvbeta2-8}\end{subfigure}
\begin{subfigure}[b]{1\textwidth}
\caption{$\alpha = 0.5, \; \gamma = -0.5, \; m = 5,\; k=4$}
\includegraphics[width=.3\linewidth]{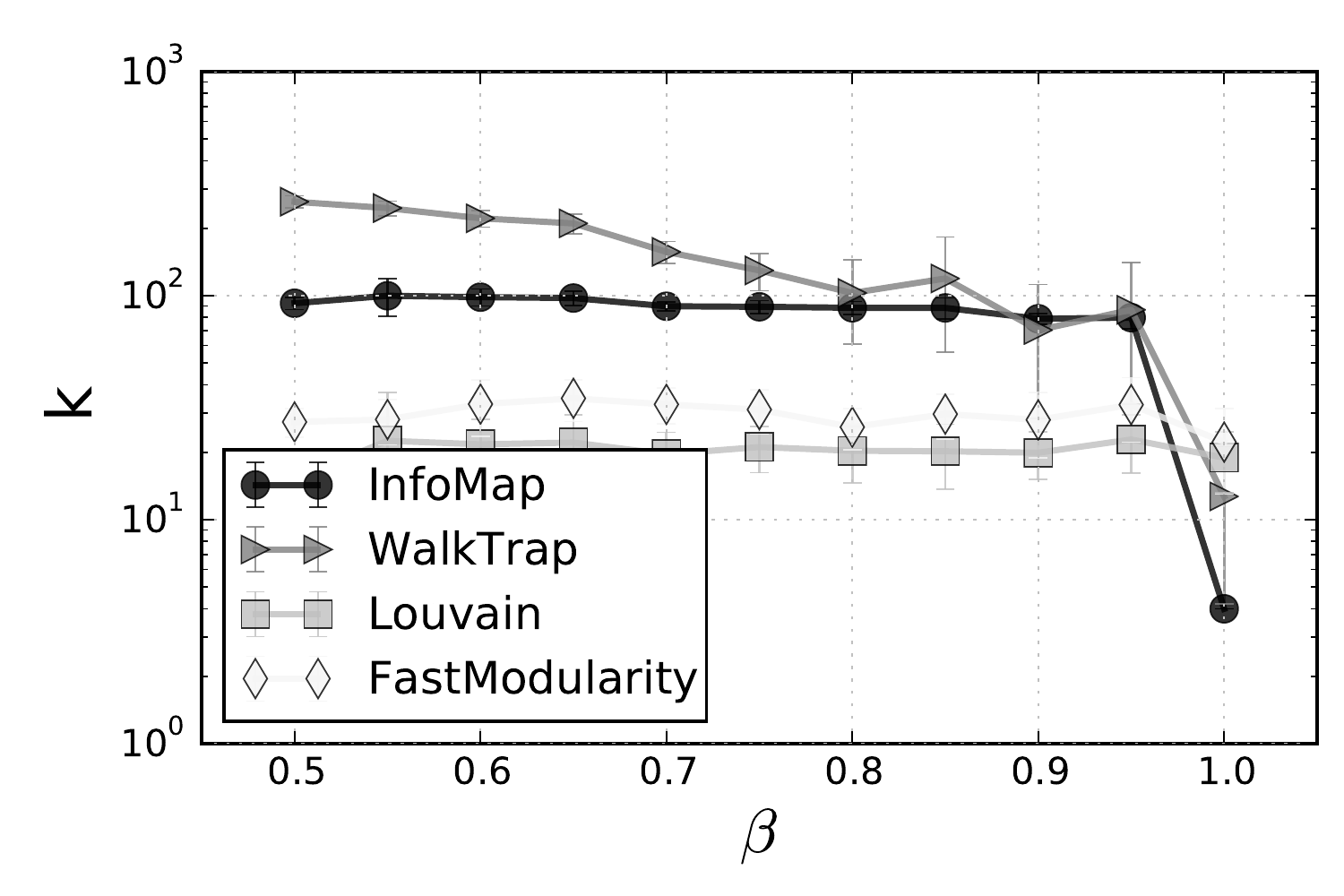}
\includegraphics[width=.6\linewidth]{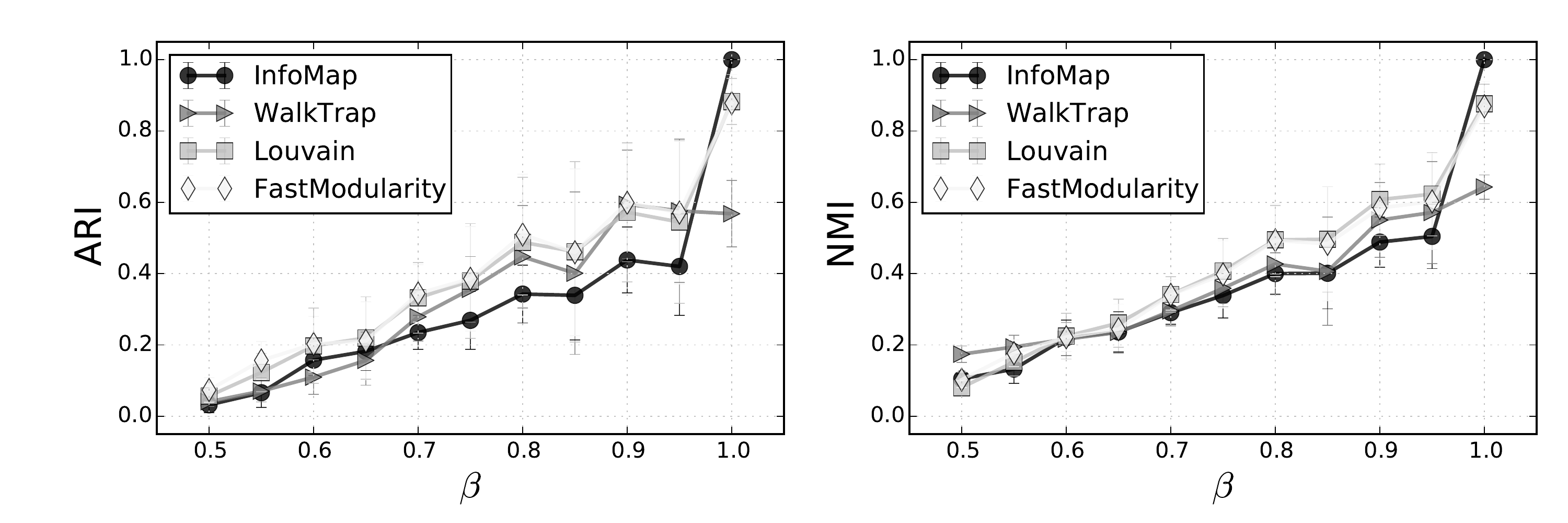}
\end{subfigure}
\end{figure*}

\begin{figure*}
\caption{Same algorithms compared on LFR, setting is the 1000B used in \cite{Lancichinetti09Comparison}, i.e. -N 1000 -k 20  -maxk 50 -t1 2 -t2 1 -minc 20 -maxc 100. }
\label{fig:farzsupp7}
\includegraphics[width=.3\linewidth]{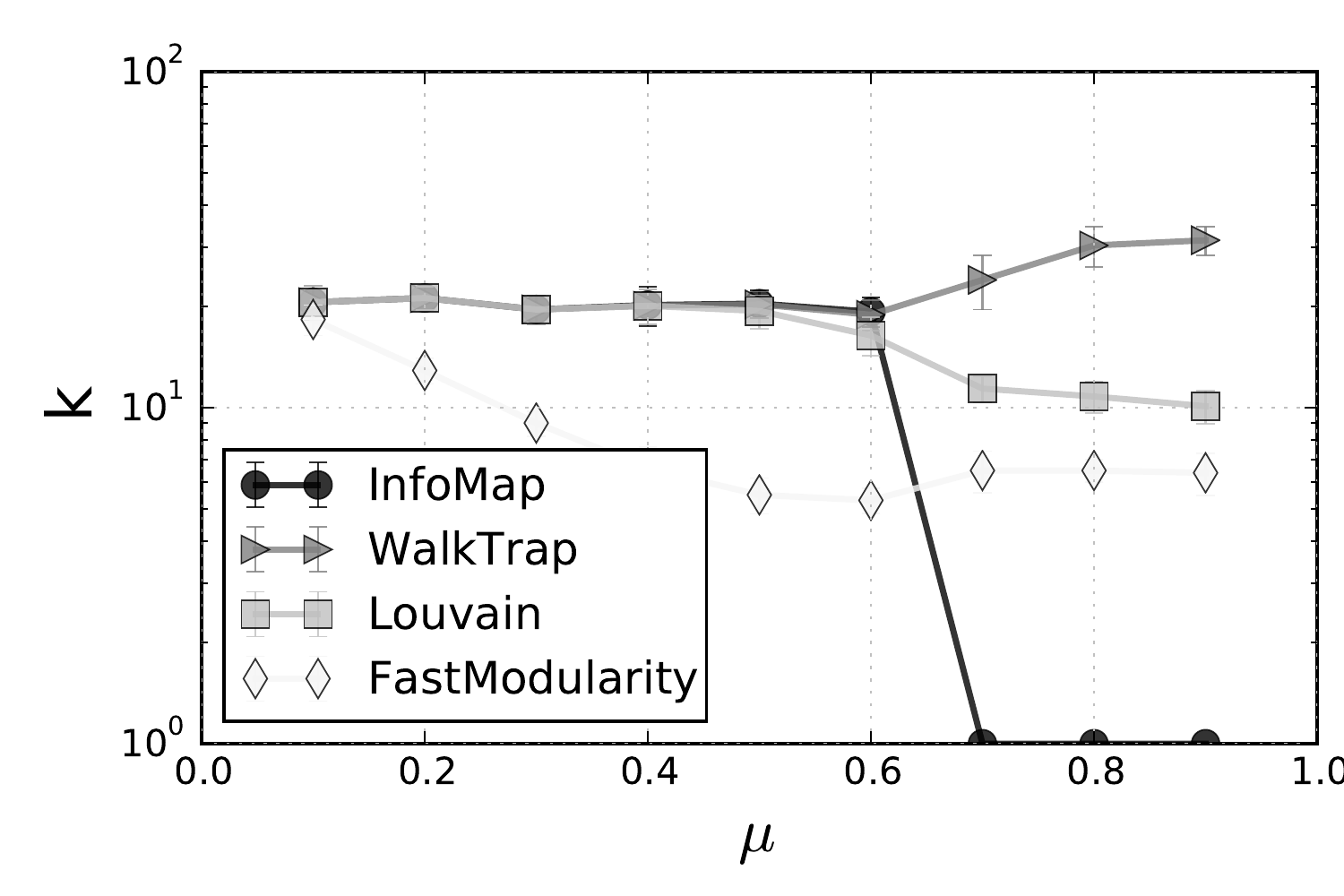}
\includegraphics[width=.6\linewidth]{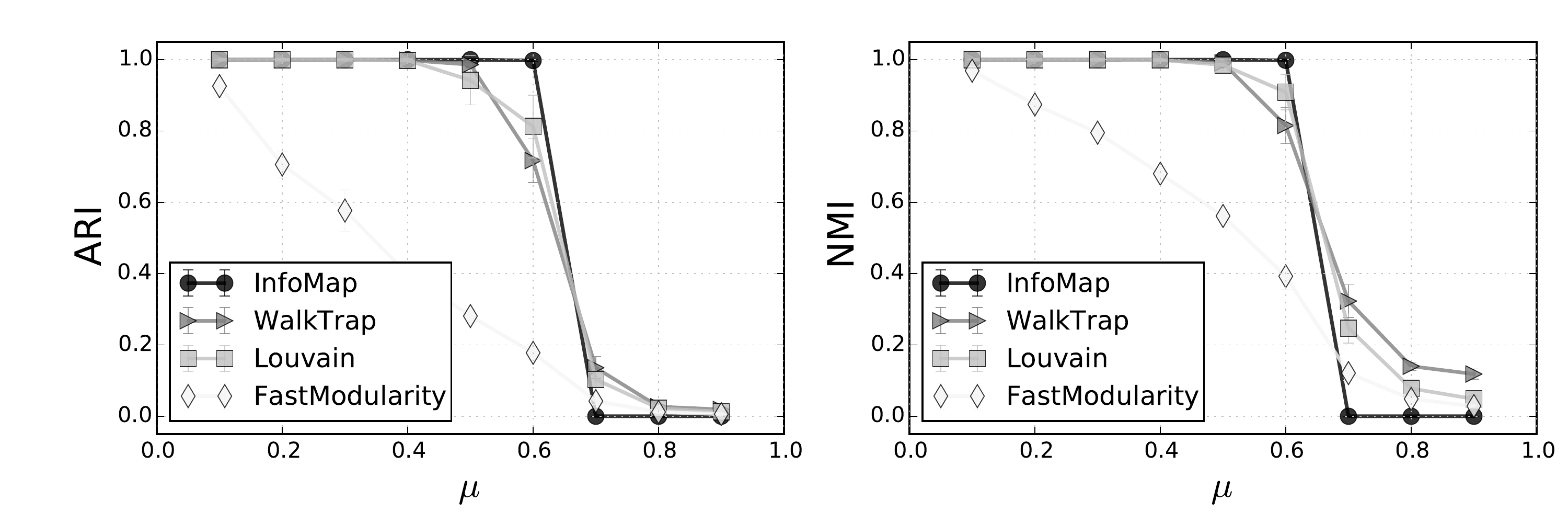}
\end{figure*}

\begin{figure*}
\caption{Comparing performance of community mining algorithms similar to Figure 6 but on \textbf{denser benchmarks ($m=6$) with more communities ($k=20$)}. In this setting Louvain clearly performs the best in particular in networks with positive degree correlation. 
The drop in $\beta =1 $ is due to the communities not linked together which makes the network disconnected and causes problem for the WalkTrap algorithm. The other random walk based method, InfoMap, also seems to have difficulty when communities are well separated, i.e. when $\beta \in [.85, .95]$ and $\gamma>0$. }
\begin{subfigure}[b]{1\textwidth}
\caption{$\alpha = 0.5, \; \gamma = 0.5, \; m = 7,\; k=20$}
\includegraphics[width=.3\linewidth]{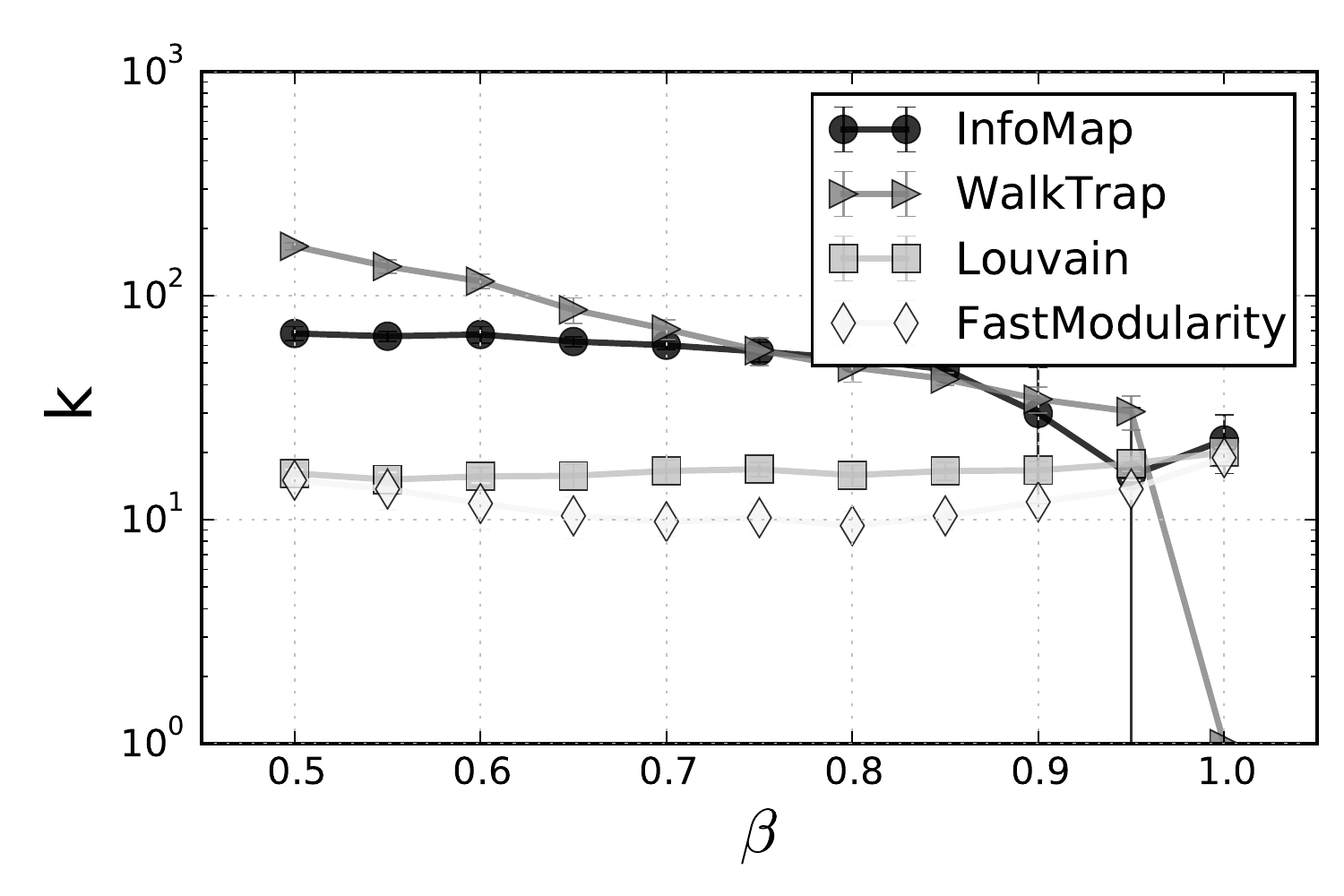}
\includegraphics[width=.6\linewidth]{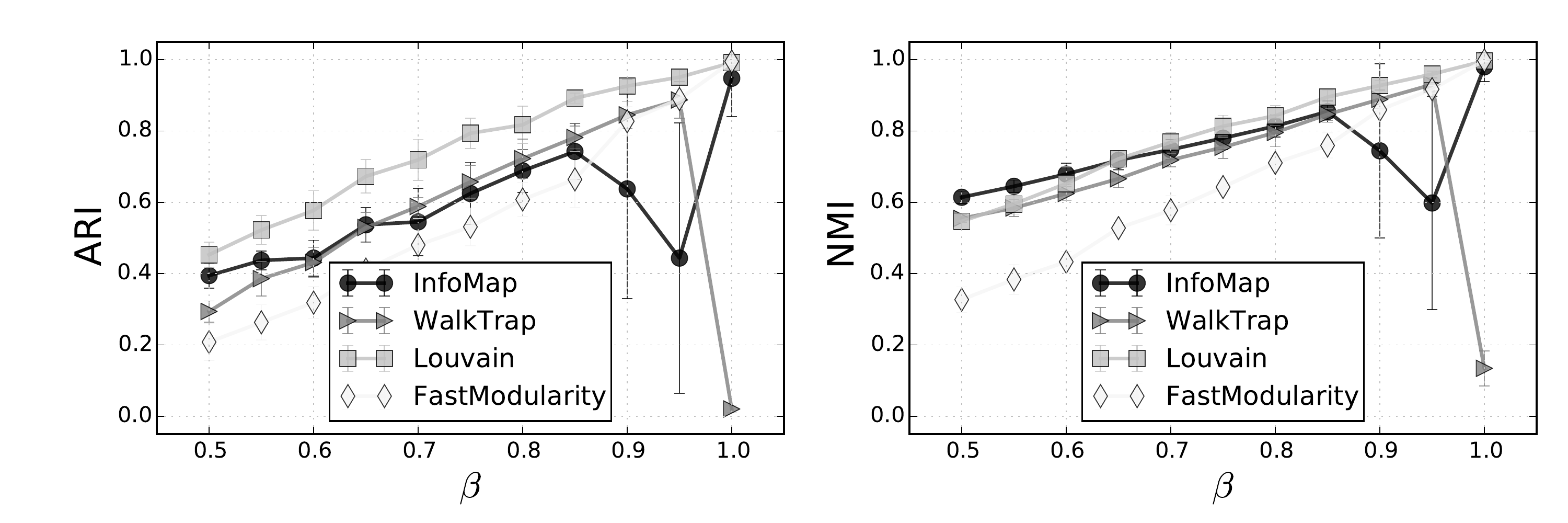}
\end{subfigure}
\begin{subfigure}[b]{1\textwidth}
\caption{$\alpha = 0.8, \; \gamma = 0.2, \; m = 7,\; k=20$}
\includegraphics[width=.3\linewidth]{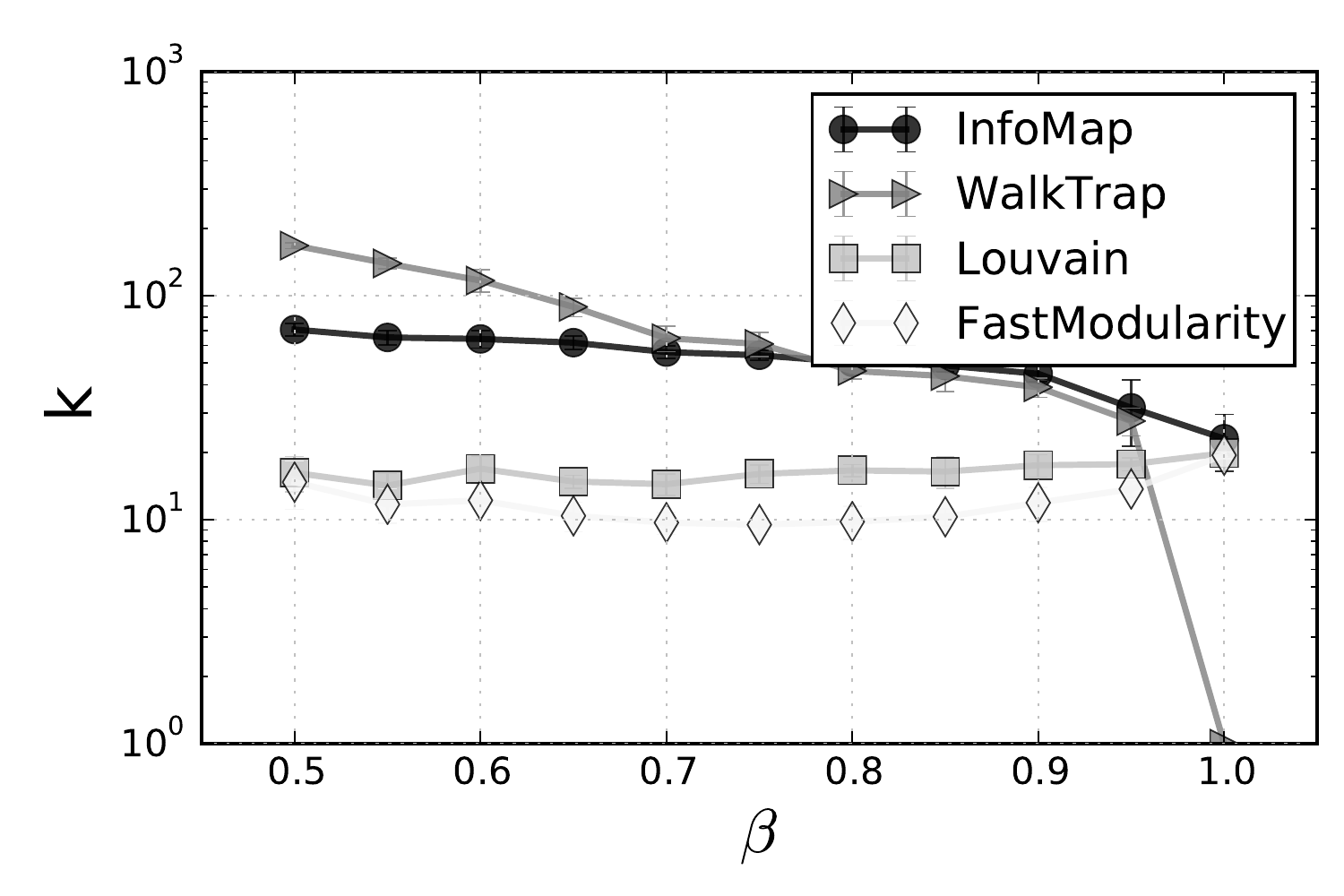}
\includegraphics[width=.6\linewidth]{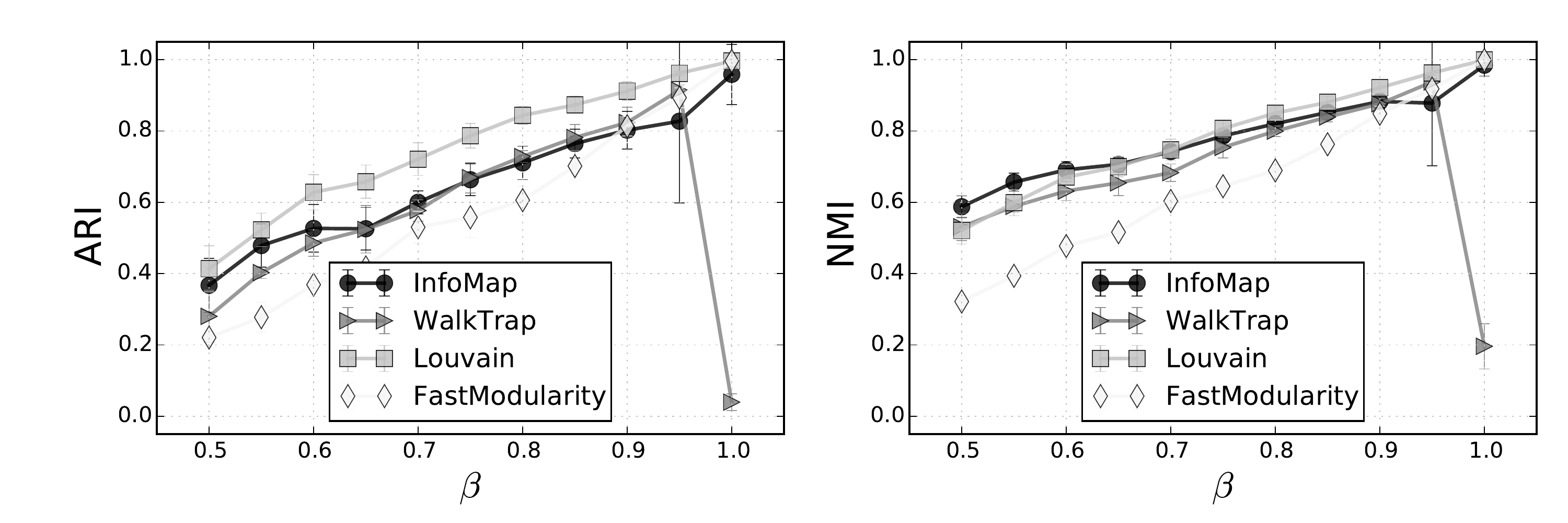}
\end{subfigure}
\begin{subfigure}[b]{1\textwidth}
\caption{$\alpha = 0.2, \; \gamma = -0.8, \; m = 7,\; k=20$}
\includegraphics[width=.3\linewidth]{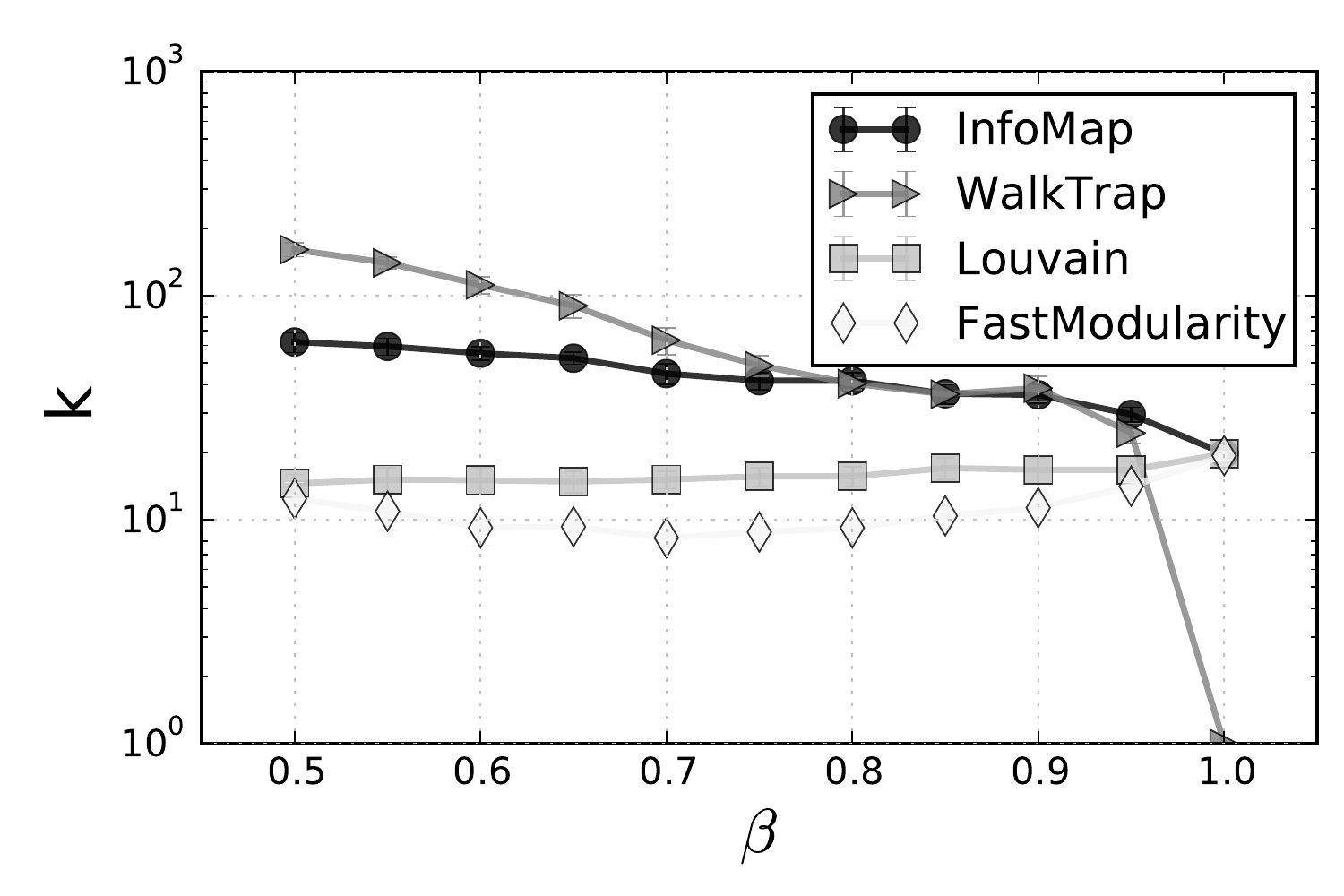}
\includegraphics[width=.6\linewidth]{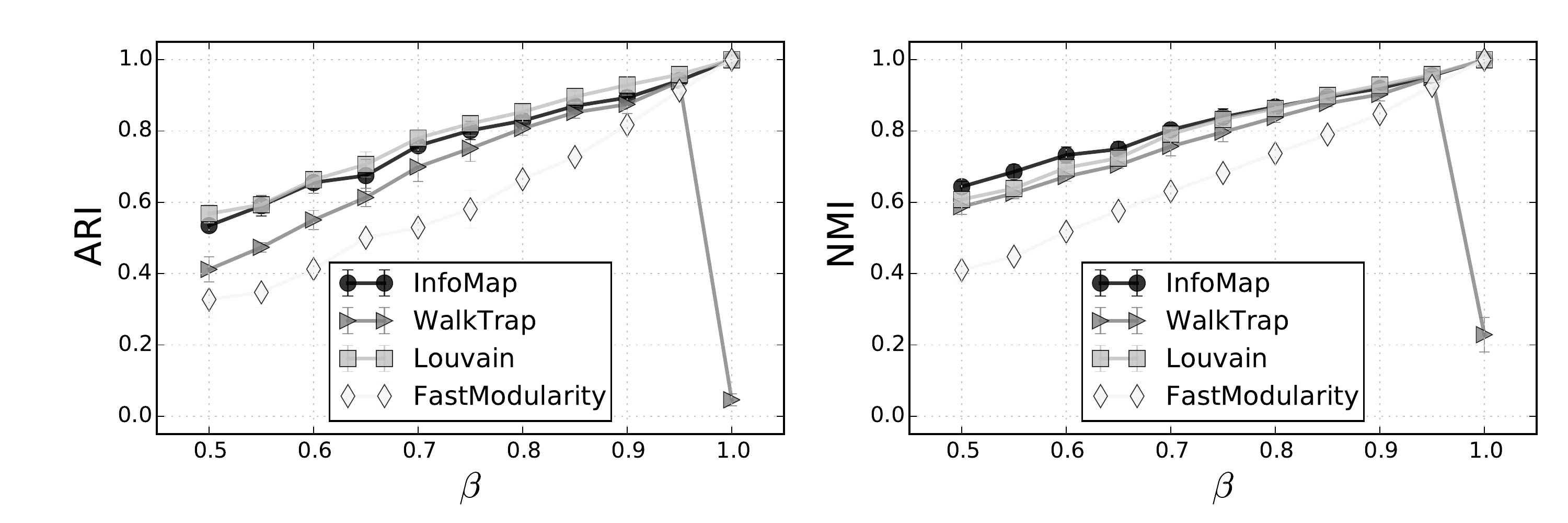}
\end{subfigure}
\begin{subfigure}[b]{1\textwidth}
\caption{$\alpha = 0.5, \; \gamma = -0.5, \; m = 7,\; k=20$}
\includegraphics[width=.3\linewidth]{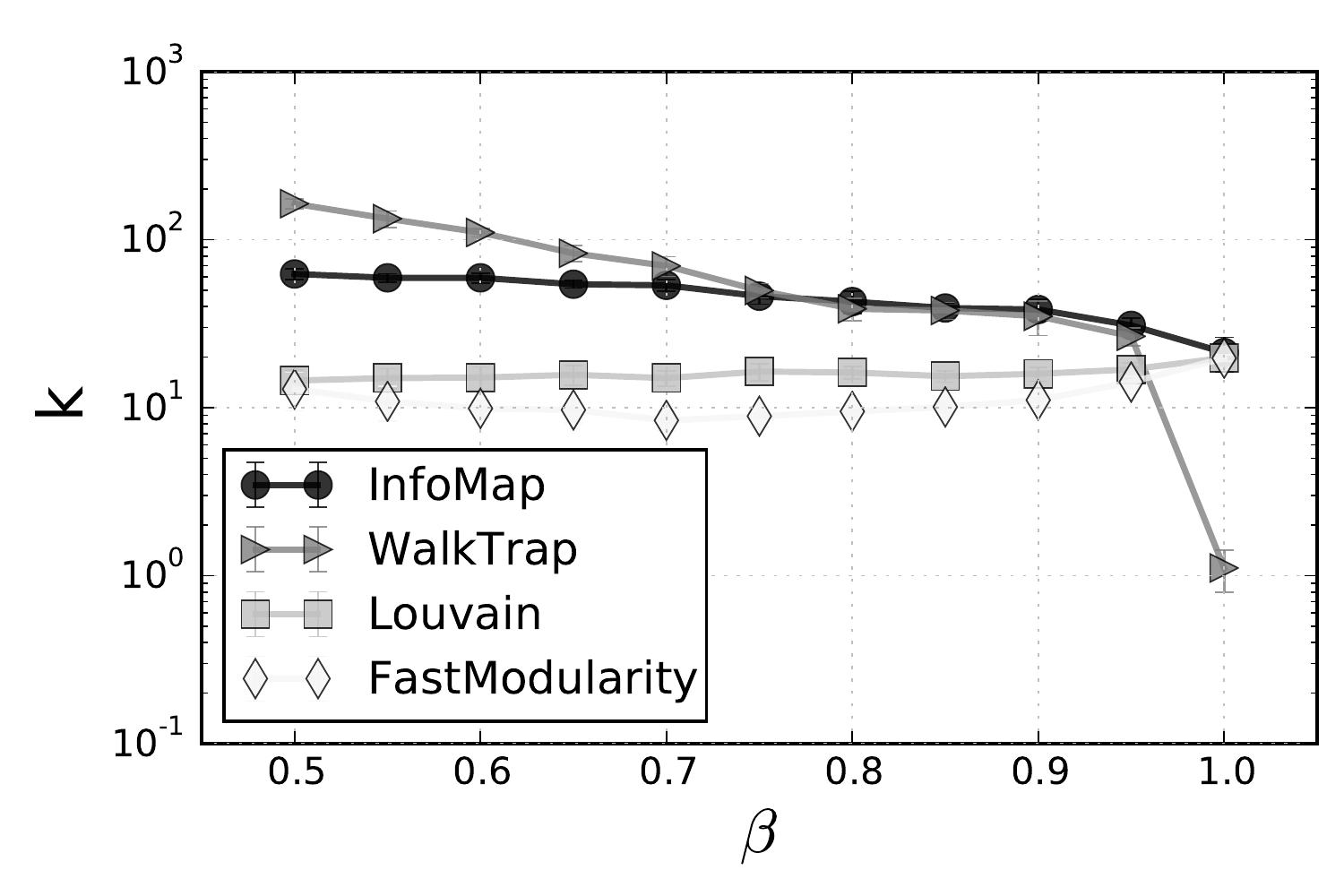}
\includegraphics[width=.6\linewidth]{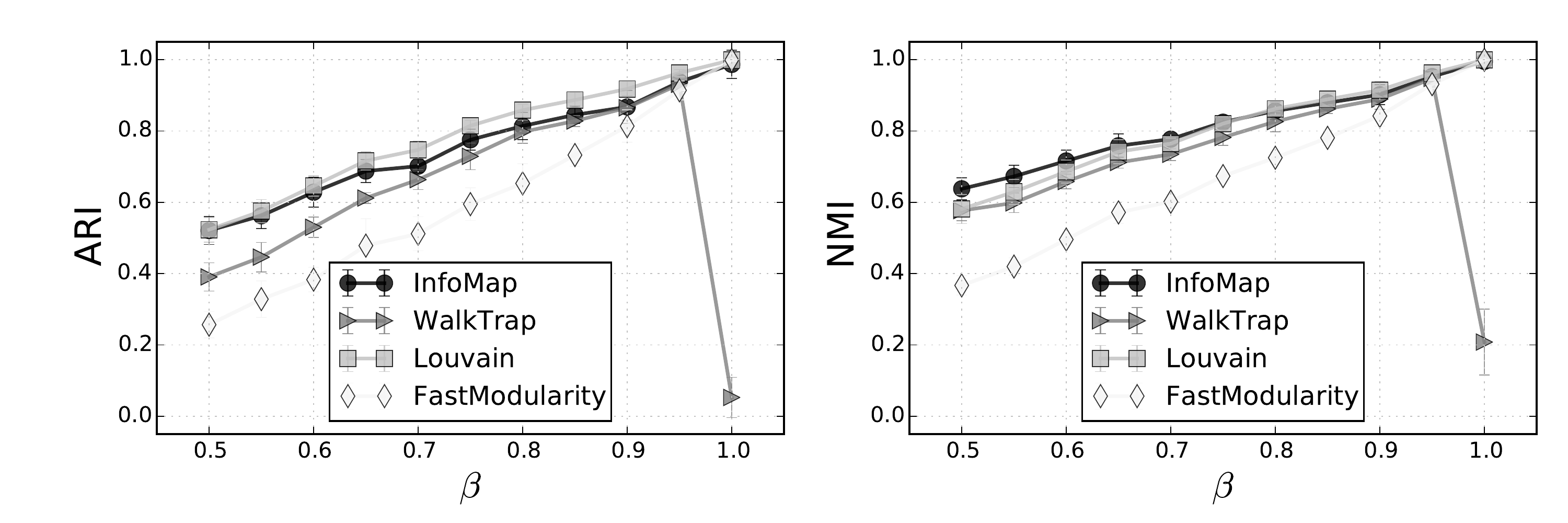}
\end{subfigure}
\end{figure*}

\begin{figure*}
\centering
\caption{Performance of community mining algorithms on benchmarks with different \textbf{number of built-in communities}, for  all the four settings.  Also reporting the number of clusters found by each method. }
\begin{subfigure}[b]{1\textwidth}
\caption{$\alpha = 0.5, \; \gamma = 0.5, \; m = 5,\; k=4$}
\includegraphics[trim={0cm 0cm 0cm 0cm},clip,width=.3\linewidth]{figs/resk55k}\centering
\includegraphics[trim={0cm 0cm 0cm 0cm},clip,width=.6\linewidth]{figs/resk55}
\end{subfigure}
\begin{subfigure}[b]{1\textwidth}
\caption{$\alpha = 0.8, \; \gamma = 0.2, \; m = 5,\; k=4$}
\includegraphics[trim={0cm 0cm 0cm 0cm},clip,width=.3\linewidth]{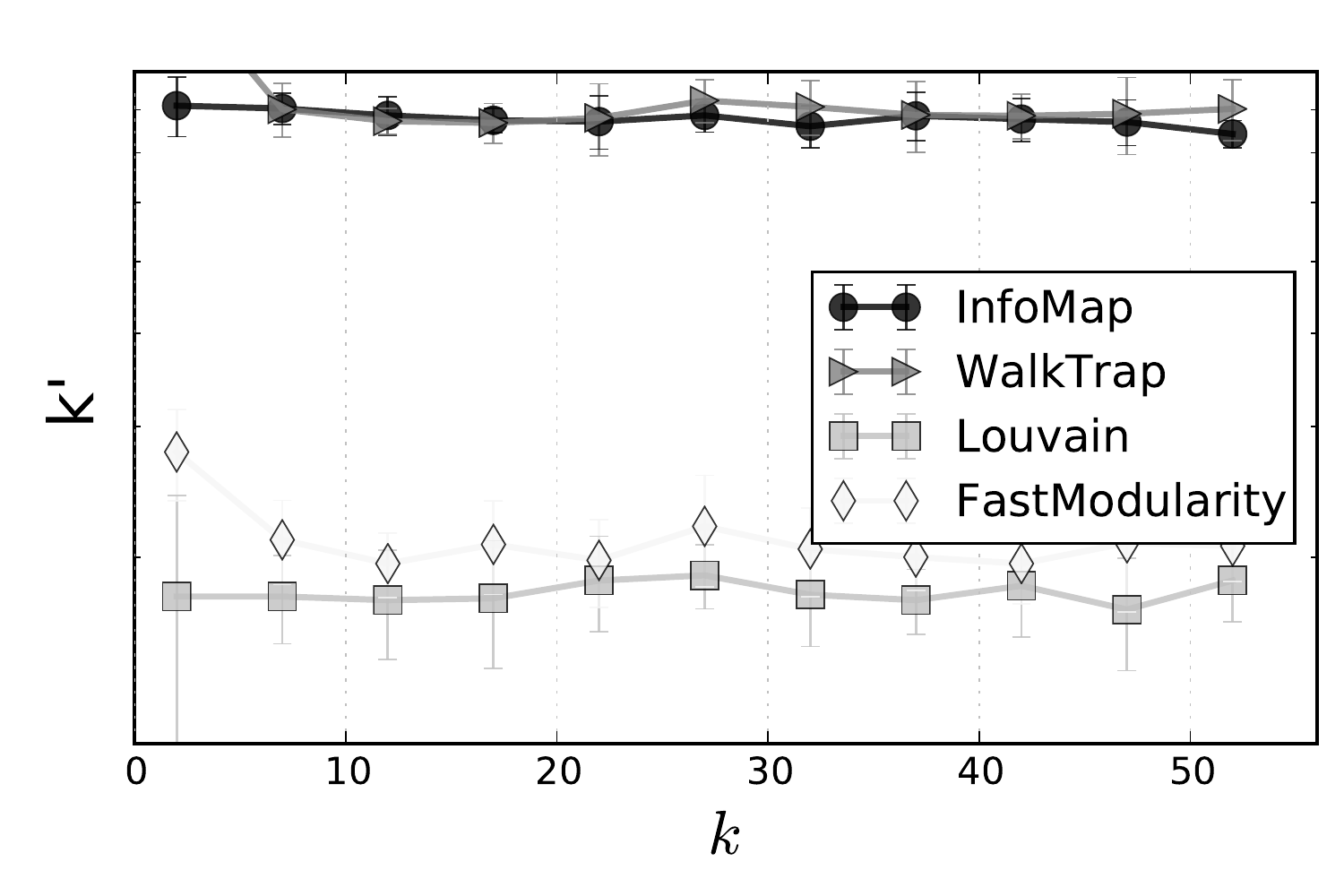}\centering
\includegraphics[trim={0cm 0cm 0cm 0cm},clip,width=.6\linewidth]{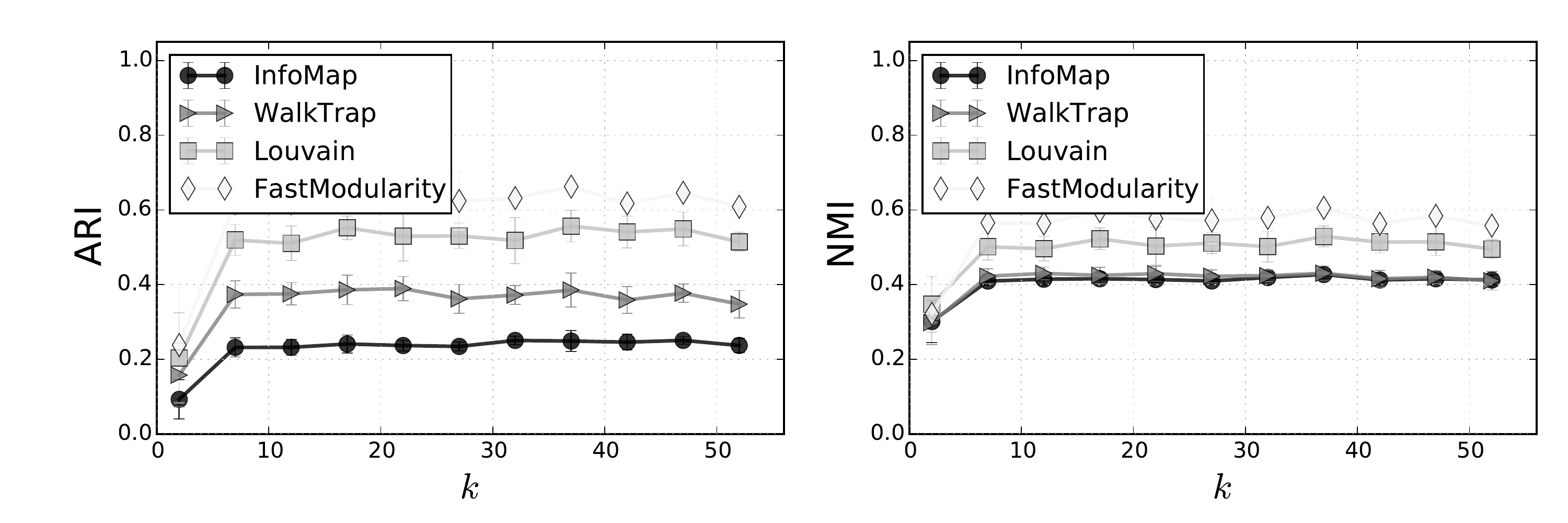}
\end{subfigure}
\begin{subfigure}[b]{1\textwidth}
\caption{$\alpha = 0.2, \; \gamma = -0.8, \; m = 5,\; k=4$}
\includegraphics[trim={0cm 0cm 0cm 0cm},clip,width=.3\linewidth]{figs/resk2-8k}\centering
\includegraphics[trim={0cm 0cm 0cm 0cm},clip,width=.6\linewidth]{figs/resk2-8}
\end{subfigure}
\begin{subfigure}[b]{1\textwidth}
\caption{$\alpha = 0.5, \; \gamma = -0.5, \; m = 5,\; k=4$}
\includegraphics[trim={0cm 0cm 0cm 0cm},clip,width=.3\linewidth]{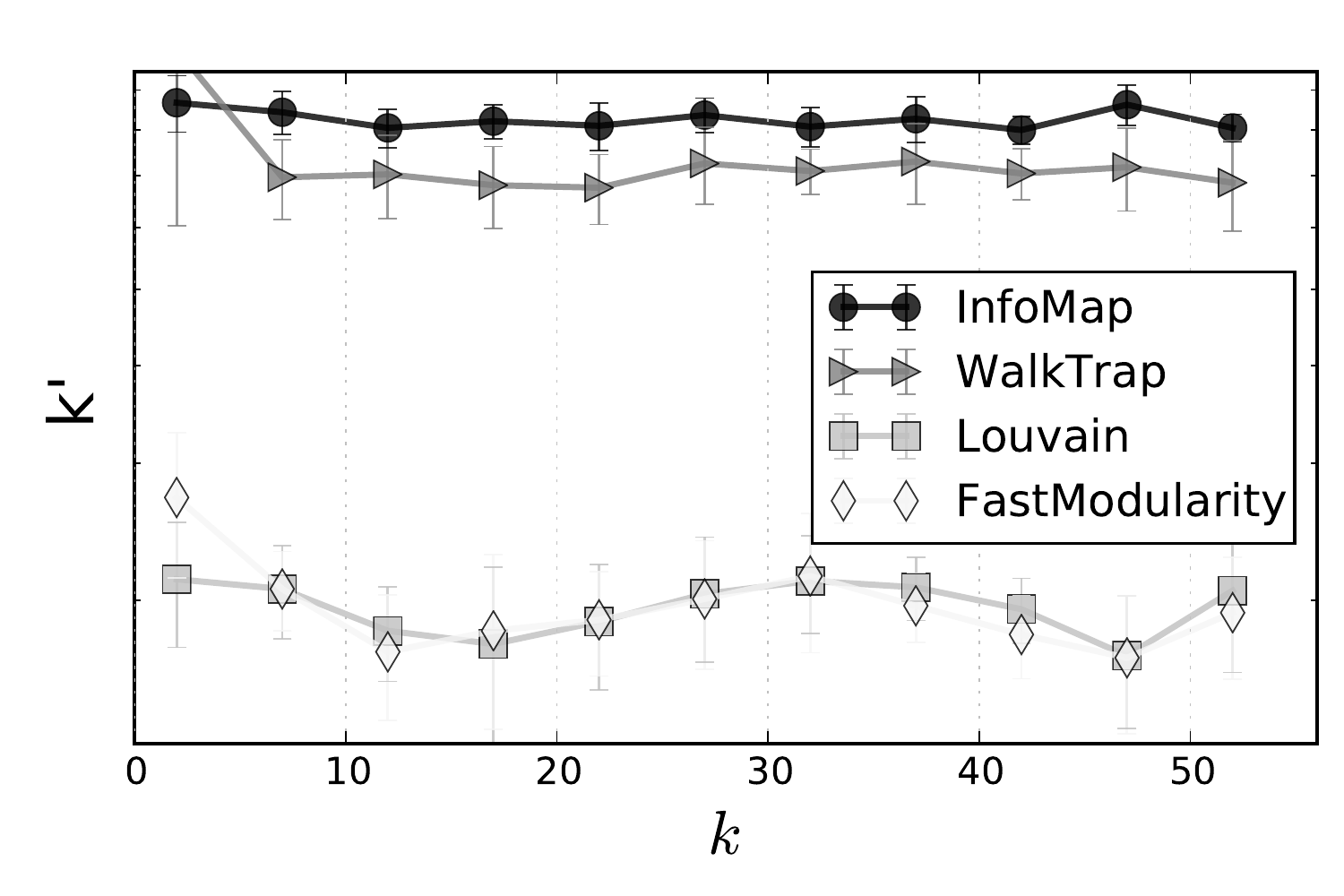}\centering
\includegraphics[trim={0cm 0cm 0cm 0cm},clip,width=.6\linewidth]{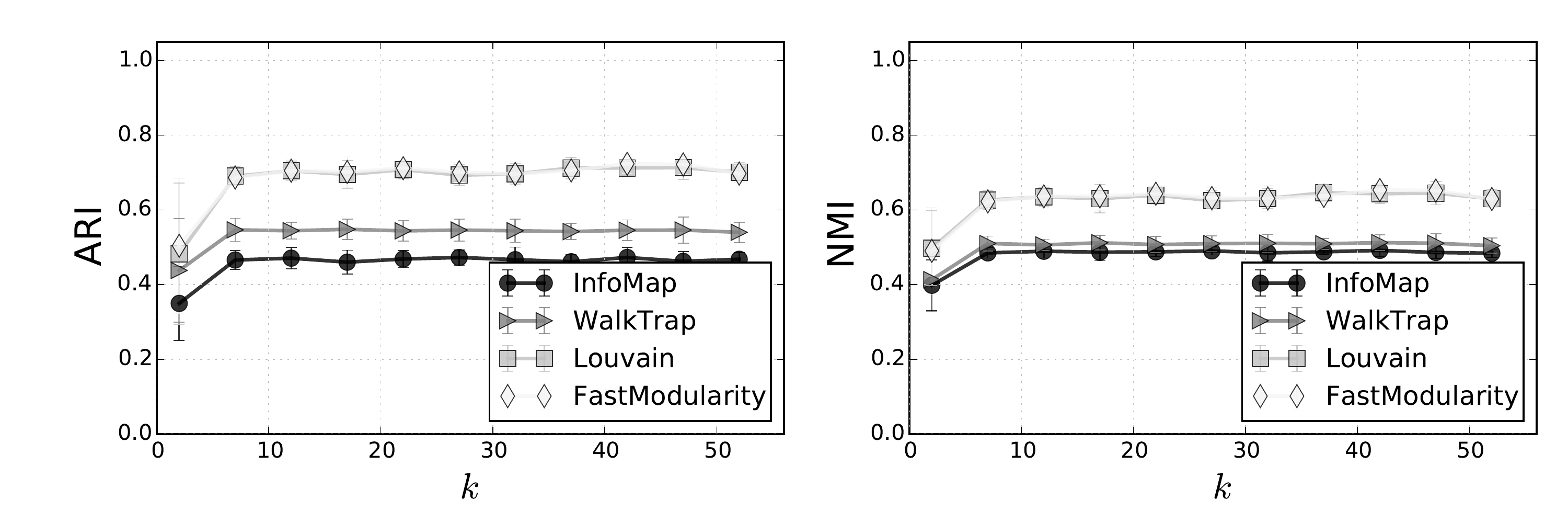}
\end{subfigure}
\end{figure*}

\begin{figure*}[h!]
\caption{Example of actual graphs generated by FARZ, used in the previous plots, $\alpha = 0.5, \; \gamma = 0.5, \; m = 5,\; k=4$. Plots visualized with Gephi toolbox using ForceAtlas2 layout, where node sizes corresponds to the degree of the nodes, and the colours of nodes to their assigned communities.}
\begin{subfigure}[b]{.32\textwidth}
\caption{$\beta = 1$}
\includegraphics[width=.99\linewidth]{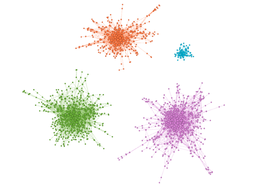}
\end{subfigure}
\begin{subfigure}[b]{.32\textwidth}
\caption{$\beta = 0.95$}
\includegraphics[width=.99\linewidth]{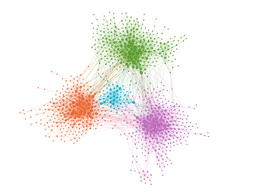}
\end{subfigure}
\begin{subfigure}[b]{.32\textwidth}
\caption{$\beta = 0.9$}
\includegraphics[width=.99\linewidth]{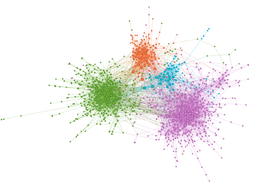}
\end{subfigure}
\begin{subfigure}[b]{.32\textwidth}
\caption{$\beta = 0.85$}
\includegraphics[width=.99\linewidth]{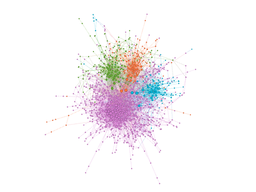}
\end{subfigure}
\begin{subfigure}[b]{.32\textwidth}
\caption{$\beta = 0.8$}
\includegraphics[width=.99\linewidth]{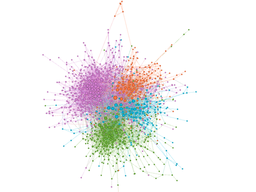}
\end{subfigure}
\begin{subfigure}[b]{.32\textwidth}
\caption{$\beta = 0.75$}
\includegraphics[width=.99\linewidth]{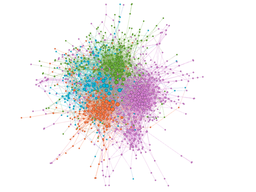}
\end{subfigure}
\begin{subfigure}[b]{.32\textwidth}
\caption{$\beta = 0.7$}
\includegraphics[width=.99\linewidth]{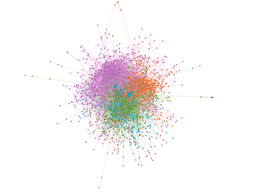}
\end{subfigure}
\begin{subfigure}[b]{.32\textwidth}
\caption{$\beta = 0.65$}
\includegraphics[width=.99\linewidth]{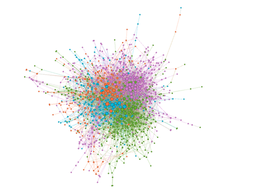}
\end{subfigure}
\begin{subfigure}[b]{.32\textwidth}
\caption{$\beta = 0.6$}
\includegraphics[width=.99\linewidth]{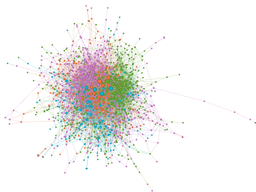}
\end{subfigure}
\begin{subfigure}[b]{.32\textwidth}
\caption{$\beta = 0.55$}
\includegraphics[width=.99\linewidth]{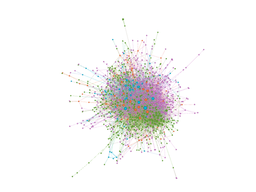}
\end{subfigure}
\begin{subfigure}[b]{.32\textwidth}
\caption{$\beta = 0.5$}
\includegraphics[width=.99\linewidth]{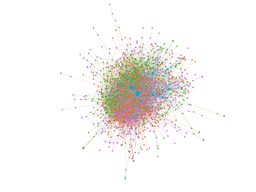}
\end{subfigure}
\end{figure*}
\begin{figure*}[h!]
\caption{Example of actual graphs generated by FARZ, used in the previous plots, $\alpha = 0.5, \; \gamma = 0.5, \; m = 7,\; k=20$.}
\begin{subfigure}[b]{.32\textwidth}
\caption{$\beta = 1$}
\includegraphics[width=.99\linewidth]{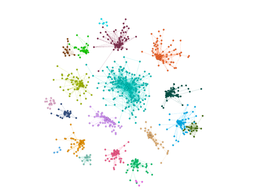}
\end{subfigure}
\begin{subfigure}[b]{.32\textwidth}
\caption{$\beta = 0.95$}
\includegraphics[width=.99\linewidth]{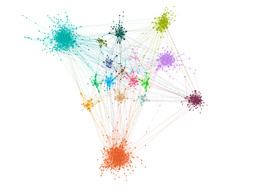}
\end{subfigure}
\begin{subfigure}[b]{.32\textwidth}
\caption{$\beta = 0.9$}
\includegraphics[width=.99\linewidth]{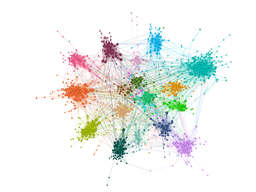}
\end{subfigure}
\begin{subfigure}[b]{.32\textwidth}
\caption{$\beta = 0.85$}
\includegraphics[width=.99\linewidth]{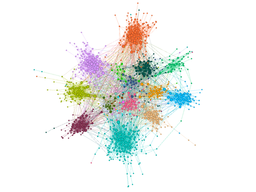}
\end{subfigure}
\begin{subfigure}[b]{.32\textwidth}
\caption{$\beta = 0.8$}
\includegraphics[width=.99\linewidth]{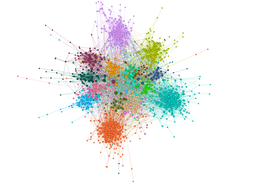}
\end{subfigure}
\begin{subfigure}[b]{.32\textwidth}
\caption{$\beta = 0.75$}
\includegraphics[width=.99\linewidth]{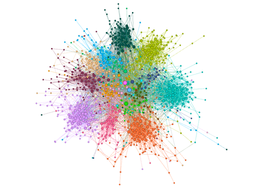}
\end{subfigure}
\begin{subfigure}[b]{.32\textwidth}
\caption{$\beta = 0.7$}
\includegraphics[width=.99\linewidth]{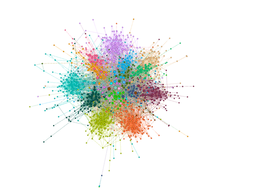}
\end{subfigure}
\begin{subfigure}[b]{.32\textwidth}
\caption{$\beta = 0.65$}
\includegraphics[width=.99\linewidth]{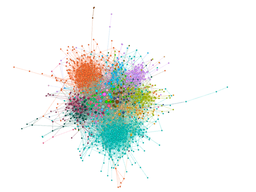}
\end{subfigure}
\begin{subfigure}[b]{.32\textwidth}
\caption{$\beta = 0.6$}
\includegraphics[width=.99\linewidth]{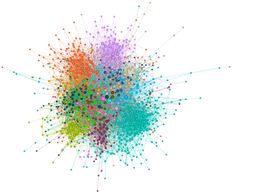}
\end{subfigure}
\begin{subfigure}[b]{.32\textwidth}
\caption{$\beta = 0.55$}
\includegraphics[width=.99\linewidth]{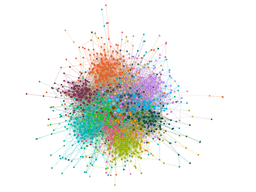}
\end{subfigure}
\begin{subfigure}[b]{.32\textwidth}
\caption{$\beta = 0.5$}
\includegraphics[width=.99\linewidth]{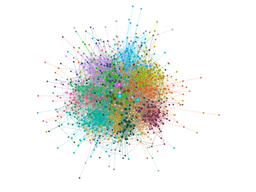}
\end{subfigure}
\end{figure*}


\begin{figure*}[h!]
\centering
\caption{Comparing performance of community mining algorithms on  benchmarks with \textbf{overlapping communities}, plotted as a function of the number of communities each node can belong to. All methods perform poorly, for when nodes are all overlapping.}
\begin{subfigure}[b]{1\textwidth}
\caption{$\alpha = 0.5, \; \gamma = 0.5, \; \beta = 0.8$}
\includegraphics[trim={1cm 0.1cm 0cm 0cm},clip,width=.6\linewidth]{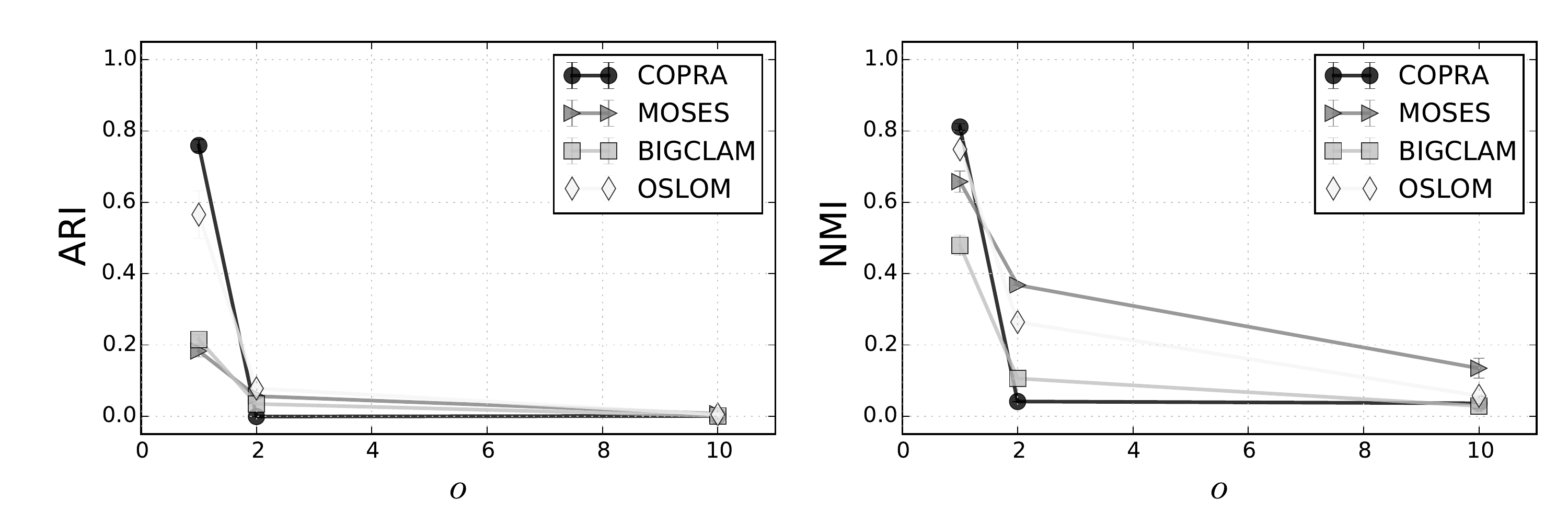}
\includegraphics[trim={0cm 0.1cm 0cm 0cm},clip,width=.3\linewidth]{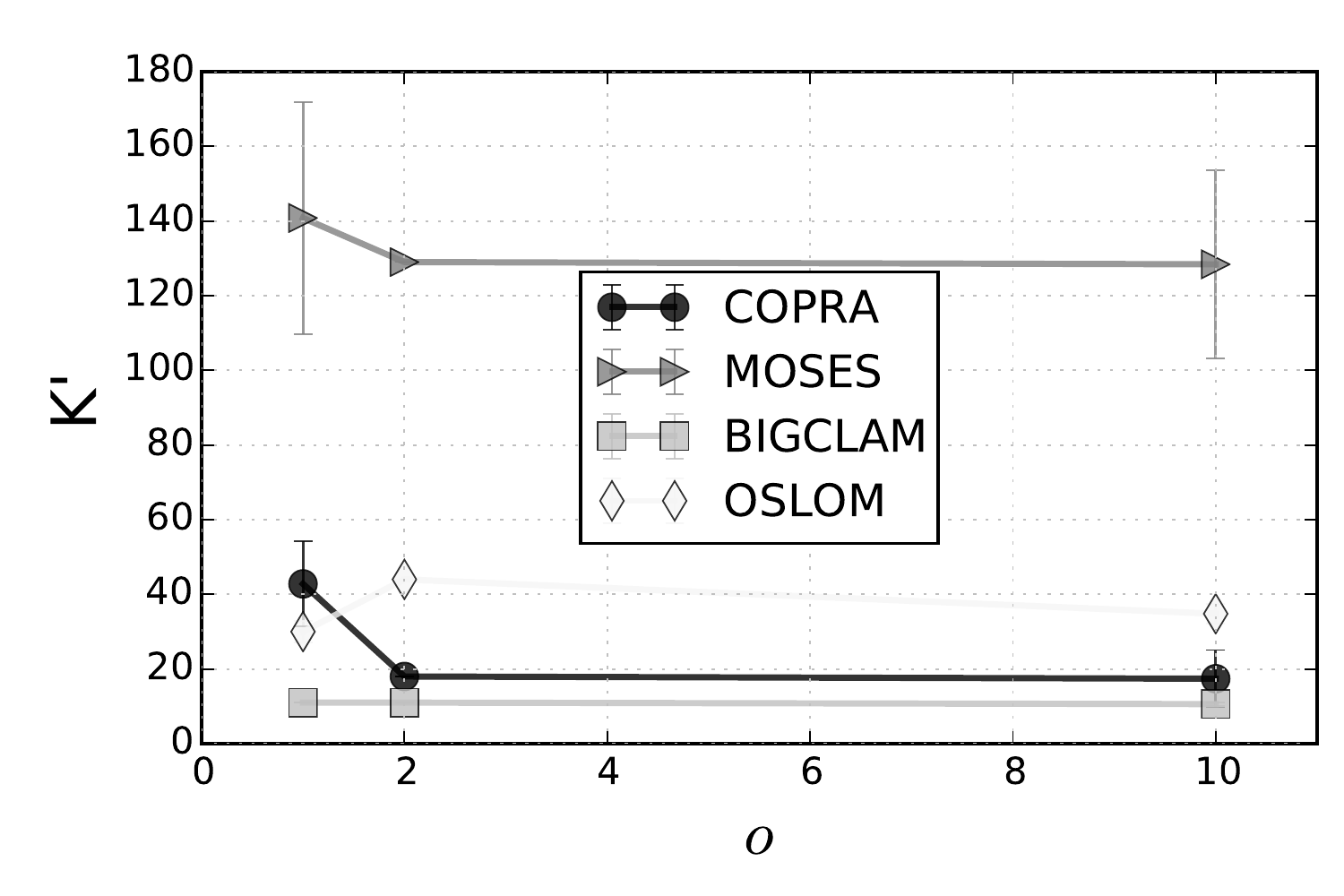}
\end{subfigure}
\begin{subfigure}[b]{1\textwidth}
\caption{$\alpha = 0.5, \; \gamma = 0.5, \; \beta = 0.9$}
\includegraphics[trim={1cm 0.1cm 0cm 0cm},clip,width=.6\linewidth]{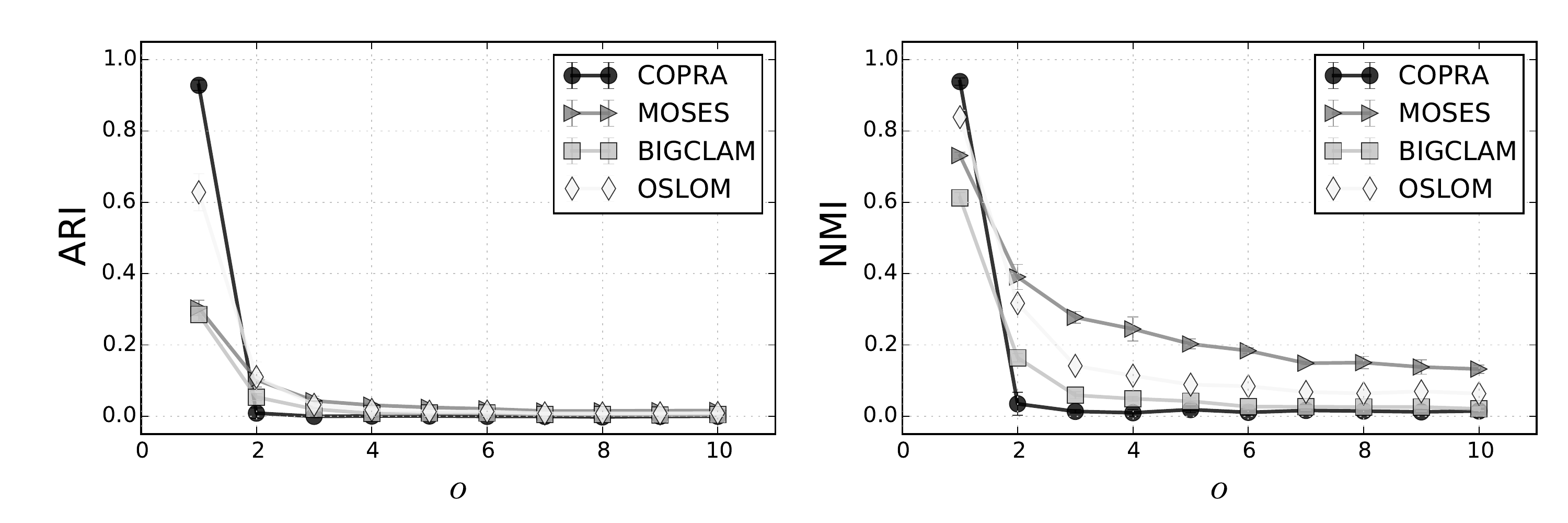}
\includegraphics[trim={0cm 0.1cm 0cm 0cm},clip,width=.3\linewidth]{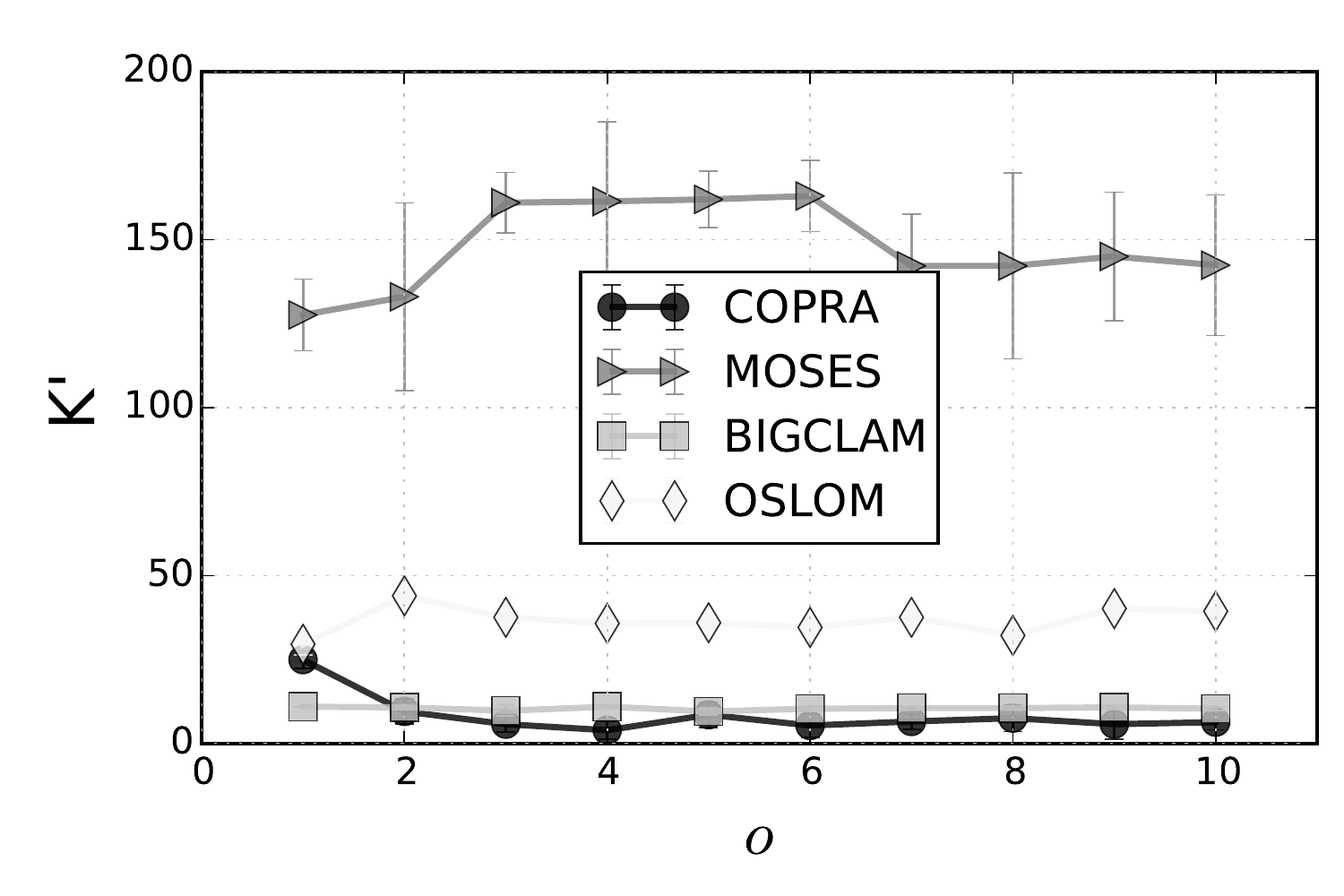}
\end{subfigure}
\begin{subfigure}[b]{1\textwidth}
\caption{$\alpha = 0.2, \; \gamma = -0.8, \; \beta = 0.8$}
\includegraphics[trim={1cm 0.1cm 0cm 0cm},clip,width=.6\linewidth]{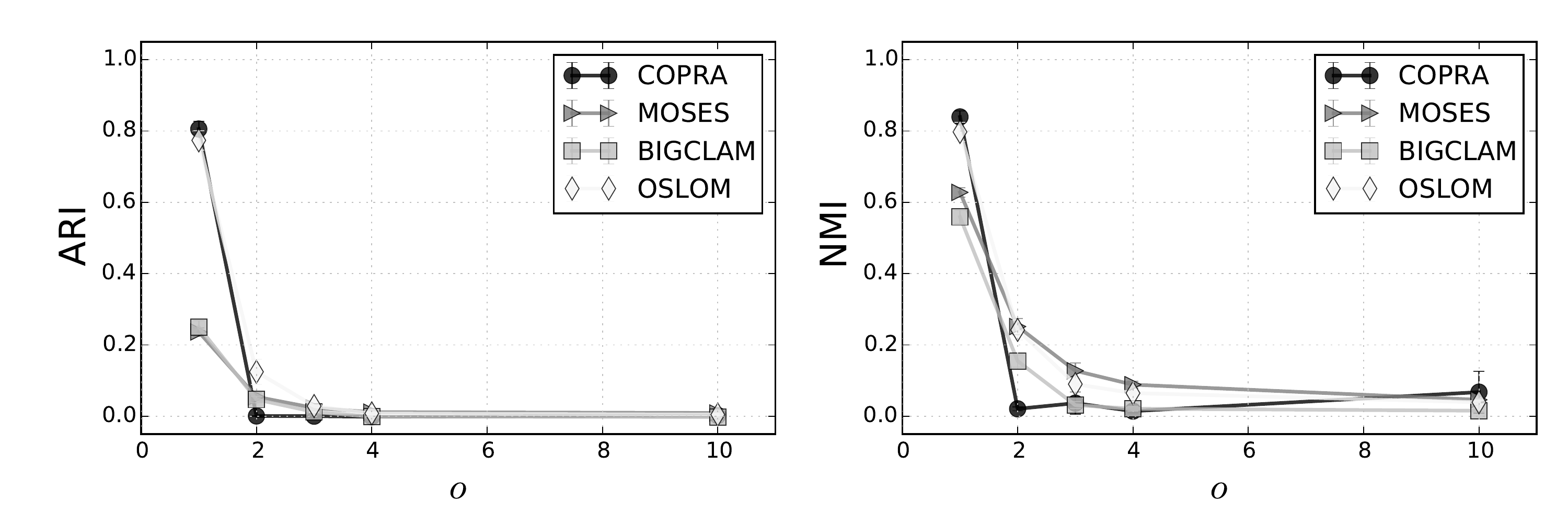}
\includegraphics[trim={0cm 0.1cm 0cm 0cm},clip,width=.3\linewidth]{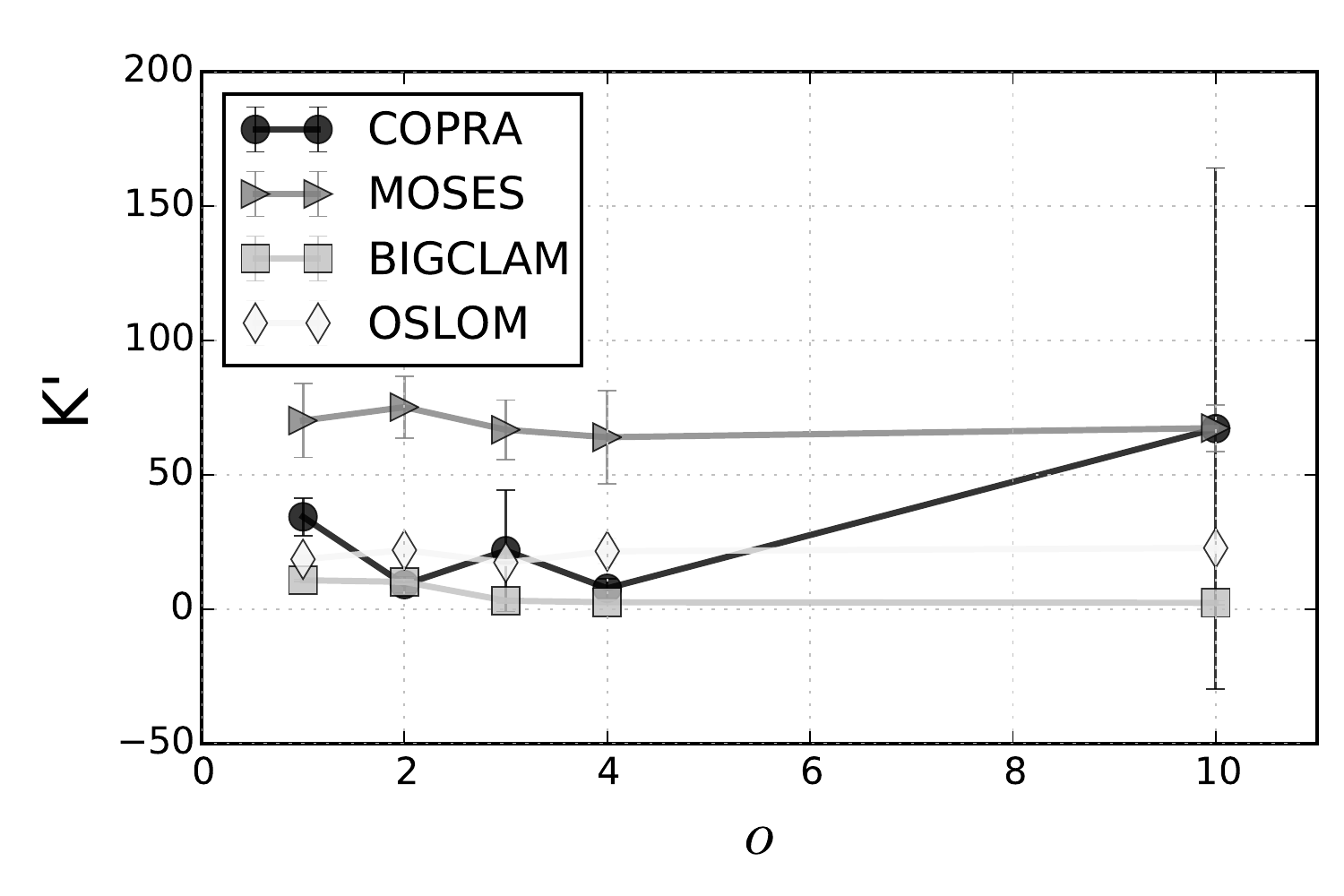}\end{subfigure}
\begin{subfigure}[b]{1\textwidth}
\caption{$\alpha = 0.2, \; \gamma = -0.8, \; \beta = 0.9$}
\includegraphics[trim={1cm 0.1cm 0cm 0cm},clip,width=.6\linewidth]{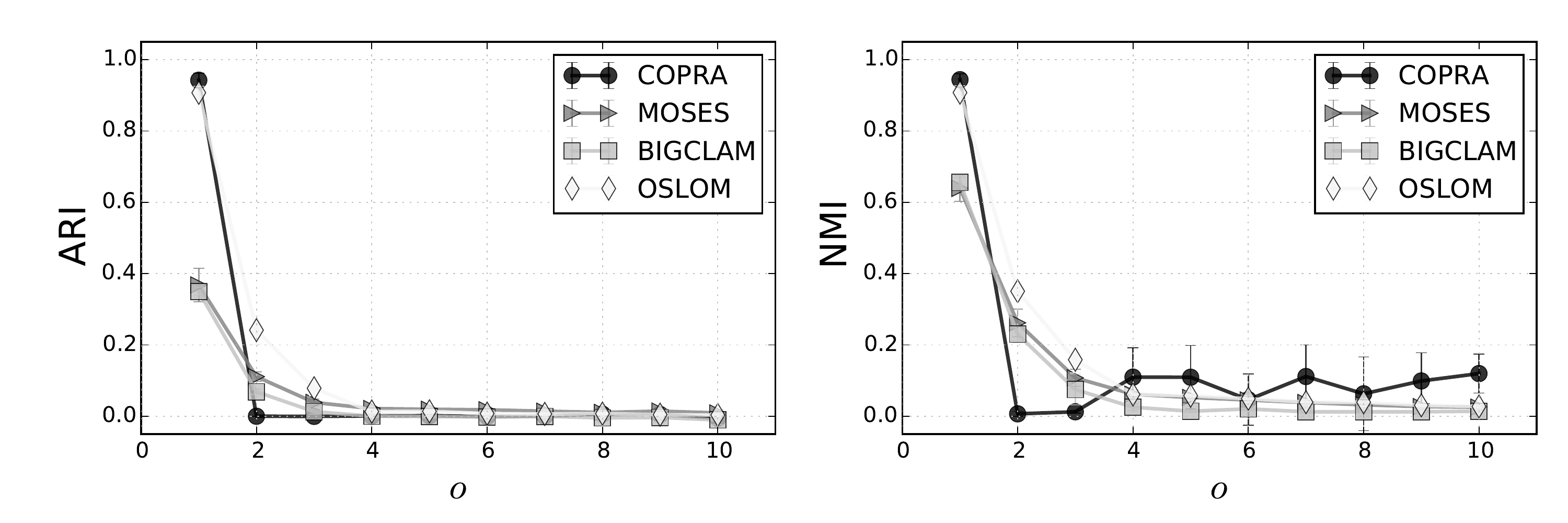}
\includegraphics[trim={0cm 0.1cm 0cm 0cm},clip,width=.3\linewidth]{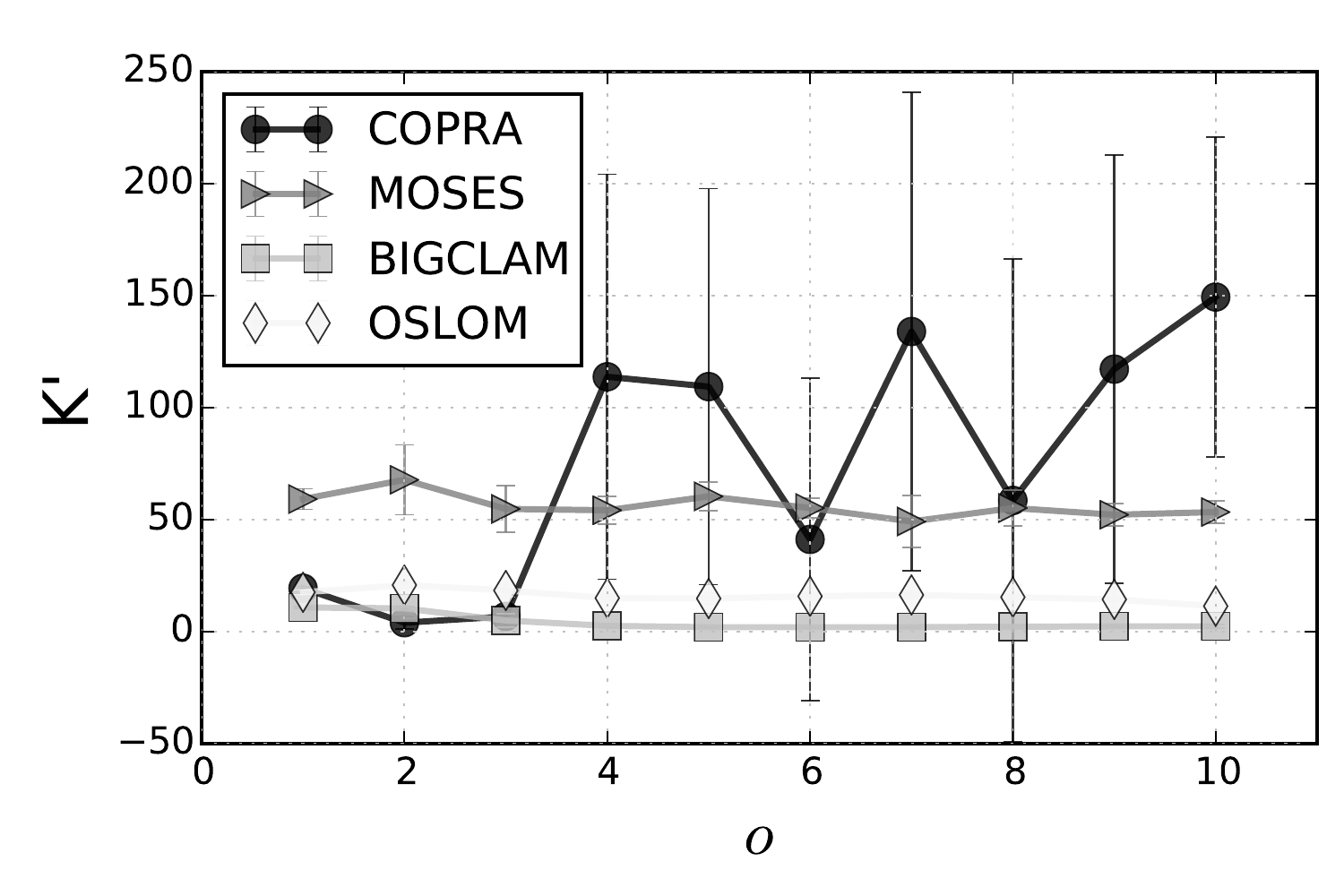}
\end{subfigure}
\end{figure*}

\begin{figure*}[h!]
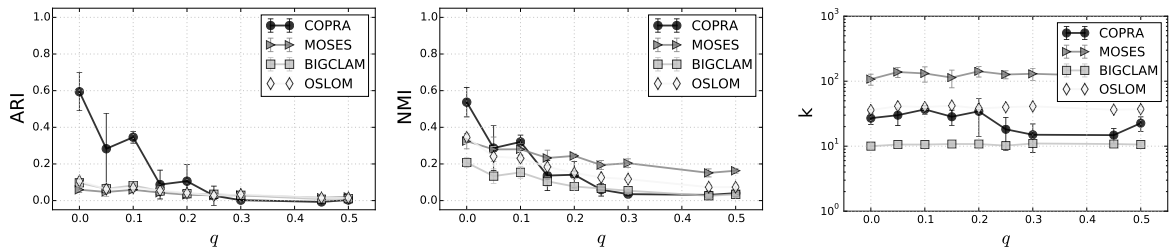

\caption{\textbf{Overlapping communities}, for setting (a) in Figure 11, where the number of communities that each node can belong to is fixed to 3, and the portion of overlapping nodes ($q$) is varied from 0.0 (no overlap), to 0.5 (half of the nodes are overlapping). Again all methods perform poorly, except COPRA, which is able to detect communities when the portion of overlapping nodes is small enough, i.e. $q<0.2$. }
\includegraphics[trim={1cm 0.1cm 0cm 0cm},clip,width=.6\linewidth]{figs/resvq55}
\includegraphics[trim={0cm 0.1cm 0cm 0cm},clip,width=.3\linewidth]{figs/resvq55k}
\end{figure*}

\begin{figure*}[h!]
\caption{Comparing \textbf{LocalT}, \textbf{LocalCM}, and \textbf{TopLeaders}  methods on FARZ benchmarks. }
\begin{subfigure}[b]{1\textwidth}
\caption{$\alpha = 0.5, \; \gamma = 0.5, \; \beta = 0.8, \; m = 5$}
\includegraphics[trim={0cm 0cm 0cm 0cm},clip,width=.3\linewidth]{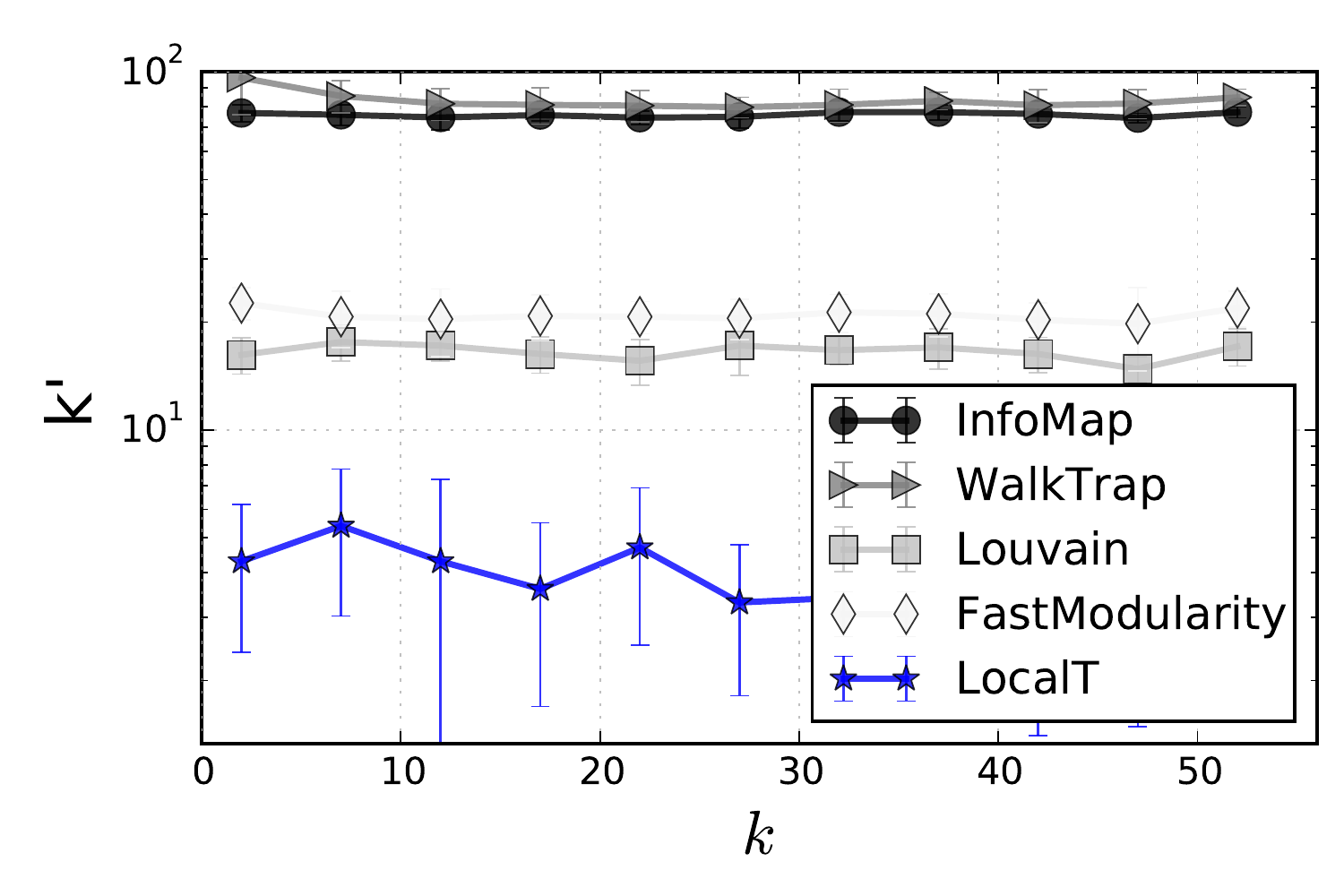}
\includegraphics[trim={0cm 0cm 0cm 0cm},clip,width=.6\linewidth]{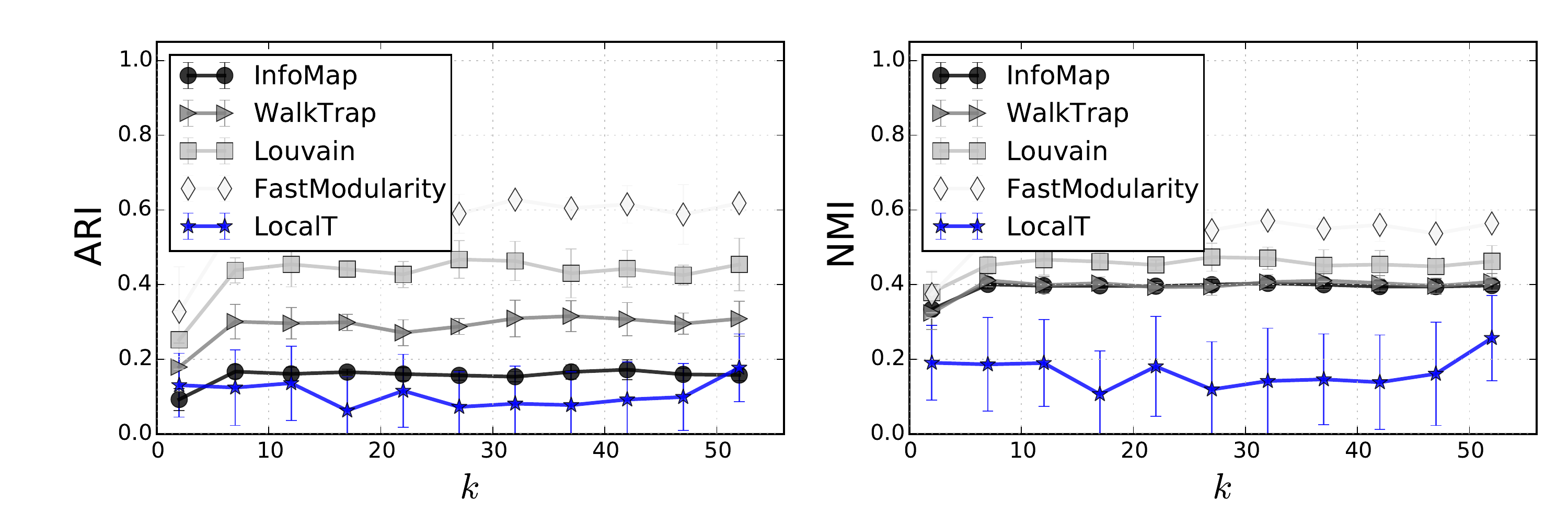}\\
\end{subfigure}
\begin{subfigure}[b]{1\textwidth}
\caption{$\alpha = 0.5, \; \gamma = 0.5, \; m = 5,\; k=4$}
\includegraphics[trim={0cm 0cm 0cm 0cm},clip,width=.3\linewidth]{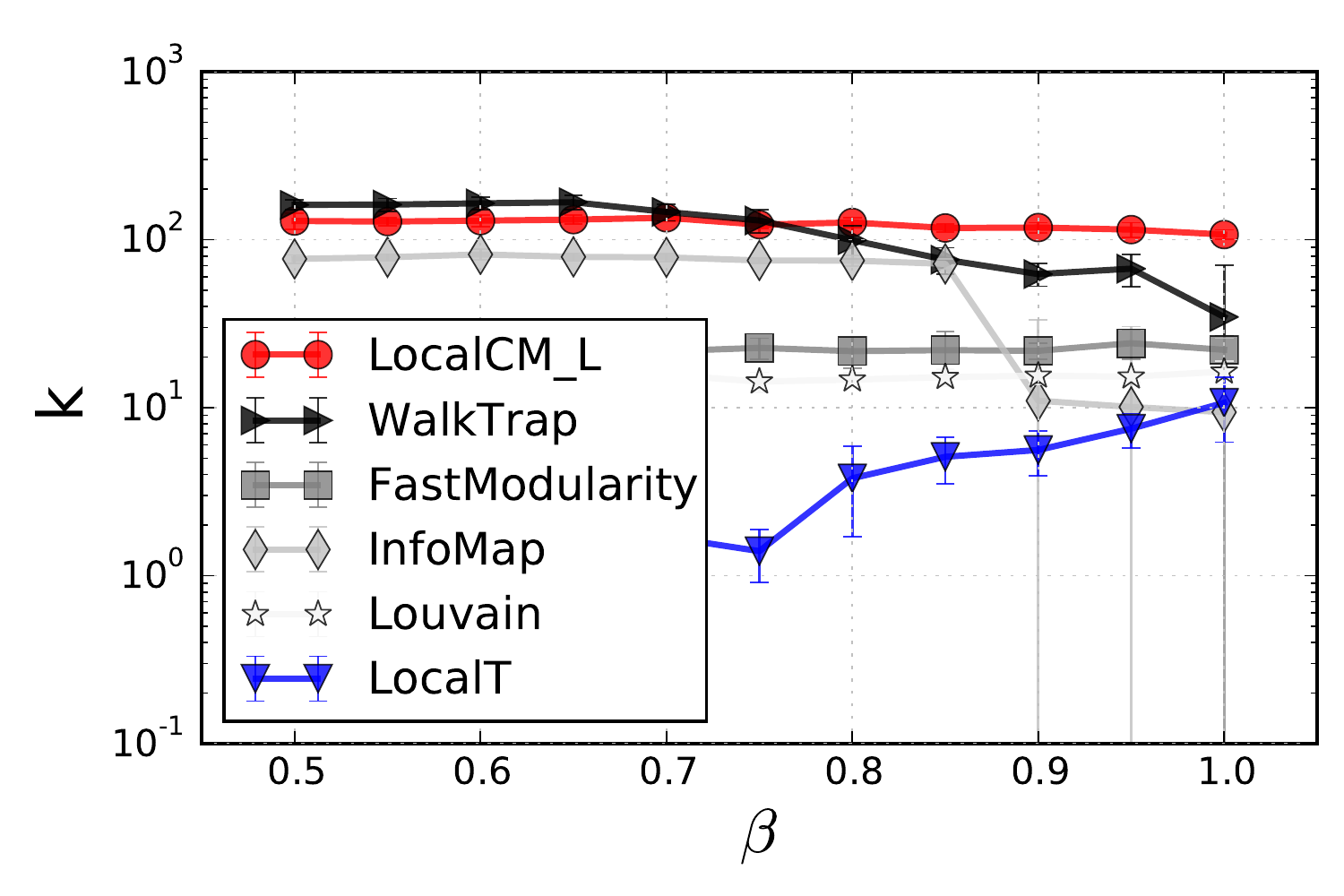}
\includegraphics[trim={0cm 0cm 0cm 0cm},clip,width=.6\linewidth]{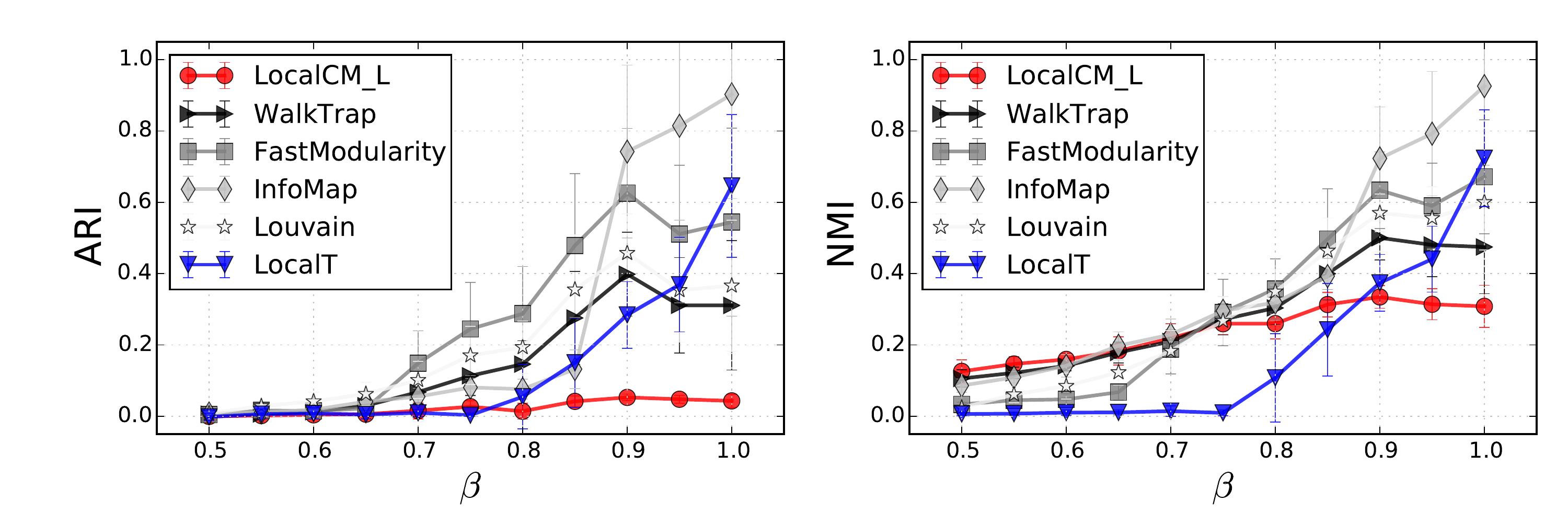}\\
\end{subfigure}
\begin{subfigure}[b]{1\textwidth}
\caption{$\alpha = 0.5, \; \gamma = 0.5, \; m = 5,\; k=4$}
\includegraphics[trim={0cm 0cm 0cm 0cm},clip,width=.3\linewidth]{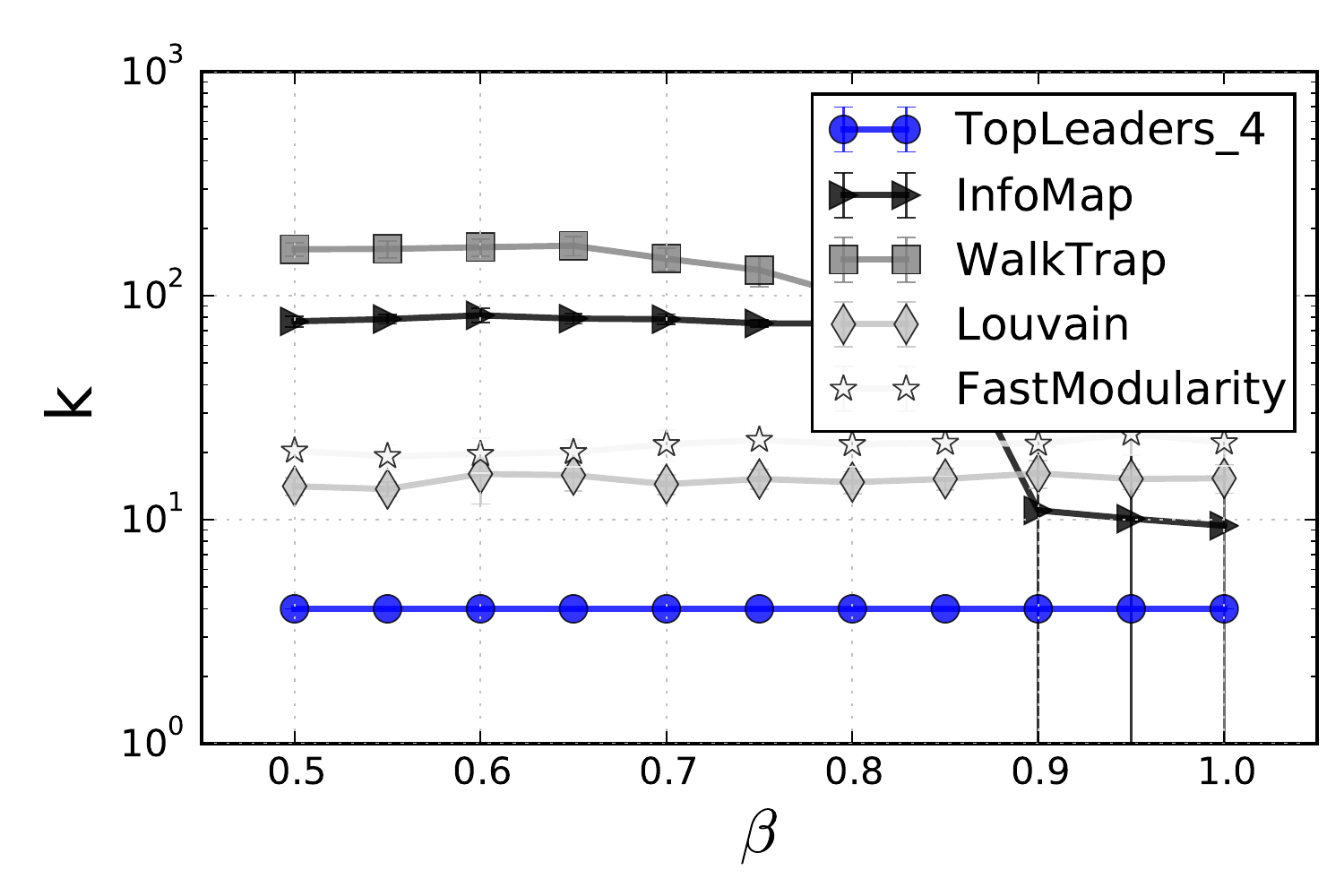}
\includegraphics[trim={0cm 0cm 0cm 0cm},clip,width=.6\linewidth]{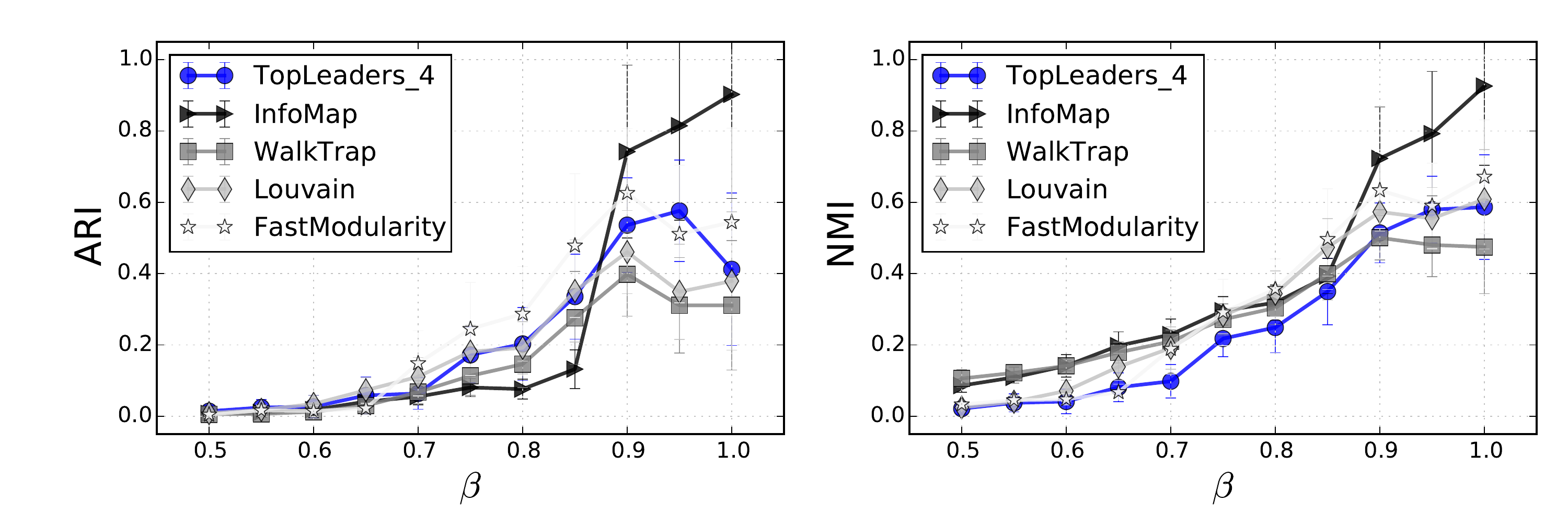}\\
\end{subfigure}
\begin{subfigure}[b]{1\textwidth}
\caption{$\alpha = 0.2, \; \gamma = -0.8, \; m = 5,\; k=4$}
\includegraphics[trim={0cm 0cm 0cm 0cm},clip,width=.3\linewidth]{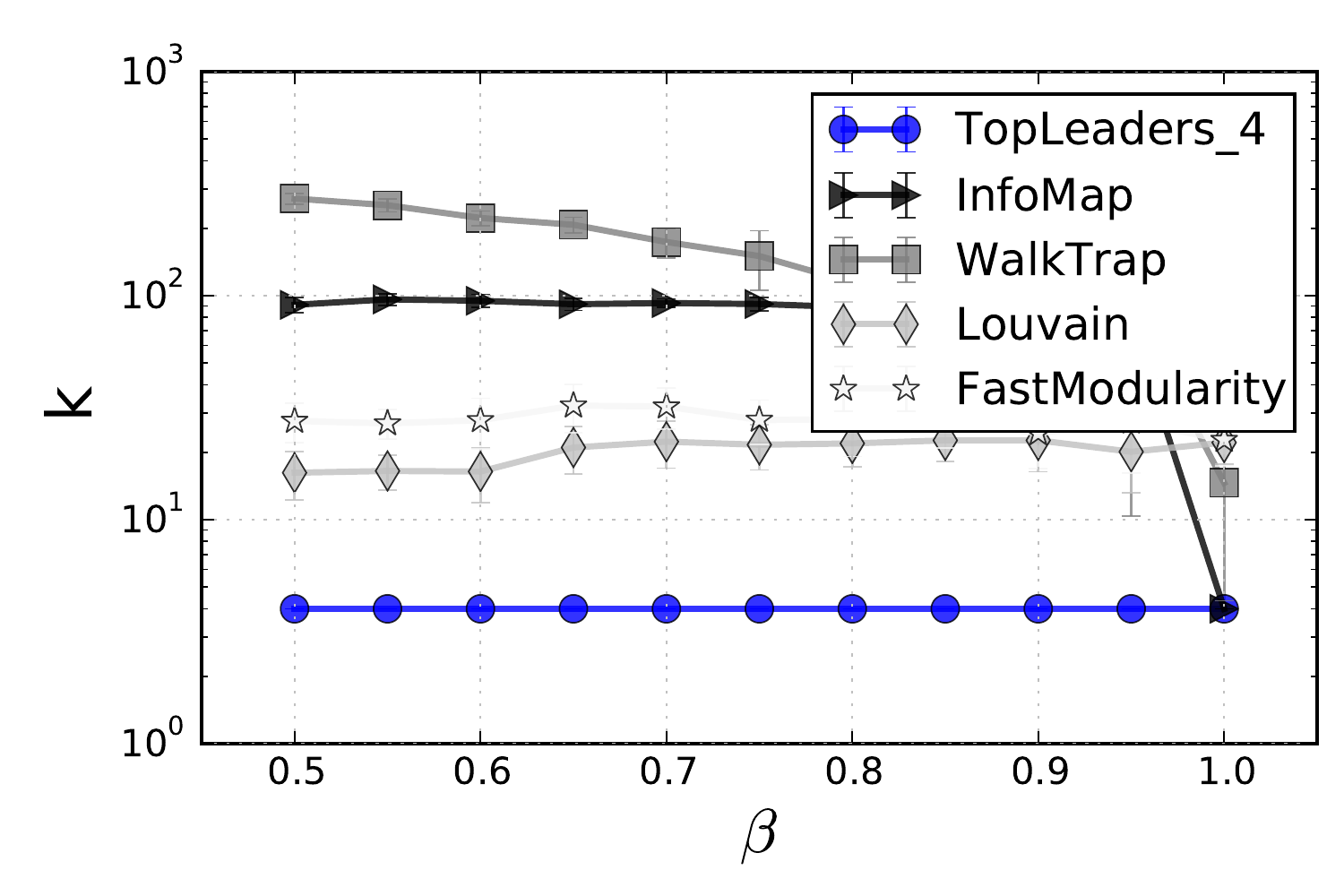}
\includegraphics[trim={0cm 0cm 0cm 0cm},clip,width=.6\linewidth]{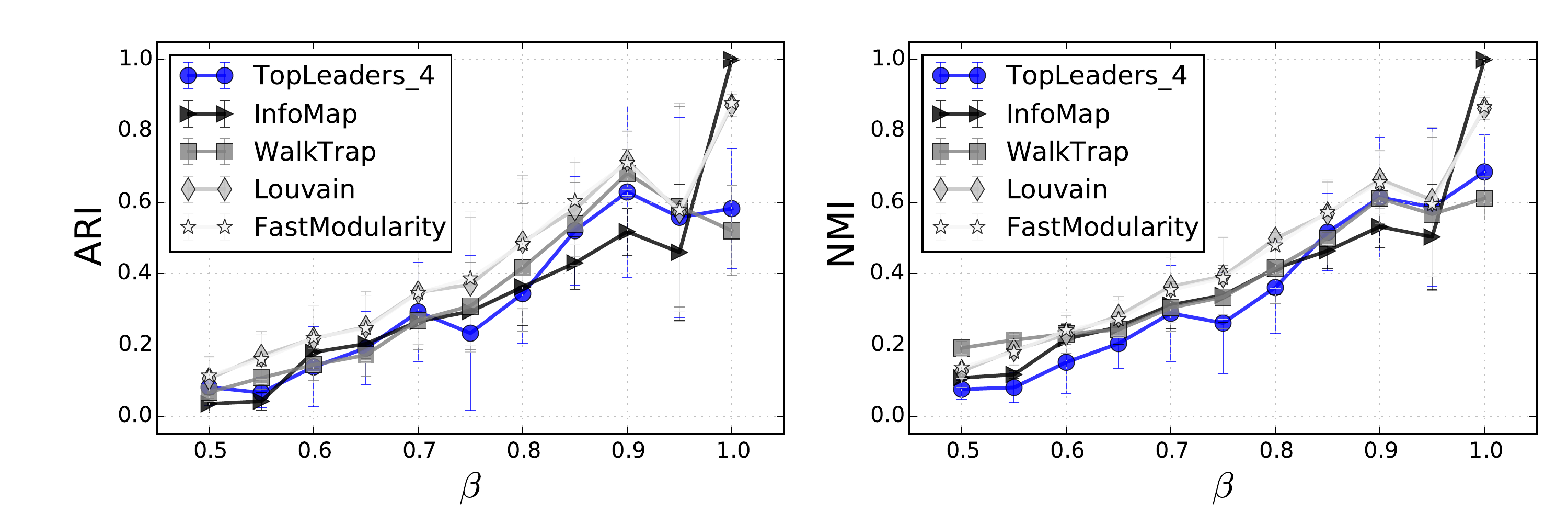}\\
\end{subfigure}
\begin{subfigure}[b]{1\textwidth}
\caption{$\alpha = 0.5, \; \gamma = 0.5, \; m = 7,\; k=20$}
\includegraphics[trim={0cm 0cm 0cm 0cm},clip,width=.3\linewidth]{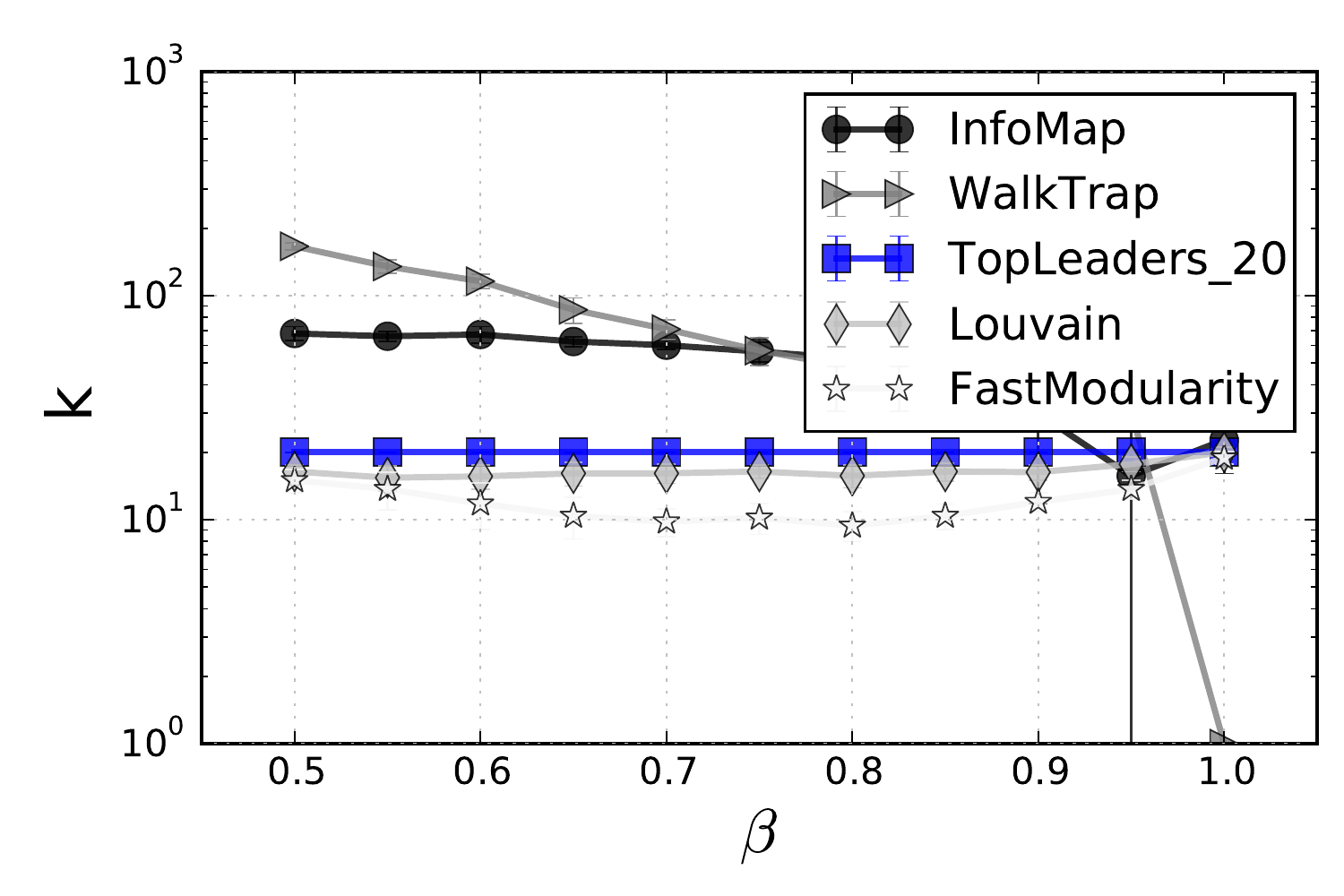}
\includegraphics[trim={0cm 0cm 0cm 0cm},clip,width=.6\linewidth]{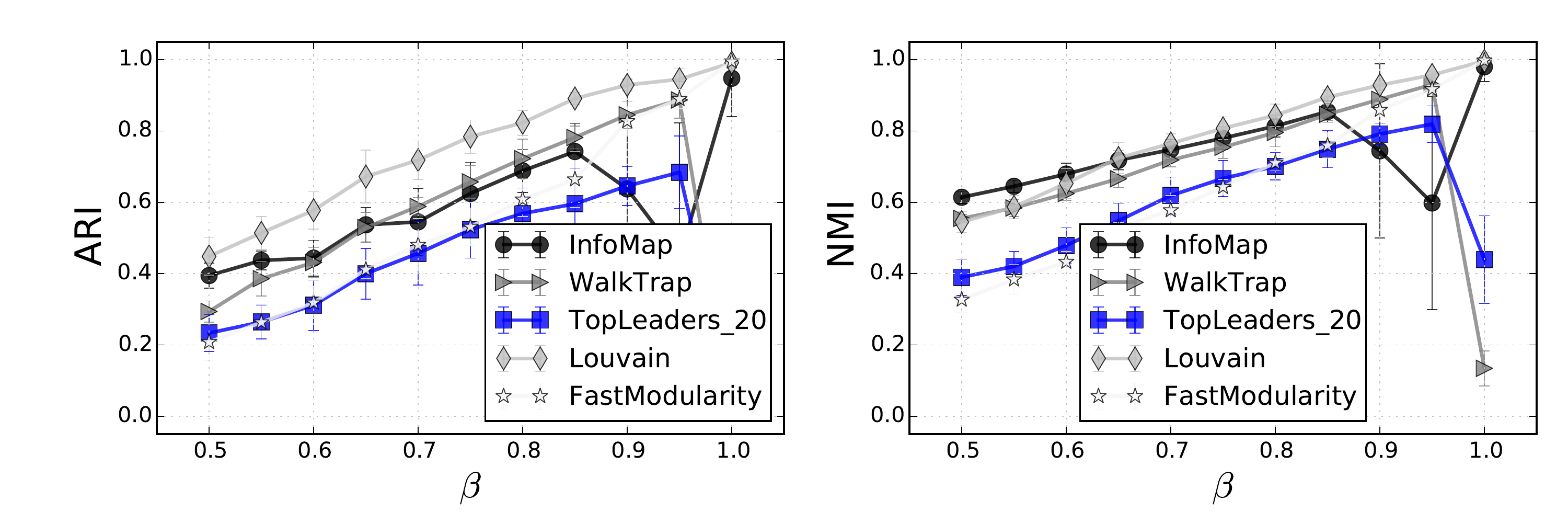}
\end{subfigure}
\end{figure*}